\documentstyle[bezier,12pt]{article}

\newcommand{\moreless}{\mbox{\scriptsize
\raisebox{1.8pt}{\raisebox{1.8pt}{$\scriptscriptstyle>$}
\raisebox{-1pt}{$\scriptscriptstyle\!\!\!\!\!\!<$}}}}      

\newcommand{\lessmore}{\mbox{\scriptsize
\raisebox{1.8pt}{\raisebox{-1pt}{$\scriptscriptstyle>$}
\raisebox{1.8pt}{$\scriptscriptstyle\!\!\!\!\!\!<$}}}}       

\newcommand{\more}{\mbox{\tiny$\scriptscriptstyle >$}}      

\newcommand{\less}{\mbox{\tiny$\scriptscriptstyle <$}}      

\begin{document}

\setcounter{equation}{0}

\title{\sf The ``Mixed'' Green's Function Approach to 
Quantum Kinetics with Initial Correlations}

\author{V.G. Morozov\thanks{Permanent address: 
Department of Physics, Moscow Institute
of Radioengineering, Electronics and Automation,
Vernadsky Prospect 78, 117454, Moscow, Russia
}\ 
\ and G. R\"opke
\\
{\small\em Department of Physics, Rostock University, 
18051~Rostock, Germany}
}

\date{}
\maketitle

{\small
A method for deriving quantum kinetic equations with initial
correlations is developed on the basis of the nonequilibrium 
Green's function formalism. 
The method is applicable to a wide range of correlated initial states
described by nonequilibrium statistical
thermodynamics. Initial correlations and the real-time evolution are 
treated by a unified technique employing many-component
``mixed'' Green's functions. 
The Dyson equation for the mixed Green's 
function leads to a set of equations for
real-time Green's functions and new (cross) components linking
initial correlations with dynamical processes. 
These equations are used to formulate
a generalized Kadanoff-Baym ansatz  for
correlated initial states. A non-Markovian short-time kinetic equation is
derived within the $T$-matrix approximation for the 
self-energies.
The properties of the memory kernels in this equation are
considered in detail in Born approximation for the 
$T$-matrices. 
The kinetic equation is demonstrated to
conserve the total energy of the system. An explicit
expression for the time-dependent correlation energy
is obtained.             
}

\vspace*{20pt}

\setcounter{equation}{0}
\renewcommand{\theequation}{1.\arabic{equation}}

\begin{center}
{I. {\large I}NTRODUCTION} 
\end{center}

The problem of memory effects and initial correlations in nonequilibrium
systems is as old as transport theory. In 1960s, 
some general aspects of this problem were studied on the basis of
fundamental principles of statistical mechanics, and formally exact
equations for the one-particle distribution function 
have been derived, involving initial correlations and 
memory~\cite{Zwanzig60}\,--\,\cite{Balescu68}.
However,  since a many-particle system ``forgets'' irrelevant
details of its initial state, the short-time memory effects and 
the evolution of initial correlations 
were long thought to be purely theoretical problems related to
the justification of Boltzmann-like kinetic equations and hydrodynamic
equations. Therefore, no serious attempts were made to construct
explicit kinetic equations describing the early stage of
evolution of a nonequilibrium system. 
Now this topic has come to more practical 
importance due to experimental studies of fast relaxation processes
caused by the interaction of very short laser pulses with
matter~\cite{Shan96,HaugJauho96}.
Another field of high interest in short-time quantum kinetics
with initial correlations
are nuclear collisions~\cite{Danielewicz84b}\,--\,\cite{Kohler96}.

Although some interesting facts showing the interplay between
collisions and correlations in a many-body system were noticed 
many years ago 
(see, e.g,~\cite{Lee70}), a detailed investigation of
short-time kinetics with initial correlations has been undertaken
only recently by Kremp {\em et al.}~\cite{Kremp97,Bonitz98}
using the density matrix approach. 
In these works, the quantum BBGKY hierarchy for the reduced
density matrices is truncated at the three-particle level.
This allows one to consider the dynamics of two-particle correlations
on the same footing as kinetic processes described by
the one-particle distribution. 
Numerical calculations~\cite{Bonitz98}
show that if the truncation procedure is consistent with
some necessary conditions, say 
the conservation of the total energy of the system, then
the density matrix approach gives physically reasonable results for
the time  behavior of the Wigner function and the 
correlation energy.
The advantage of this method is that one is dealing with
the single-time reduced density matrices which are the natural
ingredients of kinetic theory. On the other hand, using the
truncation procedure, one has from the beginning to
work with approximate equations. Then the problem arises
how to justify and improve these equations in a systematic way.

It is well known that the real-time Green's function formalism,
which to a large extend is based on the fundamental ideas
of Kadanoff and Baym~\cite{KadanoffBaym62} and 
Keldysh~\cite{Keldysh64},
provides an alternative to the density matrix approach to
kinetic theory. Within this formalism, the kinetic equation
can be derived as a component of the exact Dyson equation
on the Keldysh directed contour in the complex time plane.
This is one of the advantages of
the Green's function formalism since the structure of the
kinetic equation follows directly from the Dyson equation
and all approximations are only introduced for the self-energy
which may be regarded as a result of partial summation
of the BBGKY hierarchy.  In the literature several 
generalizations of the Kadanoff-Baym-Keldysh formalism
to arbitrary initial states are available. The most successful
attempts are due to Hall~\cite{Hall75a,Hall75b}, Kukharenko and 
Tikhodeev~\cite{Kukharenko82}, Danielewicz~\cite{Danielewicz84a},
Wagner~\cite{Wagner91}, and Semkat {\em et al.}~\cite{SemkatKremp99}.
Physically, all these approaches are equivalent, but they differ in the
description of initial states and in mathematical formulation.
Up to now the main goal of such theories was to derive generalized
Kadanoff-Baym equations  or to construct a modified 
Keldysh diagram technique
for the Green's functions in the presence of initial correlations.
However, transport theory is conveniently formulated in terms
of a  kinetic equation for the Wigner function. 
The question then arises how to go over from the system of
equations for the {\em double-time\/} Green's functions
to a closed kinetic equation for the {\em single-time\/}
Wigner function.
This question arises in any version of the
real-time Green's function formalism and is known as
the {\em reconstruction problem\/} or the 
{\em problem of ansatz\/}. The task is to 
express the double-time correlation functions in terms
of the Wigner function. 
Within the framework of the standard Kadanoff-Baym formalism 
(i.e., for non-correlated initial states), Lipavsk\'y 
{\em et al.}~\cite{Lipavsky86} were able to derive exact
integral equations which allow one, in principle,    
to solve the reconstruction problem by iteration.
Based on these equations, they also formulated 
a simple, but rather successful relation 
between the time correlation
functions and the Wigner function, 
which  is called the {\em generalized Kadanoff-Baym (GKB) ansatz\/}  
and now is extensively used in quantum kinetic 
theory (see, e.g.~\cite{HaugJauho96}). 
Clearly, for applications of the Green's function formalism
to short-time quantum kinetics 
the reconstruction problem should be solved with taking account
of initial correlations. 
This is, however, a difficult task since 
the derivation of the GKB ansatz~\cite{Lipavsky86} 
rests directly on the Dyson equation for the path-ordered Green's
function, which in turn implies validity of Wick's theorem.
For arbitrary initial states, Wick's theorem is not valid
so that the real-time Green's function does not obey the
Dyson equation. The possibility of deriving the Dyson equation
for arbitrary initial states by introducing
path-ordered Green's functions on an extended Keldysh
contour  was noted by Danielewicz~\cite{Danielewicz84a} and then used by
Wagner~\cite{Wagner91} to formulate a diagram expansion of
such Green's functions. Unfortunately, this scheme
has not been worked out to an extent that explicit kinetic
equations can be derived.

The purpose of this paper is to develop a  Green's function approach to
short-time quantum kinetics, which applies to arbitrary 
initial  states
and provides the basis for the derivation of explicit kinetic equations. 
We shall show the theory ``in action'' and give some examples
of such kinetic equations.
The important ingredient
of our approach is the extension of the GKB ansatz~\cite{Lipavsky86}
to correlated initial states.  

The paper is outlined as follows. In Sec.~II, we 
touch briefly on the description of initial correlated states
and discuss a link between this problem and nonequilibrium
statistical thermodynamics. 
We also introduce the 
{\em thermodynamic Green's functions\/} describing initial
correlations. 

In Sec.~III, the path-ordered {\em mixed Green's functions\/} 
are defined on the deformed Keldysh contour in the $(x,t)$-plane, 
where the $x$-variable is associated with the ``imaginary evolution''
caused by initial correlations  and the $t$-variable corresponds
to the real-time evolution governed by the Hamiltonian.
The mixed Green's functions have different components depending on 
the branch of the contour on which the arguments are situated. 
In addition to the thermodynamic 
Green's functions and the real-time Green's functions, 
the formalism involves new auxiliary objects, namely the 
{\em cross Green's functions\/} with different types of evolution.    
The introduction of these functions is 
necessary to ensure that the full mixed Green's function obeys 
the Dyson equation on the extended contour. In many 
features our scheme is similar to those considered by
Danielewicz~\cite{Danielewicz84a} and Wagner~\cite{Wagner91}.
However, we do not refer directly to the diagram technique
because the Feynman rules for evaluating Green's functions depend on the
initial statistical ensemble describing the system.
This is why we prefer to formulate the theory in an algebraic
language, where the self-energies are related to many-particle 
Green's functions.

In Sec.~IV the matrix Dyson equation for the mixed Green's function 
is used to derive 
a coupled system of equations for the real-time and cross Green's 
functions. In particular, we obtain generalized 
Kadanoff-Baym equations including contributions from initial 
correlations. These equations are shown to be equivalent to  
analogous equations derived previously by diagram
expansions~\cite{Danielewicz84a}.
  
Section~V is concerned with the problem of reconstruction of
two-time correlation functions from the  Wigner
function.  In the case under
consideration, this problem is closely allied to
the evolution of initial correlations. 
Starting from the Dyson equation for the mixed Green's function,     
we derive a set of exact integral equations providing
a generalization of the  reconstruction 
procedure developed by Lipavsk\'y {\em et al.}~\cite{Lipavsky86}. 

In Sec.~VI we consider the $T$-matrix approximation 
for the mixed Green's functions and obtain explicit expressions for
the components of the self-energy  in terms of
the $T$-matrices. An essential point is that there exist 
exact relations between the cross components of the $T$-matrix
and its real-time components. These relations are used to
eliminate the cross components and derive a
generalized optical theorem including the contribution from
initial correlations.

Then, in Sec.~VII we derive for the one-particle distribution
function a short-time kinetic equation in the  $T$-matrix approximation.
The collision integral shows retardation and contains
extra terms due to initial correlations.
As a special case, the properties of the 
collision integral and the spectral function 
are investigated in Born approximation 
for the real-time $T$-matrices. 
We show that the kinetic equation conserves the total
energy and find an expression for the time-dependent
correlation energy. Our results are
compared with results obtained  within the framework 
of the density matrix method.
        
\setcounter{equation}{0}
\renewcommand{\theequation}{2.\arabic{equation}}

\vspace*{15pt}

\begin{center}
II. {\large D}ESCRIPTION OF {\large C}ORRELATED 
{\large I}NITIAL   {\large S}TATES 
\end{center}

\vspace*{5pt}

To put our analysis into more straightforward
language, we shall consider a non-relativistic quantum 
system with the Hamiltonian (we are using units in which $\hbar=1$)
\begin{eqnarray}
& &
\hspace*{-40pt}
\hat H=\hat H^0 +\hat H'=
\frac{1}{2m}\int dr^{}_1\,
\nabla^{}_1\psi^{\dagger}(r^{}_1)\cdot
\nabla^{}_1 \psi(r^{}_1)
\nonumber\\[3pt]
& &
\hspace*{-10pt}
{}+{1\over 2}\int
dr^{}_1\, dr^{}_2\, dr^{\prime}_1\, dr^{\prime}_2\,
\langle r^{\prime}_1 r^{\prime}_2| V |r^{}_1 r^{}_2\rangle\,
\psi^{\dagger}(r^{\prime}_2)
\psi^{\dagger}(r^{\prime}_1)
\psi(r^{}_1)
\psi(r^{}_2),
\label{2.1}
\end{eqnarray} 
where $r^{}_i$ denotes a complete set
of single-particle quantum numbers, say the position
vector ${\bf r}^{}_i$ and the spin variable $\sigma^{}_i$.
Integration over $r$
implies summation over discrete quantum numbers.
The field operators satisfy the usual
commutation relations
\begin{equation}
\psi(r)\psi^{\dagger}(r')-\eta
\psi^{\dagger}(r')\psi(r)=
\delta(r-r'),
\qquad
\psi(r)\psi(r')-\eta
\psi(r')\psi(r)=0,
\label{2.2} 
\end{equation}
where the parameter $\eta$ is equal to $-1$ for fermions
and $+1$ for bosons. From here on the delta function
$\delta(r-r')$
includes the Kronecker delta with respect to
discrete quantum numbers. If, for instance, $r=({\bf r},\sigma)$, then 
$\delta(r-r')=\delta({\bf r}-{\bf r'})\,\delta^{}_{\sigma\sigma'}$.

The second term in the Hamiltonian~(\ref{2.1}) describes pairwise
interactions between the particles. For simplicity, we shall
assume that the potential $v(|{\bf r}^{}_1-{\bf r}^{}_2|)$ does not depend
on spin. In that case the interaction amplitude has the form
\begin{equation}
\langle r^{\prime}_1 r^{\prime}_2| V |r^{}_1 r^{}_2\rangle=
v(|{\bf r}^{}_1-{\bf r}^{}_2|)\,
\delta({r}^{}_1-{r}'_1)\,
\delta({r}^{}_2-{r}'_2).
\label{2.2x}
\end{equation} 
In general, the Hamiltonian contains additional terms
describing interaction of the system with external fields.
Taking the Hamiltonian in the form~(\ref{2.1}), we thus restrict
ourselves to situations where the system is prepared in some initial
(nonequilibrium) state at time $t^{}_0$ and we are interested in the
evolution of the system at times $t>t^{}_0$. However,
the theory can easily be generalized to the case where
the system is subjected into a strong alternating external field.  

In formulating the short-time kinetics, the first thing one must
do is to specify the initial state of the system. This can be
done most directly by a full many-particle statistical operator 
$\varrho(t^{}_0)$, where $t^{}_0$ is an initial time. 
Another possible way to describe the initial state is
by the reduced  $n$-particle density matrices 
$\varrho^{}_{12\cdots n}(t^{}_0)$. 
In the $r$-representation they are defined as
\begin{equation}
\langle r^{}_1\cdots r^{}_n|\varrho^{}_{12\cdots n}(t^{}_0)|
r^{\prime}_1\cdots r^{\prime}_n\rangle=
{\rm Tr}\left\{
\psi^{\dagger}(r^{\prime}_n)\cdots \psi^{\dagger}(r^{\prime}_1) 
\psi^{}(r^{}_1)\cdots
\psi^{}(r^{}_n)\,\varrho(t_0)
\right\}.
\label{2.2zzz}
\end{equation}
Mathematically, these descriptions are equivalent if {\em all\/}
reduced density matrices  are given and both of them
are used in the Green's function formalism with initial correlations. 
It is interesting to note that, in their pioneering 
work~\cite{KadanoffBaym62}, Kadanoff and Baym start
with the {\em correlated\/} state described by
the equilibrium statistical operator
\begin{equation}
\varrho(t^{}_0)=\varrho^{}_{\rm eq}=
{\rm e}^{-\beta(\hat H -\mu \hat N)}
\left/
{\rm Tr}\,{\rm e}^{-\beta(\hat H -\mu \hat N)},
\right.
\label{1.1}
\end{equation}
where $\beta=1/T$ is the inverse temperature and $\mu$ is the
chemical potential. Here the correlation effects are incorporated
through the interaction term in the Hamiltonian $\hat H$. 
However, Kadanoff and Baym did not consider the evolution of 
correlations since their task was quite opposite.
To eliminate the influence of initial correlations,
they take the limit $t^{}_0\to -\infty$ and assume that
the nonequilibrium evolution of the system is caused by some 
auxiliary external field acting on the particles.
Further the most part of works devoted to quantum dynamics
with initial correlations, except for
the works of Danielewicz~\cite{Danielewicz84a} and 
Wagner~\cite{Wagner91}, was based on the description
of the initial state in terms of reduced density matrices. 

In this paper, we will assume that the initial statistical ensemble
is specified by the corresponding statistical operator 
$\varrho(t^{}_0)$. In thermal equilibrium
$\varrho^{}(t^{}_0)$ is given by Eq.~(\ref{1.1}), but if 
the initial state deviates from thermal equilibrium,
the problem of a representative statistical ensemble
becomes nontrivial.  
The knowledge of the Hamiltonian does not suffice to 
determine $\varrho(t^{}_0)$ and, generally speaking, 
one has to consider the past evolution of the 
system at times $t<t^{}_0$. Clearly this is not the problem 
which we intend to study. Therefore, it is reasonable to construct 
the initial statistical operator using a physical information about
the system at $t=t^{}_0$.  
 It is known from
nonequilibrium statistical mechanics that a large class
of nonequilibrium ensembles can be derived from  
Jaynes' maximum entropy principle~\cite{Janes57,Jaynes63} 
which is a straightforward
extension of Gibbs' ensemble method to nonequilibrium systems.
Without going into detailed discussion of this principle and its
wide use in theory of irreversible processes 
(see, e.g.,~\cite{ZubMorRoep1,ZubMorRoep2}), we only
recapitulate here the main ideas  which
will be of importance in our subsequent consideration. 

Let us assume that the state of a many-body system at $t=t^{}_0$ is
characterized by a set of observable quantities (state parameters)
which are  the mean values $\langle\hat P_m\rangle^{t_0}$
of some dynamical variables $\hat P_m$. Following Jaynes' principle, 
the corresponding statistical operator, usually
referred to as the {\em relevant statistical operator}, 
is found from a maximum of the entropy 
$S=-{\rm Tr}\left\{\varrho'\,\ln\varrho'\right\}$  
for given  
$\langle \hat P^{}_m\rangle^{t_0}={\rm Tr}\left\{\varrho'\,\hat P_m\right\}$, 
where $\varrho'$ is a trial statistical operator.
>From the extremum condition $\delta S=0$ one immediately finds
that the relevant statistical operator has the form
\begin{equation}
\varrho(t_0)=
{\rm e}^{-\hat S}\left/
{\rm Tr}\left\{{\rm e}^{-\hat S}\right\}
\right.
\label{1.2}
\end{equation}
with 
\begin{equation}
\hat S=\sum_m \lambda^{}_m\hat P_m.
\label{1.3}
\end{equation}      
The Lagrange multipliers $\lambda^{}_m$ are to be evaluated 
from the self-consistency conditions
\begin{equation}
\langle \hat P^{}_m\rangle^{t_0}= {\rm Tr}\left\{
\varrho(t^{}_0)\,\hat P_m
\right\}
\label{1.4} 
\end{equation}
for given $\langle \hat P_m\rangle^{t_0}$. These conditions
play the role of nonequilibrium equations of state.
The operator $\hat S$ given by Eq.~(\ref{1.3})
 may be called the {\em entropy operator\/} because its expectation 
value determines the entropy of the initial ensemble~\cite{ZubMorRoep1}. 
Clearly, the equilibrium  distributions are a special case
of the relevant statistical operators where the
$\hat P_m$ correspond to additive integrals of motion. 
Note also that the definition of the nonequilibrium
entropy as the entropy in the relevant ensemble
leads to a natural extension of thermodynamic relations
to nonequilibrium states characterized by 
macroscopic fluxes, partial equilibrium in subsystems, 
a nonuniform distribution of local thermodynamic quantities, 
etc.~\cite{ZubMorRoep1}. A more detailed 
description of many-particle correlations is achieved by the use 
of operators $\hat P_m$ corresponding to reduced density
matrices $\varrho^{}_{12\cdots n}$.  
Here we will not go into a discussion of 
the physics contained in the description of many-particle
correlations in terms of statistical operators~(\ref{1.2})
(see, e.g.~\cite{ZubMorRoep1,ZubMorRoep2}).
The only fact necessary for our purposes is that
each of the relevant dynamical variables $\hat P_m$ in the entropy
operator~(\ref{1.3}) can always be written as a cluster 
decomposition in terms of the field operators. 
Thus, the general form of the entropy operator is     
\begin{equation}
\hat S=
\sum_{k\geq 1} {1\over k!}
\int dr^{}_1\cdots dr'_k\,  
\lambda^{}_k(r^{\prime}_1\ldots r^{\prime}_k, r^{}_1\ldots r^{}_k)
\psi^{\dagger}(r'_k)\cdots
\psi^{\dagger}(r'_1)\,
\psi(r^{}_1)\cdots
\psi(r^{}_k).
\label{2.3} 
\end{equation}
The quantities 
$\lambda^{}_k(r^{\prime}_1\ldots r^{\prime}_k, r^{}_1\ldots r^{}_k)$
are some functions of the Lagrange multipliers
in Eq.~(\ref{1.3}). 
To find these functions, one has to specify the
relevant dynamical variables $\hat P_m$. In writing 
Eq.~(\ref{2.3}), we have assumed that the initial statistical
operator commutes with a particle number operator. For superfluids
and superconductors, however, particle number non-conserving terms 
must be included~\cite{Roepke94}. 

The entropy operator~(\ref{2.3}) can be represented as a
sum $\hat S=\hat S^{0} +\hat S'$, where $\hat S^{0}$ is 
bilinear in the field operators ($k=1$), and $\hat S'$
collects all higher-order terms. 
In what follows, we shall illustrate the general theory
assuming the one-particle density matrix 
$\varrho^{}_1(t^{}_0)$ and the two-particle 
density matrix $\varrho^{}_{12}(t^{}_0)$
to be independent state parameters. In that case
the entropy operator has the form      
\begin{eqnarray}
& &
\hspace*{-50pt}
\hat S=\hat S^0+ \hat S'=
\int dr^{}_1\,dr^{\prime}_1\, \lambda^{}_1(r^{\prime}_1,r^{}_1)\,
\psi^{\dagger}(r^{\prime}_1)\psi(r^{}_1)
\nonumber\\[3pt]
& &
\hspace*{-20pt}
{}+{1\over 2}
\int dr^{}_1\,dr^{}_2\,dr^{\prime}_1\,dr^{\prime}_2\,
\lambda^{}_2(r^{\prime}_1 r^{\prime}_2,r^{}_1 r^{}_2)\,
\psi^{\dagger}(r^{\prime}_2)
\psi^{\dagger}(r^{\prime}_1)
\psi^{}(r^{}_1)\psi(r^{}_2),
\label{2.4} 
\end{eqnarray}
where the functions $\lambda^{}_1$ and $\lambda^{}_2$ are determined
by the equations of state
\begin{eqnarray}
& &
\langle r^{}_|\varrho^{}_1(t^{}_0)|r^{\prime}_1\rangle=
{\rm Tr}
\left\{ 
\psi^{\dagger}(r^{\prime}_1)\psi^{}(r^{}_1)\,
\varrho(t_0)
\right\},
\nonumber\\[6pt]
& &
\langle r^{}_1 r^{}_2 |\varrho^{}_{12}(t^{}_0)
|r^{\prime}_1 r^{\prime}_2\rangle= 
{\rm Tr}\left\{
\psi^{\dagger}(r^{\prime}_2)\psi^{\dagger}(r^{\prime}_1) 
\psi^{}(r^{}_1)
\psi^{}(r^{}_2)\,\varrho(t_0)
\right\}.
\label{2.5} 
\end{eqnarray}
Generalization to other entropy operators would be quite 
straightforward.

For the relevant ensembles, it is possible to develop a Green's
function formalism which provides a way of calculating 
many-particle averages and solving nonequilibrium equations of 
state by means of a diagram technique~\cite{ZubMorRoep2}. 
To summarize some basic relations and definitions
(see also~Appendix~A), let us  introduce the ``evolution operator'' 
\begin{equation}
{\cal U}(x,x')={\rm e}^{-(x-x')\hat S},
\label{2.6} 
\end{equation}
where $x$, $x'$ are real variables, and define dynamical variables
$\hat A$ in the ``Heisenberg picture''
\begin{equation}
\hat A^{}_H(x)={\cal U}(0,x)\,\hat A\,{\cal U}(x,0)=
{\rm e}^{x\hat S}\,\hat A\,{\rm e}^{-x\hat S}.
\label{2.7} 
\end{equation}
The one-particle and the $n$-particle 
{\em thermodynamic Green's functions\/} are then defined as
[from here on the symbol $\langle\ldots \rangle$ means averaging with the
initial statistical operator~(\ref{1.2})]
\begin{eqnarray}
& &
\hspace*{-30pt}
{\cal G}(1,1')= - \left\langle
T^{c}_x\left(\psi^{}_H(1)\psi^{\dagger}_H(1')
\right)
\right\rangle,
\label{2.8}\\[7pt] 
& &
\hspace*{-30pt}
{\cal G}^{(n)}(1\cdots n,1'\cdots n')= (-1)^s \left\langle
T^{c}_x\left(\psi^{}_H(1)\cdots\psi^{}_H(n)
\psi^{\dagger}_H(n')\cdots\psi^{\dagger}_H(1')
\right)
\right\rangle,
\label{2.9}  
\end{eqnarray}
where the labels $(k)$ and $(k')$ indicate respectively
$(r^{}_k,x^{}_k)$ and $(r'_k,x'_k)$.
$T^c_x$ is the ``chronological'' ordering operator with
respect to the values of $x$; for Fermi systems, $T^c_x$
has the usual sign convention for permutations of 
the field operators. In the special case of thermal equilibrium,
the entropy operator is 
$\hat S=\beta\left(\hat H -\mu \hat N\right)$, so that
the thermodynamic Green's 
functions~(\ref{2.8}) and~(\ref{2.9}) go over to the well-known 
Matsubara-Green's functions~\cite{FetterWalecka71},
if one introduces a new variable $\tau=\beta x$ 
corresponding in a sense to the imaginary-time evolution.

The one-particle thermodynamic Green's function 
${\cal G}(1,1')$ is a function of the difference $x^{}_1-x'_1$
and satisfies the Dyson equation (see~Appendix~A)
\begin{equation}
\int_{x^{}_0}^{x^{}_0+1} d1''\left\{ {\cal G}^{-1}_0(1,1'')-
{\cal K}(1,1'')\right\}{\cal G}(1'',1')=\delta(1-1'),
\label{2.10} 
\end{equation}  
where 
\begin{equation}
{\cal G}^{-1}_0(1,1')=-\left[
\delta(r^{}_1 -r^{\prime}_1)\frac{\partial}{\partial x^{}_1}
+\lambda^{}_1(r^{}_1,r^{\prime}_1)
\right]\delta(x^{}_1-x^{\prime}_1)
\label{2.11} 
\end{equation}
is the generator of ``free evolution'', and 
${\cal K}(1,1')$ is the {\em thermodynamic self-energy}.
The value of $x^{}_0$ may be prescribed arbitrarily. 
A special choice of this parameter will be discussed below. 
Equation~(\ref{2.10}) and its adjoint can be written in the matrix form 
\begin{equation}
\left({\cal G}^{-1}_0 - {\cal K}\right){\cal G}=I,
\qquad
{\cal G}\left({\cal G}^{-1}_0-{\cal K}\right)=I,
\label{2.12} 
\end{equation}
where the multiplications are defined as 
matrix multiplication with respect to single-particle
quantum numbers plus integration over $x$ in the interval
from $x=x^{}_0$ to $x=x^{}_0+1$. The identity matrix is
$I(1,1')=\delta(1-1')$.

\setcounter{equation}{0}
\renewcommand{\theequation}{3.\arabic{equation}}

\vspace*{15pt}

\begin{center}
III. {\large T}HE ``{\large M}IXED'' {\large G}REEN'S {\large F}UNCTIONS
\end{center}

\vspace*{5pt}

In kinetic theory, the thermodynamic 
Green's functions are auxiliary 
quantities which can be used  for evaluating
characteristics of initial correlations. 
To study the evolution of the system, 
one has to consider real-time
Green's functions. Within the standard
formalism~\cite{Danielewicz84a,Botermans90}, 
the one-particle real-time Green's functions 
of interest can be put into a matrix Green's function
defined on the Keldysh contour $C$ (see Fig.~1):
\begin{equation}
G(1,1')=\left(
\begin{array}{ll}
 G^{++}(1,1') & G^{+-}(1,1')\\
 G^{-+}(1,1') & G^{--}(1,1')
\end{array}
\right)
=\left(
\begin{array}{ll}
 g^{c}(1,1') & g^{\less}(1,1')\\
 g^{\more}(1,1') & g^{a}(1,1')
\end{array}
\right),
\label{3.1} 
\end{equation}
where $(1)=(r^{}_1,t^{}_1)$, $(1')=(r'_1,t'_1)$,
and the field operators are taken in the usual Heisenberg
representation with the Hamiltonian $\hat H$.
The components of $G(1,1')$ are the causal 
Green's function $g^c(1,1')$, the anti-causal 
Green's function $g^a(1,1')$, and the correlation
functions $g^{\moreless}(1,1')$, which are defined as
\begin{eqnarray}
& 
\displaystyle{
g^{c/a}(1,1')=-i\left\langle
T^{c/a}_t\left(\psi^{}_H(1) \psi^{\dagger}_H(1')\right)\right\rangle,
             }
\label{3.2}\\[7pt]
& 
\displaystyle{
g^{\more}(1,1')=
-i\left\langle
\psi^{}_H(1)\psi^{\dagger}_H(1')
\right\rangle,
\qquad
g^{\less}(1,1')=
-i\eta\left\langle
\psi^{\dagger}_H(1')\psi^{}_H(1)
\right\rangle,
}
\label{3.3} 
\end{eqnarray}
where $T^c_t$ ($T^a_t$) is the chronological (anti-chronological)
time ordering operator and the averages are calculated with the
initial statistical operator $\varrho(t^{}_0)$. 
The matrix Green's function~(\ref{3.1}) can also be written in 
a compact form
\begin{equation}
G(1,1')=
-i\left\langle T^{}_C\left(\psi^{}_H(1) \psi^{\dagger}_H(1')\right)
\right\rangle,
\label{3.4} 
\end{equation}
where $T^{}_C$ is the path-ordering operator on the Keldysh
contour $C$. The real-time Green's functions have the symmetry properties
\begin{equation}
\left[g^{c/a}(1,1')\right]^*=-g^{a/c}(1',1),
\qquad
\left[g^{\moreless}(1,1')\right]^*= - g^{\moreless}(1',1),
\label{3.5}  
\end{equation}
which are valid for any initial statistical operator $\varrho(t_0)$. 

The diagonal parts (with respect to time) of the correlation 
functions $g^{\moreless}$ are of special importance in kinetic theory 
due to relations
\begin{equation}
i\eta\,g^{\less}(1,1')\Big|^{}_{t_1=t'_1}= 
\langle r^{}_1|\varrho^{\less}_1(t_1)|r'\rangle,
\qquad
ig^{\more}(1,1')\Big|^{}_{t_1=t'_1}= 
\langle r^{}_1|\varrho^{\more}_1(t_1)|r'\rangle,
\label{3.6}
\end{equation}
where
\begin{eqnarray}
& &
\langle r^{}_1|\varrho^{\less}_1(t)|r'\rangle\equiv
\langle r^{}_1|\varrho^{}_1(t)|r'\rangle=
\left\langle \psi^{\dagger}_H(r'_1t)\psi^{}_H(r^{}_1t)
\right\rangle,
\label{3.8x}\\[6pt]
& &
\langle r^{}_1|\varrho^{\more}_1(t_1)|r'\rangle=
\delta(r_1-r'_1) +\eta \,\varrho^{}_1(r_1,r'_1;t)
\label{3.8xx}
\end{eqnarray}
are the nonequilibrium one-particle density matrices in the $r$-representation.

The crucial point in  the standard real-time Green's function
formalism is that, for a {\em non-correlated
initial state}, the function~(\ref{3.4}) obeys
the Dyson equation on the contour $C$. 
Although in most applications the existence of
the Dyson equation is not considered at all, 
this is a nontrivial fact because it implies that the 
Green's function $G(1,1')$
has the unique inverse function $G^{-1}(1,1')$ on $C$.
The original derivation of the Dyson equation for 
the nonequilibrium real-time Green's functions
was based on a perturbation expansion in the interaction
picture and Wick's theorem~\cite{Keldysh64,Craig68}. 
More recently it was pointed out by Kremp 
{\em et al.}~\cite{Kremp85} that,
in order to justify the existence of the Dyson equation
for $G(1,1')$, one has to specify the initial 
(or boundary) condition for 
the two-particle Green's function
\begin{equation}
G^{(2)}(12,1'2')=
(-i)^2
\left\langle T^{}_C\left(\psi^{}_H(1)\psi^{}_H(2)
\psi^{\dagger}_H(2') \psi^{\dagger}_H(1')\right)
\right\rangle.
\label{s3.8}
\end{equation}
For instance, it may be the condition of weakening of initial
correlations in a distant past~\cite{Kremp85}:
\begin{equation}
\lim_{t_1\to -\infty}G(12,1'2')
\Big|^{}_{\mbox{\scriptsize$\begin{array}{l}
t_1=t_2\\
t'_1=t'_2=t^+_1
\end{array}$
}}\!\!=
G(1,1')\, G(2,2')+\eta G(1,2')\, G(2,1').
\label{ss3.8}
\end{equation} 
Clearly this boundary condition
implies that the Keldysh contour is deformed
such as $t^{}_0\to -\infty$ and the whole past evolution
of the system must be involved, except when
at $t=t^{}_0$ the initial state of the system is non-correlated
and, hence, the boundary condition~(\ref{ss3.8}) 
may be replaced by the analogous initial condition of
complete weakening of correlations.
Since, in general, the evolution starts from some correlated state,
the statistical operator $\varrho(t_0)$ 
does not admit Wick's decomposition,  so that 
the existence of a Dyson equation for  
$G(1,1')$ cannot be justified
by a perturbative expansion in $\hat H'$. Nevertheless, 
using the cluster decomposition of initial correlations in 
terms of the reduced density matrices 
$\varrho^{}_{12\cdots n}(t^{}_0)$, it is possible to 
develop a modified diagram technique for 
the matrix real-time Green's function and study some general properties of 
equations of motion for its 
components~\cite{Hall75a}~--~\cite{Danielewicz84a}.
We will follow, however, the idea advocated by
Danielewicz~\cite{Danielewicz84a} (see also Wagner's work~\cite{Wagner91}). 
We introduce a 
matrix Green's function $\underline G$ on the extended contour $\underline C$
from Fig.~2 involving the
real-time evolution and the ``imaginary evolution'' governed by
the entropy operator $\hat S$ as in the case of the thermodynamic
Green's functions. The structure of $\underline G$
is more complicated than that of the real-time
Green's function $G$ defined on the Keldysh contour.
Nevertheless, by going to the ``interaction picture'' on
the extended contour, it is possible to obtain for 
$\underline G$ an expression which
is quite analogous to the expression in the standard
real-time formalism with averaging over a non-correlated 
statistical ensemble~(see~Appendix~B).  
Then, upon applying Wick's theorem
to every term in the series expansion of the Green's
function $\underline G$ in $\hat H'$ and $\hat S'$,
the Dyson equation can be derived. 
In Appendix~B we show that the above procedure can be 
performed on a contour $\underline{C}$ (Fig.~2)
with an arbitrary value of the parameter $x^{}_0$.
We will use this fact to simplify the formalism by setting
$x^{}_0=0$, which leads to the contour displayed in 
Fig.~3. This choice has the advantage that now
$\underline G$ has the minimal number of components.
In what follows the extended contour $\underline{C}$ 
will always be assumed to have the form shown in Fig.~3. 

Let us start with some definitions.
First of all, we introduce the variable $\xi=(t,x)$
that specifies a point on the contour $\underline{C}$ from
Fig.~3 and define integrals along the contour by the rule
\begin{equation}
\int_{\underline C} d\xi\, \underline{F}(\xi)=
\int_{t_0}^{\infty} 
dt\,F(t,0)\Big|^{}_{{\rm on}\ C^+}
-
\int_{t_0}^{\infty} 
dt\,F(t,0)\Big|^{}_{{\rm on}\ C^-}
+\int_{0}^{1} 
dx\,F(t_0,x)\Big|^{}_{{\rm on}\ C^{}_x}.
\label{m3.5}  
\end{equation}
The first two integrals correspond to the chronological ($C^+$)
and the anti-chronological ($C^-$) branch of the Keldysh contour 
$C$, respectively. To shorten notation, we shall use the underlining
of functions and operators defined on the 
contour $\underline{C}$. The labels $(k)$ in such
functions and operators mean $(k)=(r^{}_k,\xi^{}_k)$. 
It is convenient to introduce the function
$\underline{\delta}(1,2)$ which plays the role of the 
delta function on the contour $\underline{C}$, i.e.,
\begin{equation}
\int_{\underline{C}} d1'\,\,
\underline{\delta}(1,1')\,\underline{F}(1')=
\underline{F}(1).
\label{m3.7}
\end{equation}
According to the integration rule~(\ref{m3.5}), 
we find that
\begin{equation}
\underline{\delta}(1,1')=
\left\{
\begin{array}{ll}
\delta^{}_C(1,1')= \delta(r^{}_1-r'_1)\,
\delta^{}_C(t^{}_1-t'_1)
 & \quad  1,2\in C,\\[3pt]
\delta(x^{}_1-x^{}_2)\,\delta(r^{}_1-r^{}_2) & \quad 
1,2\in C^{}_x,\\[3pt]
0 & \quad  {\rm otherwise},
\end{array}
\right.
\label{m3.8}
\end{equation}
where $\delta^{}_C(t^{}_1- t^{\prime}_1)$ 
is the delta function
on the Keldysh contour~\cite{Danielewicz84a,Botermans90}:
\begin{equation}
\delta^{}_C(t^{}_1-t^{\prime}_1)=
\left\{
\begin{array}{ll}
\delta(t^{}_1-t^{\prime}_1) & \quad 
t^{}_1, t'_1 \in C^+,\\[3pt]
{}-\delta(t^{}_1-t^{\prime}_1) & \quad 
t^{}_1, t'_1 \in C^-,\\[3pt]
0 & \quad  {\rm otherwise}.
\end{array}
\right.
\label{m3.9}
\end{equation}

We next introduce the Heisenberg picture for operators
on the contour $\underline{C}$:
\begin{equation}
\underline{{\hat A}}^{}_H(\xi)=\underline{U}(\xi^{}_0,\xi)\,
\hat A\,\underline{U}(\xi,\xi^{}_0),
\label{3.10}
\end{equation}
$\xi^{}_0=(t_0,0)$ being the point at the
junction of the parts $C$ and $C^{}_x$ of the
contour $\underline{C}$ (see Fig.~3). 
The operator $\underline{U}(\xi^{}_1,\xi^{}_2)$ 
is defined in such a way as to describe the real-time
evolution on $C$ and the imaginary evolution on $C^{}_x$ :
\begin{equation}
\underline{U}(\xi^{}_1,\xi^{}_2)=
T^{}_{\underline{C}}\,\exp\left\{
-i\int_{\xi^{}_2}^{\xi^{}_1}
\underline{{\hat{\cal H}}}(\xi)\,d\xi
\right\},
\label{3.11}
\end{equation}
where $T^{}_{\underline{C}}$ is the path-ordering operator
on $\underline{C}$, and the ``effective Hamiltonian'' is given by
\begin{equation}
\underline{{\hat{\cal H}}}(\xi)=
\left\{
\begin{array}{ll}
\hat H &\quad \xi\in C,\\[3pt]
-i\hat S &\quad \xi\in C^{}_x.
\end{array}
\right.
\label{3.12}
\end{equation}
With the definition~(\ref{3.10}) of the Heisenberg
picture on the contour $\underline{C}$, we can construct
the corresponding one-particle and many-particle
Green's functions. We shall call these functions
the {\em mixed Green's functions\/}~\cite{ZubMorRoep2},
because they involve the real-time Green's 
functions (on the Keldysh contour $C$) as well
as the thermodynamic Green's functions 
(on the contour $C^{}_x$). The one-particle
mixed Green's function is defined as
\begin{equation}
\underline{G}(1,1')=
-i
\left\langle
T^{}_{\underline{C}}
\left(
\underline{\psi}^{}_{H}(1)\,
\underline{\psi}^{\dagger}_{H}(1')
\right)
\right\rangle.
\label{3.14}
\end{equation}
The factor $(-i)$ is introduced in order to have
the usual definition for the real-time components
[cf.~Eq.(\ref{3.4})].
Similarly, the $n$-particle mixed Green's function
is defined as
\begin{equation}
\underline{G}^{(n)}(1\cdots n,1'\cdots n')=
(-i)^n 
\left\langle
T^{}_{\underline{C}}
\left(
\underline{\psi}^{}_{H}(1)\cdots \underline{\psi}^{}_{H}(n)\,
\underline{\psi}^{\dagger}_{H}(n')
\cdots
\underline{\psi}^{\dagger}_{H}(1')
\right)
\right\rangle.
\label{3.15}
\end{equation}
The difference between the mixed Green's functions and
the usual real-time Green's functions is conveniently
illustrated by expressing the one-particle 
Green's function $\underline{G}$ in terms of its components.
Recalling the definition of the contour $\underline{C}$
(see Fig.~3), we write
\begin{equation}
\underline{G}(1,1')=
\left\{
\begin{array}{rl}
G(1,1') & \quad
1,1'\in C,\\[3pt]
{\cal G}^{\more}(1,1') & \quad
1\in C^{}_x,1'\in C,\\[3pt]
{\cal G}^{\less}(1,1') & \quad
1\in C,1'\in C^{}_x,\\[3pt]
i\,{\cal G}(1,1') & \quad
1, 1'\in C^{}_x,
\end{array}
\right.
\label{3.16}
\end{equation}
where $G(1,1')$ and ${\cal G}(1,1')$
are the matrix real-time Green's 
function~(\ref{3.4}) and the thermodynamic Green's
function~(\ref{2.8}), respectively. 
It is significant that 
we also have to consider the {\em cross Green's functions},
${\cal G}^{\moreless}(1,1')$, which are 
constructed from Heisenberg operators with different types of 
evolution:
\begin{eqnarray}
& &
{\cal G}^{\more}(1,1')=
{\cal G}^{\more}(r^{}_1x^{}_1,r'_1 t'_1)=
-i\left\langle
\psi^{}_H(r^{}_1x^{}_1)\,
\psi^{\dagger}_H(r'_1t'_1)
\right\rangle,
\label{3.17}\\[5pt]
& &
{\cal G}^{\less}(1,1')=
{\cal G}^{\less}(r^{}_1t^{}_1,r'_1x'_1)=
-i\eta\left\langle
\psi^{\dagger}_{H}(r'_1 x'_1)\,
\psi^{}_{H}(r^{}_1t^{}_1)
\right\rangle.
\label{3.18}
\end{eqnarray}
These functions satisfy the obvious boundary conditions
\begin{eqnarray}
& &
\hspace*{-30pt}
{\cal G}^{\more}(1,1')\Big|^{}_{t'_1=t^{}_0}=
i{\cal G}(1,1')\Big|^{}_{x'_1=0},
\qquad
{\cal G}^{\less}(1,1')\Big|^{}_{t_1=t^{}_0}=
i{\cal G}(1,1')\Big|^{}_{x^{}_1=0},
\label{3.19}\\[5pt]
& &
\hspace*{-30pt}
{\cal G}^{\more}(1,1')\Big|^{}_{x^{}_1=0}=
g^{\more}(1,1')\Big|^{}_{t^{}_1=t^{}_0},
\qquad
{\cal G}^{\less}(1,1')\Big|^{}_{x'_1=0}=
g^{\less}(1,1')\Big|^{}_{t'_1=t^{}_0},
\label{m3.20}
\end{eqnarray}
which relate them to the thermodynamic and real-time Green's 
functions. As we shall see later, the cross Green's functions
play the crucial role in the theory.

\setcounter{equation}{0}
\renewcommand{\theequation}{4.\arabic{equation}}

\vspace*{15pt}

\begin{center}
IV. {\large D}YSON {\large E}QUATION ON THE {\large E}XTENDED 
{\large C}ONTOUR
\end{center}

\vspace{5pt}  

As already noted, the one-particle mixed Green's function~(\ref{3.14}) 
satisfies the Dyson equation
on the extended contour $\underline{C}$ (Fig.~3).
This equation and its adjoint can be written in the form
\begin{eqnarray}
& &
\int_{\underline{C}} d1''\,
\underline{G}^{-1}_{\,0}(1,1'')\,\underline{G}(1'',1')=
\underline{\delta}(1,1')
+ 
\int_{\underline{C}} d1''\,
\underline{\Sigma}(1,1'')\,\underline{G}(1'',1'),
\label{m4.12}\\[8pt]
& &
\int_{\underline{C}} d1''\,
\underline{G}(1,1'')\,
\underline{G}^{-1}_{\,0}(1'',1')\,=
\underline{\delta}(1,1')
+ 
\int_{\underline{C}} d1''\,
\underline{G}(1,1'')\,
\underline{\Sigma}(1'',1').
\label{m4.13}
\end{eqnarray}
The operator $\underline{G}^{-1}_{\,0}$ is defined as
\begin{equation}
\hspace*{10pt}
\underline{G}^{-1}_{\,0}(1,1')=
\left\{
\begin{array}{cl}
G^{-1}_0(1,1') 
&
\quad \ \ 
1,1'\in C,\\[3pt]
-i\,{\cal G}^{-1}_0(1,1')
 &
\quad \ \ 
1,1'\in C^{}_x,\\[3pt]
0
 &
\quad \ \ 
{\rm otherwise},
\end{array}
\right.
\label{m4.8}
\end{equation}  
where
\begin{equation}
G^{-1}_0(1,1')
=
\left(i\,\frac{\partial}{\partial t_1}+
\frac{\nabla^2_1}{2m}\right) \delta^{}_C(1,1'),
\label{m4.8a}
\end{equation}
and ${\cal G}^{-1}_0(1,1')$ is given by Eq.~(\ref{2.11}).
The components of the matrix
self-energy $\underline{\Sigma}(1,1')$ in 
Eqs~(\ref{m4.12}) and~(\ref{m4.13}) will be denoted by
 \begin{equation}
{\underline\Sigma}(1,1')=
\left\{
\begin{array}{rl}
\Sigma(1,1') & 
\quad 1,1'\,\in C,\\[3pt]
{\cal K}^{\less}(1,1') &
\quad 1\in C,\, 1'\in C^{}_x,\\[3pt]
{\cal K}^{\more}_{}(1,1') &
\quad 1\in C^{}_x,\, 1'\in C,
\\[3pt]
-i\,{\cal K}(1,1') & 
\quad 1,1'\,\in C^{}_x,
\end{array}
\right.
\label{m4.4}  
\end{equation} 
where $\Sigma(1,1')$ is the matrix self-energy on the
Keldysh contour $C$~\cite{Danielewicz84a,Botermans90}:
\begin{equation}
\Sigma=
\left(
\begin{array}{ll}
 \Sigma^{++}   & \Sigma^{+-}\\
 \Sigma^{-+}   & \Sigma^{--}
\end{array}
\right)
=
\left(
\begin{array}{ll}
 \Sigma^{c}       & \Sigma^{{\less}}\\
 \Sigma^{{\more}} & \Sigma^{a}
\end{array}
\right).
\label{m4.4a}  
\end{equation}
In the last line of Eq.~(\ref{m4.4}) the factor $(-i)$ is chosen 
from the requirement that the thermodynamic component of the mixed 
Green's function satisfy Eqs.~(\ref{2.12})  with the self-energy
${\cal K}$ (see below).  
The Dyson equations~(\ref{m4.12}) and~(\ref{m4.13}) will serve as the basis
for our study of short-time dynamics with correlated initial 
states. First we wish to discuss some general consequences of these
equations, which do not depend on the explicit form
of the self-energy $\underline{\Sigma}$. For this purpose,
we shall consider Eqs.~(\ref{m4.12}) 
and~(\ref{m4.13}) for different components of the
mixed Green's function $\underline{G}$.\\

\noindent
{\em A. Real-time components of the mixed Green's function}\\

Suppose that in Eqs.~(\ref{m4.12}) 
and~(\ref{m4.13}) the arguments $1$ and $1'$ correspond to
the Keldysh part $C$ of the contour $\underline{C}$.
Then, since the part $C^{}_x$ is later along $\underline{C}$ than
$C$ (see Fig.~3), we have
\begin{eqnarray}
& &
\hspace*{-70pt}
\left(i\,\frac{\partial}{\partial t^{}_1} +
{\nabla^2_1\over 2m}   
\right) G(1,1')=\delta^{}_C(1,1') 
+ 
\int_C d1''\,\Sigma(1,1'')\,G(1'',1')
\nonumber\\[5pt]
& &
\hspace*{150pt}
{}+
\int_{C^{}_x} d1''\,
{\cal K}^{\less}_{}(1,1'')\,{\cal G}^{\more}(1'',1'),
\label{m4.15}\\[10pt] 
& &
\hspace*{-70pt}
\left(-i\,\frac{\partial}{\partial t^{\prime}_1} +
{\nabla^2_{1'}\over 2m}   
\right) G(1,1')=\delta^{}_C(1,1') 
+ 
\int_C d1''\, G(1',1'')\,\Sigma(1'',1')
\nonumber\\[5pt]
& &
\hspace*{150pt}
{}+
\int_{C^{}_x} d1''\,
{\cal G}^{\less}(1,1'')\,{\cal K}^{\more}_{}(1'',1').
\label{m4.16} 
\end{eqnarray}
These are still matrix equations because 
$G(1,1')$ and $\Sigma(1,1')$
have different components depending
on the position of the arguments  $1$ and $1'$ on the Keldysh
contour $C$. Taking the $(\pm)$\,-\,components of Eqs.~(\ref{m4.15})
and~(\ref{m4.16}), and then using expressions~(\ref{3.1}), 
(\ref{m3.9}), and (\ref{m4.4a}) together with the integration rule 
\begin{equation}
\int_C d1\, F(1)=
\int_{t^{}_0}^\infty d1\,
\left\{
F(1)\Big|^{}_{{\rm on}\ C^+}
-
F(1)\Big|^{}_{{\rm on}\ C^-}
\right\},
\label{m4.17}
\end{equation}
one can easily obtain a system of equations for
$g^{c/a}$ and $g^{\moreless}$. It is convenient,
however, to use retarded ($g^R,\Sigma^R$) and
advanced ($g^A,\Sigma^A$) functions instead of the 
causal ($g^c,\Sigma^c$) and  and anti-causal  ($g^a,\Sigma^a$) ones. 
The conventional  definitions~\cite{Danielewicz84a,Botermans90} are
\begin{eqnarray}
& &
g^R=g^c-g^{\less}=g^{\more}-g^a,
\qquad
g^A=g^c-g^{\more}=g^{\less}-g^a,
\label{m4.18}\\[5pt]
& &
\Sigma^R=\Sigma^c-\Sigma^{\less}=\Sigma^{\more}-\Sigma^a,
\qquad
\Sigma^A=\Sigma^c-\Sigma^{\more}=\Sigma^{\less}-\Sigma^a.
\label{m4.19}
\end{eqnarray}
Note also that 
\begin{equation}
g^{R/A}(1,1')= \pm\,\theta[\pm(t^{}_1-t'_1)]
\left\{g^{\more}(1,1') -g^{\less}(1,1')\right\}.
\label{m4.19a}
\end{equation}
Further manipulations with Eqs.~(\ref{m4.15}) and~(\ref{m4.16}) 
are straightforward and similar to those in the 
standard real-time Green's function 
formalism~(see, e.g.,~\cite{Botermans90}). The only point of
importance is that, by definition, 
the cross functions ${\cal G}^{\moreless}$
and ${\cal K}^{\moreless}$ do not depend on the position
of their time argument on the Keldysh contour $C$.
Taking this property into account, 
Eq.~(\ref{m4.15}), when written in terms of 
$g^{\moreless}$ and $g^{R/A}$, becomes
\begin{eqnarray}
& &
\hspace*{-40pt}
\left(i\,\frac{\partial}{\partial t^{}_1} +
{\nabla^2_1\over 2m}\right)   
g^{\moreless}(1,1')=
\int_{t_0}^{\infty} d1''\,
\left\{
\Sigma^R(1,1'')\,g^{\moreless}(1'',1')+
\Sigma^{\moreless}(1,1'')\,g^A(1'',1') 
\right\}
\nonumber\\[5pt]
& &
\hspace*{150pt}
{}+
\int_{C^{}_x} d1''\,
{\cal K}^{\less}_{}(1,1'')\,
{\cal G}^{\more}(1'',1'),
\label{m4.20}\\[10pt] 
& &
\hspace*{-40pt}
\left(i\,\frac{\partial}{\partial t^{}_1} +
{\nabla^2_1\over 2m}\right)   
g^{R/A}(1,1')=\delta(1-1')
+
\int_{t_0}^{\infty} d1''\, 
\Sigma^{R/A}(1,1'')\,g^{R/A}(1'',1').
\label{m4.21}   
\end{eqnarray}
The adjoint equations follow from Eq.~(\ref{m4.16}):
\begin{eqnarray}
& &
\hspace*{-30pt}
\left(-i\,\frac{\partial}{\partial t^{\prime}_1} +
{\nabla^2_{1'}\over 2m}\right)g^{\moreless}(1,1')=
\int_{t_0}^{\infty} d1''\,
\left\{
g^R(1,1'')\,\Sigma^{\moreless}(1'',1') 
+g^{\moreless}(1,1'')\,\Sigma^A(1'',1')
\right\}
\nonumber\\[5pt]
& &
\hspace*{150pt}
{}+
\int_{C^{}_x} d1''\,
{\cal G}^{\less}(1,1'')\,
{\cal K}^{\more}_{}(1'',1'),
\label{m4.22}\\[10pt] 
& &
\hspace*{-30pt}
\left(-i\,\frac{\partial}{\partial t^{\prime}_1} 
+{\nabla^2_{1'}\over 2m}\right)g^{R/A}(1,1')=\delta(1-1')
+
\int_{t_0}^{\infty} d1''\, g^{R/A}(1,1'')\,\Sigma^{R/A}(1'',1').
\label{m4.23}  
\end{eqnarray} 
Again it is convenient to go over to a compact matrix notation.
We define for functions of $t$ (on the real-time axis) 
and $x$ the  multiplication as integration 
over the intervals $t^{}_0<t<\infty$ and $0<x<1$,
respectively, plus matrix multiplication with respect
to single-particle quantum numbers. Then the above
equations for $g^{\moreless}$ and $g^{R/A}$ take the form  
\begin{eqnarray}
& &
\left(g^{-1}_0 - \Sigma^{R}\right)g^{\moreless}=
\Sigma^{\moreless} g^A + {\cal K}^{\less} {\cal G}^{\more},
\label{m4.24}\\[7pt] 
& &
\left(g^{-1}_0 - \Sigma^{R/A}\right)g^{R/A}=I,
\label{m4.25}\\[7pt]
& &
g^{\moreless} \left(g^{-1}_0 - \Sigma^A\right)
=g^R \Sigma^{\moreless} + {\cal G}^{\less}{\cal K}^{\more},
\label{m4.26}\\[7pt] 
& &
g^{R/A}\left(g^{-1}_0 - \Sigma^{R/A}\right)= I,
\label{m4.27}  
\end{eqnarray}   
where $I(1,1')=\delta(1-1')$ plays the role of the identity matrix, and
\begin{equation}
g^{-1}_0(1,1')=\left( i\,\frac{\partial}{\partial t^{}_1}
+ \frac{\nabla^2_1}{2m}\right)
\delta(1-1').
\label{m4.28}  
\end{equation}
Equations~(\ref{m4.25}) and~(\ref{m4.27}) 
are formally identical with equations for
the retarded and advanced Green's functions in the standard
real-time formalism~\cite{Danielewicz84a,Botermans90}, while
Eqs.~(\ref{m4.24}) and~(\ref{m4.26}), which are
modified Kadanoff-Baym equations, contain explicit 
contributions from initial correlations. 
Note that the terms associated with initial correlations,
${\cal K}^{\less}{\cal G}^{\more}$ and 
${\cal G}^{\less}{\cal K}^{\more}$, enters into
Eqs.~(\ref{m4.24}) and~(\ref{m4.26})
as source terms. On the other hand, a perturbative
diagram expansion~\cite{Danielewicz84a} 
leads to modified Kadanoff-Baym equations in which
initial correlations appear as some corrections to the
self-energies $\Sigma^{\moreless}$. In Appendix~C we
show that, in fact, these two descriptions are equivalent.\\

\noindent
{\em B. The thermodynamic component of the mixed Green's function}\\
 
Let us turn back to the Dyson equation~(\ref{m4.12})
and take the arguments
$1=(r^{}_1,x^{}_1)$ and $1'=(r^{\prime},x^{\prime}_1)$ 
on the part $C^{}_x$ of the contour $\underline C$ 
(Fig.~3). Then we arrive at the equation
\begin{equation}
\int_{C_x} d1''\,
\left\{ {\cal G}^{-1}_{\,0}(1,1'') -
{\cal K}(1,1'')
\right\}
{\cal G}(1'',1') =
\delta(1-1')+
\int_C d1''\, 
{\cal K}^{\more}_{}(1,1'')\,{\cal G}^{\less}(1'',1'). 
\label{m4.29} 
\end{equation}
Since the values of 
${\cal K}^{\more}_{}(1,1'')$ and
${\cal G}^{\more}(1'',1')$
do not depend on whether the argument 
$1''$ corresponds
to the chronological ($C^{+}$) or the anti-chronological ($C^{-}$)
branch of the Keldysh contour,
the last term in Eq.~(\ref{m4.29})
is zero by virtue of the integration rule~(\ref{m4.17}).
We see that the thermodynamic component of $\underline{G}$
satisfies the  Dyson equation which is not connected with
the real-time evolution on the Keldysh contour $C$.
The above conclusion, however, is almost trivial because 
${\cal G}(1,1')$ is 
the thermodynamic Green's function~(\ref{2.8}) with the imaginary
evolution governed by $\hat S$.
Nevertheless, the point to remember is that the 
existence of a closed Dyson 
equation for ${\cal G}(1,1')$
means that it has an inverse function ${\cal G}^{-1}(1,1')$ on  
$C_x$  and that
the component ${\cal K}$ of the self-energy~(\ref{m4.4})
is not connected with the real-time evolution of the system.\\

\noindent
{\em C. The cross components of the mixed Green's function}\\

To complete our general discussion of the Dyson equation
for $\underline G$, we now derive equations of motion for
the cross components, ${\cal G}^{\moreless}$. Again we
turn to Eq.~(\ref{m4.12}) and 
take $1\in C$, $1'\in C^{}_x$. Then we obtain
\begin{equation}
\int_{C} d1''
\left\{
G^{-1}_0(1,1'')-  \Sigma(1,1'')
\right\}
{\cal G}^{\less}(1'',1')=
i\int_{C^{}_x} d1''\, 
{\cal K}^{\less}_{}(1,1'')\,
{\cal G}(1'',1').
\label{m4.30}  
\end{equation}
Here the argument $t^{}_1$ may be assigned to
either of the two branches of 
the Keldysh contour $C$. It can easily be checked that
in both cases we get the same equation.
In the matrix form, it reads
\begin{equation}
\left(
g^{-1}_0- \Sigma^{R}
\right)
{\cal G}^{\less}= 
i{\cal K}^{\less}\,{\cal G}.
\label{m4.31} 
\end{equation} 
This equation describes the evolution of 
${\cal G}^{\less}(1,1')={\cal G}^{\less}(r^{}_1t^{}_1,r'_1x'_1)$
with respect to its real-time argument. 
To derive the equation describing the ``imaginary'' 
evolution of ${\cal G}^{\less}$, we use Eq.~(\ref{m4.13}) where
we take $1\in C$ and $1'\in C^{}_x$. This gives 
\begin{equation}
\int_{C_x} d1''\,
{\cal G}^{\less}(1,1'')\left\{
{\cal G}^{-1}_0(1'',1') - {\cal K}(1'',1')\right\}
=
i\int_{C} d1''\,
G(1,1'')\,
{\cal K}^{\less}_{}(1'',1').
\label{m4.32} 
\end{equation}
Again, the time argument $t_1$ may be taken on either of the  
two branches of the Keldysh contour. Assuming for definiteness
that $t^{}_1\in C^+$, we get the equation which is written
in the matrix form as
\begin{equation}
{\cal G}^{\less}
\left(
{\cal G}^{-1}_0 - {\cal K}
\right)
=
ig^{R}\,{\cal K}^{\less}.
\label{m4.33}    
\end{equation} 
The analogous procedure can  be repeated for  
${\cal G}^{\more}(1,1')={\cal G}^{\more}(r^{}_1x^{}_1,r'_1t'_1)$.
As a result we have two equations
\begin{eqnarray}
& &
{\cal G}^{\more}\!
\left(
g^{-1}_0 - \Sigma^A
\right)
=i{\cal G}\,{\cal K}^{\more},
\label{m4.34}\\[5pt] 
& &
\left(
{\cal G}^{-1}_0 - {\cal K}
\right)
{\cal G}^{\more}=
i{\cal K}^{\more}g^A,
\label{m4.35}  
\end{eqnarray}
which describe the real-time and imaginary
evolution of ${\cal G}^{\more}$.\\

\noindent
{\em D. Relation between $\underline\Sigma$ and the two-particle mixed
Green's function}\\

To obtain explicit expressions for the components of
the self-energy $\underline{\Sigma}$,
one has to specify the form of the Hamiltonian $\hat H$ and 
the entropy operator $\hat S$. If these operators
are given by Eqs.~(\ref{2.1}) and~(\ref{2.4}), then the matrix
self-energy $\underline{\Sigma}$ can be expressed in terms of the
two-particle mixed Green's function $\underline{G}^{(2)}$
(see Appendix~D):
\begin{eqnarray}
\hspace*{-30pt}
{\underline\Sigma}(1,1')&=&
i\eta
\int_{\underline{C}} d2\,d1''\,d2''\,d1'''\,
\underline{V}(12,1'' 2'')\,
\underline{G}^{(2)}(1''2'',1''' 2^+)\,
\underline{G}^{-1}(1''',1')
\nonumber\\[10pt]
\hspace*{-30pt}{}&=&
i\eta
\int_{\underline{C}} d2\,d1''\,d2''\,d1'''\,
\underline{G}^{-1}(1,1''')\,
\underline{G}^{\,(2)}(1''' 2^-, 1'' 2'')\,
\underline{V}(1'' 2'', 1' 2), 
\label{m4.14}    
\end{eqnarray}
where the symbols $k^+=(r^{}_k,\xi^+_k)$  and 
$k^-=(r^{}_k,\xi^-_k)$ denote
points infinitesimally later (earlier) on the contour 
$\underline{C}$ than $k=(r^{}_k,\xi^{}_k)$. 
The ``interaction matrix'' 
$\underline{V}$ on the contour $\underline{C}$ is defined as
\begin{equation}
\underline{V}(12,1'2')=
\left\{
\begin{array}{ll}
V(12,1'2') &
\quad\
1,\,2,\,1',\,2' \in C,\\[5pt]
i\,{\cal V}(12,1'2') &
\quad\
1,\,2,\,1',\,2' \in  C^{}_x,\\[5pt]
0 &
\quad\
{\rm otherwise},
\end{array}
\right.
\label{m4.9}
\end{equation}
where
\begin{eqnarray}
& &
V(12,1'2')=
\langle r^{}_1r^{}_2|V|r'_1r'_2\rangle\,
\delta^{}_{C}(t^{}_1-t^{}_2)\,
\delta^{}_{C}(t^{}_1-t^{\prime}_1)\,
\delta^{}_{C}(t^{\prime}_1-t^{\prime}_2),
\label{m4.10}\\[5pt]
& &
{\cal V}(12,1'2')=
-\lambda^{}_2(r^{}_1r^{}_2,r'_1r'_2)\,
\delta(x^{}_1-x^{}_2)\,
\delta(x^{}_1-x^{\prime}_1)\,
\delta(x^{\prime}_1-x^{\prime}_2).
\label{m4.11} 
\end{eqnarray}
Note that the matrix $\underline{V}$ has the  symmetry property
\begin{equation}
\underline{V}(12,1'2')=\underline{V}(21,2'1'),
\label{m4.11x}
\end{equation}
which reflects  the fact that $\hat H$ and $\hat S$ are 
self-adjoint operators.

\setcounter{equation}{0}
\renewcommand{\theequation}{5.\arabic{equation}}

\vspace*{15pt}

\begin{center}
V. {\large T}HE {\large P}ROBLEM OF {\large A}NSATZ FOR 
{\large C}ORRELATED
{\large I}NITIAL {\large S}TATES
\end{center}

\vspace*{5pt}

A precursor of kinetic equations can be obtained from the generalized
Kadanoff-Baym equations~(\ref{m4.24}) and~(\ref{m4.26}) for
$g^{\less}$. Subtracting these equations, we have
\begin{equation}
\left[g^{-1}_0, g^{\less}\right]=
\Sigma^R\,g^{\less} - g^{\less}\,\Sigma^A
+\Sigma^{\less}\,g^A - g^R\,\Sigma^{\less}
+{\cal K}^{\less}\,{\cal G}^{\more}
- {\cal G}^{\less}\,{\cal K}^{\more}.
\label{c5.0}
\end{equation}  
The first step to a kinetic
equation for the one-particle density matrix
is to take the diagonal part of Eq.~(\ref{c5.0})
with respect to time variables. Then the left-hand side
can exactly be expressed in terms of 
$\varrho^{}_1(t)$, while the right-hand side still contains
double-time correlation functions $g^{\moreless}$.
In order to convert Eq.~(\ref{c5.0})  
into a closed equation for $\varrho^{}_{1}(t)$, one has to
find $g^{\moreless}$ as functionals of the 
one-particle density matrix.  In other words, 
this is the point where the problem of ansatz arises.
The simple ansatz proposed by Kadanoff and Baym 
(KB ansatz)~\cite{KadanoffBaym62}  works only in situations 
where the quasiparticle picture is
adequate to describe the evolution of the system and memory
effects can be neglected. In succeeded years efforts were
directed towards refining the KB ansatz for special problems
of kinetic theory. As already mentioned in Introduction, 
an important result along this line has been
obtained by Lipavsk\'y {\em et al.}~\cite{Lipavsky86}.
These authors have derived exact integral equations which allow,
in principle, to find $g^{\moreless}$ in terms of $\varrho^{}_{1}$
by iteration  or, at least, to find corrections to the
KB ansatz in a self-consistent way, including memory effects and
dynamical correlations. It should be noted, however, that the
derivation of the
generalized Kadanoff-Baym ansatz 
(the GKB ansatz) in Ref.~\cite{Lipavsky86}
is based on the boundary condition of complete weakening
of correlations in a distant past. Therefore we have to re-formulate
the GKB ansatz for $g^{\moreless}$ in such a way as to take 
initial correlations into account. 

Another new feature of the short-time kinetics
is the appearance 
of the cross Green's functions in Eq.~(\ref{c5.0}). 
There are two possible approaches to these functions. 
The first is to use Eqs.~(\ref{m4.31}) and~(\ref{m4.34}) as independent 
equations together with a kinetic     
equation for $\varrho^{}_{1}(t)$. 
The second is to eliminate these functions by means
of some ansatz. In this section we will derive exact equations
which allow one to formulate a generalized ansatz for the
cross Green's functions in just the same way as for the
correlation functions $g^{\moreless}$.  

We start with the correlation functions $g^{\moreless}$ and   
follow the line of reasoning which is
close to that of the work of 
Lipavsk\'y {\em et al.}~\cite{Lipavsky86}.
In order to preserve causality, it is convenient to introduce
two auxiliary correlation functions (denoting only the time arguments)
\begin{equation}
G^{\moreless}_R(t_1,t_2)=\theta(t_1-t_2)\,g^{\moreless}(t_1,t_2),
\qquad
G^{\moreless}_A(t_1,t_2)=\theta(t_2-t_1)\,g^{\moreless}(t_1,t_2).
\label{c5.1}
\end{equation}
Then, for $t_1\not=t_2$, we have
\begin{equation}
g^{\moreless}(t_1,t_2)=
G^{\moreless}_R(t_1,t_2)
+G^{\moreless}_A(t_1,t_2).
\label{c5.2}
\end{equation}  
Strictly  speaking,
this relation does not determine $g^{\moreless}$
for $t_1=t_2$ because $G^{\moreless}_{R/A}$ are discontinuous at this
point. One can complete, however, the definition of
$g^{\moreless}$ at $t_1=t_2$ by the limits $t_1\to t_2\pm 0$  
which give the same result. Since the representation~(\ref{c5.2})
will always be used in integrals over time, the above
refinement is irrelevant. 

Let us now derive the equation of motion for $G^{\moreless}_R$.
Differentiation with respect to $t_1$ gives
\begin{equation}
i\,\frac{\partial}{\partial t_1}\,G^{\moreless}_R(t_1,t_2)=
i\delta(t_1-t_2)\,g^{\moreless}(t_2,t_2)
+\theta(t_1-t_2)\,i\,
\frac{\partial}{\partial t_1}\,g^{\moreless}(t_1,t_2).
\label{c5.3}
\end{equation}
The time derivative in the last term can be eliminated with
the help of the Kadanoff-Baym equation~(\ref{m4.20}). 
Using also the representation~(\ref{c5.2}) in the time integral, 
Eq.~(\ref{c5.3}) becomes
\begin{eqnarray}
& &
\hspace*{-20pt}
\left(
i\,\frac{\partial}{\partial t_1}
+\frac{\nabla^2_1}{2m}
\right)
G^{\moreless}_R(t_1,t_2)
-\int_{t_0}^{\infty} dt_3\,
\Sigma^R(t_1,t_3)\,G^{\moreless}_R(t_3,t_2)
=i\delta(t_1-t_2)\,g^{\moreless}(t_2,t_2)
\nonumber\\[8pt]
& &
\hspace*{30pt}
{}+\theta(t_1-t_2)
\int_{t_0}^{\infty} dt_3
\left\{
\Sigma^{\moreless}(t_1,t_3)\,g^A(t_3,t_2)
+ \Sigma^R(t_1,t_3)\,G^{\moreless}_A(t_3,t_2)
\right\}
\nonumber\\[8pt]
& &
\hspace*{90pt}
{}+ \theta(t_1-t_2)\,{\cal K}^{\less}(t_1)\,{\cal G}^{\more}(t_2).
\label{c5.4}
\end{eqnarray}
A formal solution of this equation satisfying the required boundary
condition at $t_1=t_2$ is given by
\begin{eqnarray}
& &
\hspace*{-30pt}
G^{\moreless}_R(t_1,t_2)=
ig^R(t_1,t_2)\,g^{\moreless}(t_2,t_2)
+ \int_{t_2}^{t_1} dt_3\,
g^R(t_1,t_3)\,{\cal K}^{\less}(t_3)\,{\cal G}^{\more}(t_2)
\nonumber\\[8pt]
& &
\hspace*{-20pt}
{}+\int_{t_2}^{t_1}
dt_3\int_{t_0}^{\infty} dt_4\,
g^R(t_1,t_3)
\left\{
\Sigma^{\moreless}(t_3,t_4)\,g^A(t_4,t_2)
+ \Sigma^R(t_3,t_4)\,G^{\moreless}_A(t_4,t_2)
\right\}.
\label{c5.5}
\end{eqnarray}
This is not of course an explicit expression for
$G^{\moreless}_R$ but a rather complicated integral equation.
Among other things, it
contains the advanced  correlation functions
$G^{\moreless}_A$ and the cross
Green's function. We can derive an analogous equation
for $G^{\moreless}_A$  following the same procedure as before.
Differentiating $G^{\moreless}_A(t_1,t_2)$ with respect to
$t_2$ and then eliminating the time derivative 
$\partial g^{\moreless}(t_1,t_2)/\partial t_2$
with the help of Eq.~(\ref{m4.22}), after some algebra we obtain 
\begin{eqnarray}
& &
\hspace*{-30pt}
G^{\moreless}_A(t_1,t_2)=
-ig^{\moreless}(t_1,t_1)\,g^A(t_1,t_2)\
+ \int_{t_1}^{t_2} dt_3\,
{\cal G}^{\less}(t_1)\,{\cal K}^{\more}(t_3)\,
g^A(t_3,t_2)\,
\nonumber\\[8pt]
& &
\hspace*{-20pt}
{}+\int_{t_1}^{t_2}
dt_3\int_{t_0}^{\infty} dt_4
\left\{
g^R(t_1,t_4)\,
\Sigma^{\moreless}(t_4,t_2)
+ G^{\moreless}_R(t_1,t_4)\, \Sigma^A(t_4,t_3)\,
\right\} g^A(t_3,t_2).
\label{c5.6}
\end{eqnarray}
In comparison with 
equations for $G^{\moreless}_{R/A}$ derived by 
Lipavsk\'y {\em et al.}~\cite{Lipavsky86},
the above equations contain new terms 
(the second terms on the right-hand sides) which
stem from initial correlations. It 
should also be noted that the 
self-energies $\Sigma^{\moreless}$ and $\Sigma^A$
are modified in the presence of
initial correlations. Physically, the explicit contributions
from initial correlations as well as the corresponding corrections 
to the self-energies are expected to go to zero in the limit
$t_0\to -\infty$, since a many-particle system ``forgets''
details of its initial state.

Before discussing Eqs.~(\ref{c5.5}) and~(\ref{c5.6}), 
let us derive analogous equations for the cross Green's functions.
First of all we note that the solution to 
Eqs.~(\ref{m4.31}) and~(\ref{m4.34}) is
\begin{equation}
{\cal G}^{\less}=ig^R {\cal K}^{\less} {\cal G}
+ 
{\cal G}^{\less}_{\rm hom},
\qquad
{\cal G}^{\more}= i{\cal G}{\cal K}^{\more} g^A
+
{\cal G}^{\more}_{\rm hom},
\label{m5.1}
\end{equation}
where ${\cal G}^{\moreless}_{\rm hom}$ satisfy the 
homogeneous equations
\begin{equation}
\left(g^{-1}_0 -\Sigma^R\right) {\cal G}^{\less}_{\rm hom}=0,
\qquad
 {\cal G}^{\more}_{\rm hom}\left(g^{-1}_0 -\Sigma^A\right)=0,
\label{m5.2}
\end{equation} 
which are to be solved with the boundary conditions~(\ref{3.19})
and~(\ref{m3.20}) for ${\cal G}^{\less}(r_1t_0,r'_1x'_1)$
and ${\cal G}^{\more}(r_1x_1,r'_1t_0)$.  
Using the properties of the retarded and advanced Green's functions,
\begin{equation}
\lim_{t_1\to t_2\pm 0}
g^{R/A}(r_1t_1,r_2t_2)=
\mp i \delta(r_1-r_2),
\label{m5.3}
\end{equation}
and denoting only the time arguments, we may write the solution
to Eqs.~(\ref{m5.2}) in the form
\begin{equation}
{\cal G}^{\less}_{\rm hom}(t_1)=
ig^R(t_1,t_0)\,{\cal G}^{\less}(t_0),
\qquad
{\cal G}^{\more}_{\rm hom}(t^{}_1)=
-i{\cal G}^{\more}(t_0)\, g^A(t_0,t^{}_1).
\label{m5.4}
\end{equation}
Inserting these results into Eqs.~(\ref{m5.1}) and going to a more
transparent notation, we obtain 
\begin{eqnarray}
& &
{\cal G}^{\less}(t_1)=
ig^{R}(t_1,t_0)\,{\cal G}^{\less}(t_0)
+ i\int_{t_0}^{t_1} dt^{}_2\, g^{R}(t_1,t_2)\,
{\cal K}^{\less}(t_2)\,{\cal G},
\nonumber\\[5pt]
& &
{\cal G}^{\more}(t_1)=
-i {\cal G}^{\more}(t_0)\,g^{A}(t^{}_0,t_1)
+ i\int_{t_0}^{t_1} dt^{}_2\, 
{\cal G}\,
{\cal K}^{\more}(t_2)\,
g^{A}(t_2,t_1).
\label{c5.x}
\end{eqnarray}
It is interesting to note that substitution of these expressions 
into Eqs.~(\ref{c5.5}) and~(\ref{c5.6}) allows one to eliminate
the cross Green's functions, since the ${\cal G}^{\moreless}(t_0)$
are determined by the initial ensemble, as seen from the
boundary conditions~(\ref{3.19}).   
However, this is not a complete solution of the problem
because the 
cross Green's functions enter into 
the self-energies ${\cal K}^{\moreless}$. 

The GKB ansatz~\cite{Lipavsky86} follows from 
Eqs.~(\ref{c5.5}) and~(\ref{c5.6}) if only the first terms on
their right-hand sides are retained. In that case, using
Eqs.~(\ref{3.6}) and~(\ref{c5.2}), we obtain
\begin{eqnarray}
& &
g^{\more}(t^{}_1,t^{}_2)=g^{R}(t^{}_1,t^{}_2)\,\varrho^{\more}_{1}(t^{}_2)
- \varrho^{\more}_{1}(t^{}_1)\,g^{A}(t^{}_1,t^{}_2),
\nonumber\\[6pt]
& &
g^{\less}(t^{}_1,t^{}_2)=
\eta\left(g^{R}(t^{}_1,t^{}_2)\,\varrho^{\less}_{1}(t^{}_2)
- \varrho^{\less}_{1}(t^{}_1)\,g^{A}(t^{}_1,t^{}_2)\right).
\label{c5.zz}
\end{eqnarray}
Recently the same ansatz has been recovered by
approximate solution of the quantum hierarchy for
reduced density matrices~\cite{Kremp97}.
The GKB ansatz is exact in the Hartree-Fock approximation for
the self-energies and removes some defects of
the conventional KB ansatz~\cite{Lipavsky86,MorawetzRoepke95}.
It should be particularly emphasized that the GKB ansatz is consistent
with the exact relation~(\ref{m4.19a}) and, consequently,
preserves the correct spectral properties of microscopic dynamics.
Nevertheless, the range of validity of the GKB ansatz is not yet
known with certainty because it is very difficult to make a
general estimate of the last terms in Eqs.~(\ref{c5.5}) and~(\ref{c5.6}).     
Qualitative  physical arguments~\cite{Lipavsky86} suggest that 
these terms may be neglected if there exist two well-separated time
scales characterized by the ``collision time'' $\tau^{}_c$
and the quasiparticle lifetime $\tau$. Then the GKB 
ansatz  provides a reasonable approximation for the correlation 
functions in the collision integral if $\tau\gg \tau^{}_c$, 
i.e., not far beyond the quasiparticle picture. 
Iteration of the integral 
equations~(\ref{c5.5}) and~(\ref{c5.6}) around the GKB ansatz
allows one to find $G^{\moreless}_{R/A}$ as functionals
of the diagonal parts of the correlation functions,
$g^{\moreless}(t,t)$, to some order in $\tau^{}_c/\tau$. 
Note, however, that such a procedure can have consequences
for some important properties of the kinetic equation,
say for the conservation laws. Later we shall return to this 
point. 

In direct analogy to the GKB ansatz for
real-time correlation functions $g^{\moreless}$, one may
expect that keeping only the first terms on the right-hand sides
of Eqs.~(\ref{c5.x}) is a reasonable ansatz for the cross
Green's functions if the description of the system does not go far beyond
the quasiparticle picture. 
Thus, the simple ansatz for the cross Green's functions reads
\begin{equation}
{\cal G}^{\less}(t)=
ig^{R}(t,t_0)\,{\cal G}^{\less}(t_0),
\qquad
{\cal G}^{\more}(t)=
-i {\cal G}^{\more}(t_0)\,g^{A}(t^{}_0,t).
\label{c5.zzz}
\end{equation}
 The advantages of this ansatz
for the cross Green's functions are the same as those
of the GKB ansatz~(\ref{c5.zz}):
(i) it preserves causality and describes fading memory associated
with quasiparticle propagation, 
(ii) effects of strong external fields
can be included in a consistent manner, 
(iii) it is not connected
with a quasiclassical approximation.
The validity of this ansatz, as well as
the validity of the GKB ansatz for $g^{\moreless}$,
must be checked in each specific situation by using the exact 
integral equations~(\ref{c5.5}), (\ref{c5.6}), and~(\ref{c5.x}).
These equations can also be used to find corrections to 
$g^{\moreless}$ and ${\cal G}^{\moreless}$ by iteration.

\setcounter{equation}{0}
\renewcommand{\theequation}{6.\arabic{equation}}

\vspace*{15pt}

\begin{center}
VI. {\large $T$}-{\large M}ATRIX {\large A}PPROXIMATION
\end{center}

\vspace*{5pt}

So far we have discussed general consequences of the Dyson equation
for the mixed Green's function, which do not depend on the form
of the Hamiltonian $\hat H$ and the entropy operator
$\hat S$. In order to proceed beyond these formal results, 
we will consider the special case where the matrix self-energy 
$\underline{\Sigma}$ is given by Eq.~(\ref{m4.14}). Then we can
obtain different approximations for the components of the self-energy
by determining approximate forms for the two-particle
mixed Green's function $\underline G^{(2)}$.

In this paper we restrict ourselves to  a much used approximation of
the many-particle theory, namely the $T$-matrix approximation. 
In the context of real-time dynamics, this approximation is quite 
suitable for treating short-range interactions between 
particles. Another important point is that the $T$-approximation 
satisfies general criteria for the conservation 
laws~\cite{Botermans90}.
The applicability of this approximation to the 
thermodynamic component of $\underline{G}^{(2)}$
depends on whether the initial
correlations can be described in terms of binary correlations.  
For simplicity, we shall assume that the $T$-approximation 
may be applied to all components of the two-particle
Green's function $\underline{G}^{(2)}$ appearing in Eq.~(\ref{m4.14}),
although in some cases it would be more reasonable to 
use different approximations for its real-time and thermodynamic
components.\\

\noindent
{\em A.\ The $T$-matrix on the extended contour}\\

In the $T$-approximation, the two-particle
Green's function is obtained by performing a ladder-type summation in
the particle-particle 
channel~\cite{Danielewicz84a,Botermans90}, which leads to the
equation
\begin{eqnarray}
& &
\hspace*{-40pt}
\underline{G}^{\,(2)}(12,1'2')=
\left\{
\underline{G}(1,1')\underline{G}(2,2')+\eta\,
\underline{G}(1,2')\underline{G}(2,1')
\right\}
\nonumber\\[5pt]
& &
\hspace*{40pt}
{}+i\,\underline{G}(1,1'')\,\underline{G}(2,2'')\,
\underline{V}(1''2'',1'''2''')\,\underline{G}^{(2)}(1'''2''',1'2'),
\label{z5.1}   
\end{eqnarray}
where repeated arguments are summed over single-particle quantum 
numbers and integrated over the contour $\underline{C}$ shown
in Fig.~3. The $\underline{T}$-matrix on the contour $\underline{C}$ is 
introduced through equations
\begin{eqnarray}
& &
\hspace*{-30pt}
\underline{T}(12,1'2')=\underline{V}(12,1'2')
+i\,\underline{V}(12,1''2'')\,\underline{G}(1'',2''')\,\underline{G}(2'',2''')\,
\underline{T}(1'''2''',1'2'),
\label{z5.2}\\[5pt]
& &
\hspace*{-30pt}
\underline{T}(12,1'2')=\underline{V}(12,1'2')
+i\,\underline{T}(12,1''2'')\,\underline{G}(1'',1''')\,\underline{G}(2'',2''')\,
\underline{V}(1'''2''',1'2').
\label{z5.3}
\end{eqnarray}
In the condensed matrix notation, Eqs.~(\ref{z5.1})~--~(\ref{z5.3})
read
\begin{eqnarray}
& &
\underline{G}^{(2)}=
\left(\underline{G} \otimes \underline{G}\right)^{}_{\rm ex}
+ i\left(\underline{G} \otimes \underline{G}\right)
\underline{V}\,\underline{G}^{\,(2)},
\label{z5.1a}\\[5pt]
& &
\underline{T}= \underline{V} 
+ i\,\underline{V} \left(\underline{G}\otimes \underline{G} \right)
\underline{T},
\label{z5.2a}\\[5pt]
& &
\underline{T}= \underline{V} 
+ i\,\underline{T} \left(\underline{G}\otimes \underline{G} \right)
\underline{V},
\label{z5.3a} 
\end{eqnarray}
where $\otimes$ stands for the direct product of matrices, and
the index ``ex'' indicates the symmetrized ($\eta=1$)
or anti-symmetrized ($\eta=-1$) matrix 
[cf.~Eq.~(\ref{z5.1})].

Equation~(\ref{z5.1a}) can be solved by iteration to yield the 
relation between
the two-particle mixed Green's function and the $\underline{T}$-matrix:
\begin{equation}
\underline{G}^{\,(2)}=
\left(\underline{G} \otimes \underline{G}\right)^{}_{\rm ex}
+ i
\left(\underline{G} \otimes \underline{G}\right)
\widetilde{\underline{T}}\left(\underline{G} \otimes \underline{G}\right),
\label{z5.4}
\end{equation}
where the symmetrized (anti-symmetrized) $\underline{T}$-matrix
is denoted by
\begin{equation}
\widetilde{\underline{T}}(12,1'2')=
\underline{T}(12,1'2') +\eta\,\underline{T}(12,2'1')=
\underline{T}(12,1'2') +\eta\,\underline{T}(21,1'2').
\label{z5.4x}
\end{equation}

Recalling the definition~(\ref{m4.9}) of  $\underline{V}$, it is easy 
to see from Eqs.~(\ref{z5.2a}) and~(\ref{z5.3a}) that  
the components $\underline{T}(12,1'2')$ of the 
$\underline{T}$-matrix  are not zero if only the arguments 1 and 2
correspond to the same part of the contour $\underline{C}$;
this is also true for the arguments $1'$ and $2'$. Taking this 
into account, the $\underline{T}$-matrix is expressed as
\begin{equation}
\underline{T}(12,1'2')=
\left\{
\begin{array}{lll}
T(12,1'2')                & \quad  1,2\in C,   & 1',2' \in C,\\
{\cal T}^{\less}(12,1'2') & \quad  1,2\in C,   & 1',2' \in C_x,\\
{\cal T}^{\more}(12,1'2') & \quad  1,2\in C_x, & 1',2' \in C,\\
i\, {\cal T}^{}(12,1'2')  & \quad  1,2\in C_x, & 1',2' \in C_x. 
\end{array}
\right.
\label{z5.6}  
\end{equation} 
The real-time $T$-matrix, $T(12,1'2')$,
has four components which can be put into a $2\times 2$ matrix
\begin{equation}
T=
\left(
\begin{array}{ll}
 T^{++} & T^{+-}\\
 T^{-+}  & T^{--}
\end{array}
\right)
=
\left(
\begin{array}{ll}
 T^{c}      & T^{\less}\\
 T^{\more}  & T^{a}
\end{array}
\right),
\label{z5.7}  
\end{equation}  
where the superscripts $(+/-)$ indicate the branch of the Keldysh
contour for the pair of time arguments. In the last line of 
Eq.~(\ref{z5.6}), ${\cal T}$ is the $T$-matrix associated with 
the ``imaginary evolution'' on $C_x$. Finally, we note
that there are two cross components of the $\underline{T}$-matrix,
${\cal T}^{\moreless}$, which relate the time evolution and
initial correlations.

The real-time
components of Eqs.~(\ref{z5.2a}) and~(\ref{z5.3a}) yield 
for $T$ the equations   
\begin{eqnarray}
& &
T=V + iV\left(G\otimes G\right) T + 
iV\left({\cal G}^{\less}\otimes {\cal G}^{\less}\right)
{\cal T}^{\more},
\label{z5.12a}\\[5pt]
& &
T=V + iT\left(G\otimes G\right)V + 
i{\cal T}^{\less}\left({\cal G}^{\more}\otimes {\cal G}^{\more}\right)
V,
\label{z5.12b}
\end{eqnarray}
which differ from the analogous equations in the standard
Green's function formalism by the last terms arising from initial
correlations. These terms contain the cross $T$-matrices
${\cal T}^{\moreless}$ and , hence, Eqs.~(\ref{z5.12a}) and~(\ref{z5.12b})
are not closed. Thus, though our main interest is with the real-time
$T$-matrix defined on the Keldysh contour, 
we have to consider equations for the other components
of the full $\underline{T}$-matrix. 

By writing  Eqs.~(\ref{z5.2a}) and~(\ref{z5.3a}) 
for $\underline{T}(12,1'2')$ with $1,2\in C^{}_x$ and 
$1',2'\in C^{}_x$, we find that
the ``thermodynamic'' $T$-matrix  satisfies the equations
\begin{equation}
{\cal T}={\cal V} + {\cal V}
 \left({\cal G}\otimes {\cal G}\right) {\cal T},
\qquad
{\cal T}={\cal V} + {\cal T}
 \left({\cal G}\otimes {\cal G}\right) {\cal V},
\label{z5.14}
\end{equation}
where no real-time quantities appear, as it must be.

In order to derive equations for the  cross
$T$-matrices, we make use of Eq.~(\ref{z5.2}), where $1,2\in C^{}_x$
and $1',2'\in C$. Then, in the matrix notation, we obtain
\begin{equation}
{\cal T}^{\more}=
{\cal V}\left({\cal G}\otimes {\cal G}\right){\cal T}^{\more}
-{\cal V}\left({\cal G}^{\more}\otimes {\cal G}^{\more}\right) T.
\label{cE.1}
\end{equation}
Similarly, taking $1,2\in C$ and $1',2'\in C^{}_x$ in Eq.~(\ref{z5.3})
yields
\begin{equation}
{\cal T}^{\less}= 
{\cal T}^{\less}\left({\cal G}\otimes {\cal G}\right) {\cal V}
-T\left({\cal G}^{\less}\otimes {\cal G}^{\less}\right) {\cal V}. 
\label{cE.2}
\end{equation}
The above equations can be solved formally for the cross $T$-matrices
to give
\begin{eqnarray}
& &
{\cal T}^{\more}=
-\left[I-{\cal V}\left({\cal G}\otimes {\cal G}\right)\right]^{-1} 
{\cal V}
\left({\cal G}^{\more}\otimes {\cal G}^{\more}\right) T,
\nonumber\\[5pt]
& &
{\cal T}^{\less}=
- T\left({\cal G}^{\less}\otimes {\cal G}^{\less}\right)
{\cal V}
\left[I-\left({\cal G}\otimes {\cal G}\right) {\cal V}\right]^{-1},
\label{cE.3}
\end{eqnarray}
where $I$ is the identity matrix. 
To simplify these expressions, we make
use of Eqs.~(\ref{z5.14}), from which it follows that
\begin{equation}
\left[I-{\cal V}\left({\cal G}\otimes {\cal G}\right)\right]^{-1} 
{\cal V}=
{\cal V}
\left[I-\left({\cal G}\otimes {\cal G}\right) {\cal V}\right]^{-1}
={\cal T}.
\label{cE.4}
\end{equation}     
Now one sees that the cross $T$-matrices~(\ref{cE.3})  
can be written in the form
\begin{equation}
{\cal T}^{\less}=
- T \left({\cal G}^{\less}\otimes {\cal G}^{\less}\right)
{\cal T},
\qquad
{\cal T}^{\more}= 
- {\cal T}\left({\cal G}^{\more}\otimes {\cal G}^{\more}\right) T.
\label{z5.13}
\end{equation}
These expressions allow us to eliminate the cross $T$-matrices
in Eqs.~(\ref{z5.12a}) and~(\ref{z5.12b}).
Then we get for the real-time $T$-matrix
the following equations:
\begin{eqnarray}
& &
T= V +iV
\left[
\left(G\otimes G\right) - 
\left({\cal G}^{\less}\otimes {\cal G}^{\less} \right)
{\cal T} 
\left({\cal G}^{\more}\otimes {\cal G}^{\more} \right)
\right] T,
\label{z5.15a}\\[5pt]
& &
T= V +iT
\left[
\left(G\otimes G\right) - 
\left({\cal G}^{\less}\otimes {\cal G}^{\less} \right)
{\cal T} 
\left({\cal G}^{\more}\otimes {\cal G}^{\more} \right)
\right] V.
\label{z5.15b}
\end{eqnarray} 
As well as being compact, they illustrate
the physics of short-time processes in the presence of
initial correlations. Let us for a moment suppose that the 
second terms in square brackets are omitted. Then 
Eqs.~(\ref{z5.15a}) and~(\ref{z5.15b}) will describe binary
collisions, $(G\otimes G)$ being a two-particle propagator
in a medium. This picture 
of binary collisions is adequate at the stage of the
evolution where initial correlations have been damped. 
The structure of the full Eqs.~(\ref{z5.15a}) and~(\ref{z5.15b})
suggests that initial correlations may be regarded as an
additional source of scattering with the corresponding $T$-matrix
represented by ${\cal T}$ and the propagators represented by 
the cross Green's functions. 
It should be noted, however, that the ``correlation scattering''
described by ${\cal T}$ and the dynamical scattering described by
the interaction matrix $V$ proceed concurrently, which can be
seen by iterating Eqs.~(\ref{z5.15a}) and~(\ref{z5.15b}).  

It is convenient to reformulate Eqs.~(\ref{z5.15a}) and~(\ref{z5.15b})
in terms of quantities which reflect more clearly
the two-particle character of the $T$-matrix approximation.
First of all, recalling the definition~(\ref{m4.10}) of
$V(12,1'2')$, we check by iteration that
Eqs.~(\ref{z5.15a}) and~(\ref{z5.15b}) imply the following structure
of $T(12,1'2')$:  
\begin{equation}
T(12,1'2')=
\langle r_1 r_2|T_{12}(t_1,t'_1)|r'_1 r'_2\rangle\,
\delta^{}_C(t_1-t_2)\,\delta^{}_C(t'_1-t'_2).
\label{z5.16}  
\end{equation}
This formula  may serve as a definition of
an operator $T^{}_{12}(t_1,t'_1)$
which will be referred to as the two-particle 
real-time $T$-matrix. Similarly, we introduce the two-particle 
thermodynamic $T$-matrix, ${\cal T}^{}_{12}(x^{}_1,x'_1)$,
and the interaction operator, $V^{}_{12}$, through the relations 
\begin{eqnarray}
& &
{\cal T}(12,1'2')=
\langle r^{}_1 r^{}_2|{\cal T}^{}_{12}(x^{}_1,x'_1)|r'_1 r'_2 \rangle\,
\delta(x^{}_1-x^{}_2)\,\delta(x'_1-x'_2),
\label{z5.16x}\\[8pt] 
& &
V(12,1'2')=
\langle r_1 r_2| V^{}_{12}|r'_1 r'_2\rangle\,
\delta^{}_C(t_1-t_2)\,\delta^{}_C(t_1-t'_1)\,
\delta^{}_C(t'_1-t'_2).
\label{z5.17} 
\end{eqnarray}   
A comparison with Eq.~(\ref{m4.10}) shows that
the matrix elements of $V^{}_{12}$ coincide with the
interaction amplitudes in the Hamiltonian~(\ref{2.1}). 
Equation~(\ref{z5.15a}) is written in terms of the above operators 
(denoting only the time arguments) as
\begin{equation}
T^{}_{12}(t,t')=
V_{12}\,\delta_C(t-t')
{}+ i\int_C dt^{\prime\prime}\,
V_{12}
\left\{
G^{}_{12}(t,t^{\prime\prime})-{\cal G}^{\less}_{12}(t)\,{\cal T}_{12}\,
{\cal G}^{\more}_{12}(t^{\prime\prime}) 
\right\}  
T_{12}(t^{\prime\prime},t'),
\label{z5.19} 
\end{equation}
where we have introduced the real-time and cross 
Green's functions describing two independent particles: 
\begin{eqnarray}
& &
G^{}_{12}(t,t')=
G_1(t,t')\,G_2(t,t'),
\label{z5.20}\\[5pt]  
& &
{\cal G}^{\moreless}_{12}(t)=
{\cal G}^{\moreless}_{1}(t)\,
{\cal G}^{\moreless}_{2}(t).
\label{z5.21}  
\end{eqnarray}
It goes without saying that Eq.~(\ref{z5.19}) is a matrix
equation since each of the arguments, $t$ and $t'$,
can be taken on either of the branches of the Keldysh contour
$C$. Therefore it makes sense  to introduce two-particle operators
$T^{\moreless}_{12}$, $T^{c}_{12}$, and $T^{a}_{12}$, which
are defined in the time interval $t_0\leq t < \infty$.
Again, it is more convenient to use the retarded
$(T^R_{12})$ and advanced $(T^A_{12})$ operators instead
of the chronological and anti-chronological ones.
Following the usual rules~\cite{Botermans90} and 
recalling that 
functions ${\cal G}^{\moreless}_{}(t)$ do not depend on whether
$t$ is taken on the branch $C^+$ or on the branch $C^-$ of the
Keldysh contour, Eq.~(\ref{z5.19}) is converted into the set of 
equations
\begin{equation}
T^{R/A}_{12}(t,t')=
V^{}_{12}\,\delta(t-t')
+i \int_{t_0}^{\infty} dt^{\prime\prime}\,
V_{12}\,F^{R/A}_{12}(t,t^{\prime\prime})\,
T^{R/A}_{12}(t^{\prime\prime},t'),
\label{z5.22} 
\end{equation}
\begin{equation}
T^{\moreless}_{12}(t,t')=
i \int_{t_0}^{\infty} dt^{\prime\prime}\,
dt^{\prime\prime\prime}\,
T^{R}_{12}(t,t^{\prime\prime})
\left\{
g^{\moreless}_{12}(t^{\prime\prime},t^{\prime\prime\prime})
- {\cal G}^{\less}_{12}(t^{\prime\prime})\,{\cal T}_{12}\,
{\cal G}^{\more}_{12}(t^{\prime\prime\prime})  
\right\}
T^{A}_{12}(t^{\prime\prime\prime},t^{\prime}),
\label{z5.23}  
\end{equation}
where the two-particle Green's functions $g^{\moreless}_{12}$
and $F^{R/A}_{12}$ are defined as
\begin{eqnarray}
& &
g^{\moreless}_{12}(t,t')=
g^{\moreless}_{1}(t,t')\,
g^{\moreless}_{2}(t,t'),
\label{z5.24}\\[5pt]
& &
F^{R/A}_{12}(t,t')=
\pm\, \theta[\pm(t-t')] 
\left\{
g^{\more}_{12}(t,t') - g^{\less}_{12}(t,t')
\right\}.
\label{z5.25}
\end{eqnarray}
Equation~(\ref{z5.22}) describes effective two-particle interactions
in a medium. It is similar to the analogous equation in the standard
Green's function formalism~\cite{Danielewicz84a,Botermans90} and goes over to
the latter equation in the limit $t_0\to - \infty$. 
Thus, the influence of
initial correlations on the $T$-matrices $T^{R/A}_{12}$
manifests itself only through the short-time behavior of the
propagators $F^{R/A}_{12}$. On the other hand,
Eq.~(\ref{z5.23}), which can be viewed as a {\em generalized optical
theorem}, contains an explicit contribution from initial correlations,
the second term in braces.  

It remains to write the cross $T$-matrices~(\ref{z5.13}) in terms
of two-particle functions. According to the definition of the matrix
multiplication, we have
\begin{equation}
{\cal T}^{\less}_{12}(t)=
-\int_{t_0}^{\infty} dt'\,
T^R_{12}(t,t')\,
{\cal G}^{\less}_{12}(t')\,
{\cal T}^{}_{12},
\qquad
{\cal T}^{\more}_{12}(t)=
-\int_{t_0}^{\infty} dt'\,
{\cal T}^{}_{12}\, {\cal G}^{\more}_{12}(t')\,
T^A_{12}(t',t).
\label{z5.26}
\end{equation}
The ``thermodynamic'' $T$-matrix, ${\cal T}^{}_{12}$, is to be
evaluated from Eqs.~(\ref{z5.14}). In that case we have to deal only with
the problem of correlations in the initial state of the system.\\

\noindent
{\em B.\ Expressions for the self-energies}\\

Substitution of the two-particle mixed Green's function 
(\ref{z5.4}) into Eqs.~(\ref{m4.14}) yields the expression 
for the self-energy $\underline{\Sigma}$ in 
terms of the $\underline{T}$-matrix:
\begin{equation}
\underline{\Sigma}(1,1')= i\eta\,
\widetilde{\underline{T}}(12,1'2')\,\underline{G}(2',2^+)=
i\eta\,\underline{G}(2^-,2')\,\widetilde{\underline{T}}(12',1'2).
\label{z5.27}
\end{equation}
For its components [see Eq.~(\ref{m4.4})], we have
\begin{eqnarray}
& &
\Sigma(1,1')= i\eta\,\widetilde{T}(12,1'2')\,G(2',2^+)=
i\eta\,G(2^-,2')\,\widetilde{T}(12',1'2),
\label{z5.28}\\[5pt]
& &
{\cal K}^{\moreless}(1,1')=
i\eta\,\widetilde{\cal T}^{\moreless}(12,1'2')\,
{\cal G}^{\lessmore}(2',2),
\label{z5.29}\\[5pt]
& &
{\cal K}(1,1')=
\eta\, \widetilde{\cal T}(12,1'2')\,{\cal G}(2',2^+)=
\eta\, {\cal G}(2^-,2')\,\widetilde{\cal T}(12',1'2).
\label{z5.30}
\end{eqnarray}
The above expression for ${\cal K}$, combined with Eqs.~(\ref{2.12})
and~(\ref{z5.14}), leads to a closed equation for the 
thermodynamic Green's function ${\cal G}$. 
However, our prime interest here is with
formulae~(\ref{z5.28}) and~(\ref{z5.29}) which have a direct
relationship to the real-time dynamics.
Going to the components $\Sigma^{\moreless}$ and $\Sigma^{R/A}$
of the real-time self-energy, we infer from Eq.~(\ref{z5.28}) that
\begin{eqnarray}
& &
\hspace*{-25pt}
\Sigma^{\moreless}_1(t,t')=
i\eta\,
\mathop{\rm Tr}\limits_{(2)}\,\widetilde{T}^{\moreless}_{12}(t,t')\,
g^{\lessmore}_2(t',t),
\label{z5.31a}\\[5pt] 
& &
\hspace*{-25pt}
\Sigma^{R/A}_1(t,t')=
i\eta\,
\mathop{\rm Tr}\limits_{(2)}
\left\{
\widetilde{T}^{R/A}_{12}(t,t')\,
g^{\less}_2(t',t)
+
\widetilde{T}^{\less}_{12}(t,t')\,
g^{A/R}_2(t',t)
\right\}.
\label{z5.31b} 
\end{eqnarray}
We now turn to Eq.~(\ref{z5.29}). The cross $T$-matrices can be
eliminated by means of Eqs.~(\ref{z5.26}). Then 
we get
\begin{eqnarray}
& &
{\cal K}^{\less}_1(t)=
-i\eta
\int_{t_0}^{\infty} dt'\,
\mathop{\rm Tr}\limits_{(2)}
\left\{
T^{R}_{12}(t,t')\,
{\cal G}^{\less}_{12}(t')\,\widetilde{{\cal T}}^{}_{12}\,
{\cal G}^{\more}_{2}(t)
\right\},
\nonumber\\[5pt] 
& &
{\cal K}^{\more}_1(t)=
-i\eta
\int_{t_0}^{\infty} dt'\,
\mathop{\rm Tr}\limits_{(2)}
\left\{
{\cal G}^{\less}_{2}(t)\,
\widetilde{{\cal T}}^{}_{12}\,{\cal G}^{\more}_{12}(t')\,
T^{A}_{12}(t',t)\,
\right\}.
\label{z5.32}   
\end{eqnarray} 
In the special case
that the entropy operator contains the one-particle
term $\hat S^0$ only, we have ${\cal T}^{}_{12}=0$ and,
consequently, the cross components 
of the self-energy vanish.

Now Eqs.~(\ref{m4.24})\,--\,(\ref{m4.27}), (\ref{m4.31}),
(\ref{m4.34}), together with the expressions for the self-energies
given in this section, provide a closed description for the
evolution of the system in the $T$-matrix approximation. Initial
correlations enter into play through the thermodynamic Green
function ${\cal G}$ and the thermodynamic $T$-matrix 
${\cal T}^{}_{12}$, which are to be calculated from
Eqs.~(\ref{2.12}) and~(\ref{z5.14}). Although this scheme is
self-consistent, it can only be realized 
by numerical calculations. 
To gain a physical picture of short-time dynamics and
understand the role  of initial correlations, it is reasonable
to go over to a simpler description based on a kinetic
equation for the one-particle density matrix. In the next section
we will apply the mixed Green's function formalism to the derivation
of explicit kinetic equations.

\setcounter{equation}{0}
\renewcommand{\theequation}{7.\arabic{equation}}

\vspace*{15pt}

\begin{center}
VII. {\large K}INETIC {\large E}QUATIONS 
\end{center}

\vspace*{5pt}

For simplicity,
we  assume the initial state of the system to be 
spatially homogeneous. Generalization to inhomogeneous systems is
straightforward but adds of course algebraic complexity.    
In the homogeneous case, it is convenient to work in momentum 
representation where all matrix elements are defined with
respect to normalized single-particle quantum states 
$|p\rangle=|{\bf p},\sigma\rangle$ in a volume $\Omega$ with  
periodic boundary conditions. As usual, the limit $\Omega\to\infty$
must be taken at the end of the calculations. 
By going from
the field operators to the creation and annihilation operators,
$a^{\dagger}_p$ and $a^{}_p$, we find that
the one-particle density matrices 
$\varrho^{\moreless}_1(t)$, given in the $r$-representation
by Eqs.~(\ref{3.8x}) and~(\ref{3.8xx}), now have the elements  
\begin{equation}
\langle p^{}_1|\varrho^{\less}_1(t)|p'_1\rangle=
\delta^{}_{p^{}_1p'_1}\,f^{}_{p^{}_1}(t),
\qquad
\langle p^{}_1|\varrho^{\more}_1(t)|p'_1\rangle=
\delta^{}_{p^{}_1p'_1}\,{\bar f}^{}_{p^{}_1}(t),
\label{z6.2}
\end{equation}
where 
$f^{}_{p}(t)=\langle a^{\dagger}_{pH}(t)\,a^{}_{pH}(t)\rangle$ 
is the one-particle distribution function and
\begin{equation}
{\bar f}^{}_{p}(t)=1 +\eta\,f^{}_{p}(t).
\label{z6.2x}
\end{equation} 
For Fermi systems ($\eta=-1$), ${\bar f}^{}_{p}(t)$ 
is the distribution function for holes.\\

\noindent
{\em A.\ Short-time kinetic equation in the $T$-matrix approximation}\\

Our starting point is the generalized kinetic equation~(\ref{c5.0}).
Taking its diagonal part with respect to time
variables and the $p$-variables, the left-hand side reduces to 
$\eta\,\partial f^{}_{p^{}_1}(t)/\partial t$. 
It is convenient to eliminate retarded and advanced
functions on the right-hand side with the aid of 
Eq.~(\ref{m4.19a}) and the relation~\cite{Danielewicz84a}
\begin{equation}
\Sigma^{R/A}(t,t')=
\Sigma^{\delta}(t,t')
\pm\,\theta[\pm(t-t')]\Big\{
\Sigma^{\more}(t,t')- \Sigma^{\less}(t,t')
\Big\},
\label{z6.3}
\end{equation}
where $\Sigma^{\delta}(t,t')$ is a singular term in the 
self-energies. In the spatially homogeneous
case, $\Sigma^{\delta}$ does not contribute to the diagonal part
of Eq.~(\ref{c5.0}), so that we arrive at the equation
\begin{eqnarray}
\frac{\partial}{\partial t}\, f^{}_{p^{}_1}(t)=
\hspace*{-20pt}
{}&{}& 
\eta
\int_{t^{}_0}^t dt'
\left\{
g^{\less}_{p^{}_1}(t,t')\,\Sigma^{\more}_{p^{}_1}(t',t)
+ 
\Sigma^{\more}_{p^{}_1}(t,t')\,g^{\less}_{p^{}_1}(t',t)
\right\}
\nonumber\\[6pt]
&{}&-\eta
\int_{t^{}_0}^t dt'
\left\{
g^{\more}_{p^{}_1}(t,t')\,\Sigma^{\less}_{p^{}_1}(t',t)
+ \Sigma^{\less}_{p^{}_1}(t,t')\,g^{\more}_{p^{}_1}(t',t)
\right\}
\nonumber\\[6pt]
&{}&+\eta
\left\{
{\cal K}^{\less}_{p^{}_1}(t)\,{\cal G}^{\more}_{p^{}_1}(t)
-{\cal G}^{\less}_{p^{}_1}(t)\,{\cal K}^{\more}_{p^{}_1}(t)
\right\},
\label{z6.4} 
\end{eqnarray}
where we have taken into account that the Green's functions
and the self-energies are diagonal matrices with respect to the one-particle
$p$-variables.

Equation~(\ref{z6.4}) is still exact.  We now take for the self-energies
approximate expressions~(\ref{z5.31a}) and~(\ref{z5.32}). 
Using also the optical theorem~(\ref{z5.23}), 
the kinetic equation can be written in the form
\begin{equation}
\frac{\partial}{\partial t}\, f^{}_{p^{}_1}(t)=
I^{(B)}_{p^{}_1}(t) +I^{(C)}_{p^{}_1}(t)
\label{z6.5}
\end{equation}
with the collision integrals 
\begin{equation}
I^{(B)}_{p^{}_1}(t)=
\sum_{p^{}_2}\left\langle p^{}_1 p^{}_2\left|
I^{(B)}_{12}(t)
\right| p^{}_1 p^{}_2\right\rangle,
\qquad
I^{(C)}_{p^{}_1}(t)=
\sum_{p^{}_2}\left\langle p^{}_1 p^{}_2\left|
I^{(C)}_{12}(t)
\right| p^{}_1 p^{}_2\right\rangle. 
\label{z6.5a}
\end{equation}
The two-particle collision operators, $I^{(B)}_{12}(t)$
and $I^{(C)}_{12}(t)$, are defined as
\begin{eqnarray}
& &
\hspace*{-20pt}
I^{(B)}_{12}(t)=
\int_{t_0}^{t}dt'\,
\left\{ 
g^{\more}_{12}(t,t')
\left({T}^{R}_{12}\,g^{\less}_{12}\,\widetilde{T}^{A}_{12}\right)\!(t',t)
+\left(\widetilde{T}^{R}_{12}\,g^{\less}_{12}\,T^{A}_{12}\right)\!(t,t')\,
g^{\more}_{12}(t',t)
\right\}
\nonumber\\[6pt]
& &
\hspace*{10pt}
{}-\int_{t_0}^{t}dt'\,
\left\{
g^{\less}_{12}(t,t')
\left({T}^{R}_{12}\,g^{\more}_{12}\,\widetilde{T}^{A}_{12}\right)\!(t',t)
+ 
\left(\widetilde{T}^{R}_{12}\,g^{\more}_{12}\,T^{A}_{12}\right)\!(t,t')\,
g^{\less}_{12}(t',t)
\right\},
\label {z6.6}\\[10pt]
& &
\hspace*{-20pt}
I^{(C)}_{12}(t)=
i\left(C^{}_{12}\,T^{A}_{12} -T^{R}_{12}\,C^{}_{12}\right)\!(t,t)
\nonumber\\[6pt]
& &
\hspace*{10pt}
{}+\int_{t_0}^{t} dt'\,
\left\{
\left(g^{\less}_{12}- g^{\more}_{12}\right)\!(t,t')
\left(T^{R}_{12}\,C^{}_{12}\,T^{A}_{12}\right)\!(t',t)
\right.
\nonumber\\[6pt]
& &
\hspace*{100pt}
\left.
{}+\left(T^{R}_{12}\,C^{}_{12}\,T^{A}_{12}\right)\!(t,t')
\left(g^{\less}_{12}- g^{\more}_{12}\right)\!(t',t)
\right\},
\label{z6.7}
\end{eqnarray} 
where we have introduced the two-particle time correlation matrix 
\begin{equation}
C^{}_{12}(t,t')=
{\cal G}^{\less}_{12}(t)\,\widetilde{{\cal T}}^{}_{12}\,
{\cal G}^{\more}_{12}(t').
\label{z6.8}
\end{equation} 
In writing Eqs.~(\ref {z6.6}) and~(\ref{z6.7}), 
we have followed the
convention that matrix multiplication like $ABC$ implies 
integration over all intermediate time arguments. 

The structure of $I^{(B)}_{12}(t)$ is similar to the structure of
quantum Boltzmann-like collision operators in the $T$-matrix
approximation~\cite{KadanoffBaym62,Danielewicz84a,Botermans90}.
The new ingredient is the operator $I^{C}_{12}(t)$ which involves
the effects of initial correlations through the matrix $C^{}_{12}(t,t')$.
However, this is not the end of the story,
since Eqs.~(\ref {z6.6}) and~(\ref{z6.7}) contain 
double-time correlation functions $g^{\moreless}_{12}(t,t')$ 
which are to be expressed in terms of the distribution function 
$f^{}_p(t)$ for Eq.~(\ref{z6.5}) to become a closed kinetic equation.
We also note that  the time evolution of the cross
correlation functions ${\cal G}^{\moreless}_{12}(t)$
in Eq.~(\ref{z6.8}) must be specified.
We shall make use of Eqs.~(\ref{c5.zz}) and~(\ref{c5.zzz}) to 
complete the kinetic equation. Then we have the following expressions for 
the two-particle quantities of interest:
\begin{eqnarray}
& &
g^{\moreless}_{12}(t,t')=
g^{R}_{12}(t,t')
\varrho^{\moreless}_{1}(t')\varrho^{\moreless}_2(t')
+ \varrho^{\moreless}_{1}(t)\varrho^{\moreless}_2(t)
g^{A}_{12}(t,t'),
\label{z6.10}\\[8pt]
& &
{\cal G}^{\less}_{12}(t)=
-g^{R}_{12}(t,t^{}_0)\,{\cal G}^{\less}_{12}(t^{}_0),
\qquad
{\cal G}^{\more}_{12}(t)=
-{\cal G}^{\more}_{12}(t^{}_0)\,g^{A}_{12}(t^{}_0,t),
\label{z6.11}  
\end{eqnarray}
where
\begin{equation}
g^{R/A}_{12}(t,t')= g^{R/A}_{1}(t,t')\,g^{R/A}_{2}(t,t')
\label{z6.12} 
\end{equation}
are two-particle propagators. We show in Appendix~E that
Eqs.~(\ref{z6.11}) provide a convenient way of writing
the time correlation matrix~(\ref{z6.8}):
\begin{equation}
C^{}_{12}(t,t')=
g^{R}_{12}(t,t^{}_0)\,\chi^{}_{12}(t^{}_0)\,g^{A}_{12}(t^{}_0,t').
\label{z6.13}   
\end{equation}
Here $\chi^{}_{12}(t^{}_0)$ is the initial two-particle 
correlation matrix $\chi^{}_{12}(t)$, which is defined as
the correlated part of the two-particle density matrix:
\begin{equation}
\chi^{}_{12}(t)=
\varrho^{}_{12}(t) - 
\left(\varrho^{}_1(t)\,\varrho^{}_2(t)\right)^{}_{\rm ex}.
\label{z6.14}
\end{equation}
In momentum representation, the matrix elements of $\varrho^{}_{12}(t)$
are given by
\begin{equation}
\langle p^{}_1p^{}_2|\varrho^{}_{12}(t)|p'_1p'_2\rangle=
\big\langle 
a^{\dagger}_{p'_2H}(t)a^{\dagger}_{p'_1H}(t)
a^{}_{p^{}_1H}(t) a^{}_{p^{}_2H}(t)
\big\rangle. 
\label{zz6.14}
\end{equation}
After inserting the expressions~(\ref{z6.10}) and~(\ref{z6.13}) into
Eqs.~(\ref{z6.6}) and~(\ref{z6.7}), the collision integrals can then be
written in terms of the one-particle distribution function
$f^{}_{p}(t)$, the two-particle $T$-matrices 
$T^{R/A}_{12}(t,t')$,
and the propagators $g^{R/A}_{p}(t,t')$. Thus, to obtain a closed
kinetic description of short-time dynamics, it remains to consider
equations for $T^{R/A}$ and $g^{R/A}$. Due to the properties
\begin{equation}
\left[g^{R}_{p^{}_1}(t,t')\right]^*=
g^{A}_{p^{}_1}(t',t),
\qquad
\langle p^{}_1p^{}_2|T^{R}_{12}(t,t')|p'_1p'_2\rangle^*=
\langle p^{\prime}_1p^{\prime}_2|T^{A}_{12}(t',t)|p^{}_1p^{}_2\rangle,
\label{z6.16xz} 
\end{equation}
we may dwell on the retarded functions only.
Equation for $T^{R}_{12}$ follows directly from Eq.~(\ref{z5.22})
if we use the ansatz~(\ref{z6.10}) to  express $F^{R}_{12}$, given by
Eq.~(\ref{z5.25}), in terms of the one-particle density matrix.
We then obtain 
\begin{equation}
T^{R}_{12}(t,t')=V^{}_{12}\,\delta(t-t')
-i\int_{t^{}_0}^{t} dt''\, V^{}_{12}\,g^{R}_{12}(t,t'')
\left(\varrho^{\less}_1 \varrho^{\less}_2 -
\varrho^{\more}_1 \varrho^{\more}_2\right)^{}_{t''}
T^{R}_{12}(t'',t').
\label{7.18}
\end{equation}
Finally, the equation for $g^{R}_{1}(t,t')$
follows from Eq.~(\ref{m4.21}) in
momentum representation. Since we consider
a  homogeneous system, we have
\begin{equation}
\left( i\,\frac{\partial}{\partial t} -\varepsilon^{}_{p^{}_1}\right)
g^{R}_{p^{}_1}(t,t')=
\delta(t-t')+\int_{t^{}_0}^{\infty} dt''\, 
\Sigma^{R}_{p^{}_1}(t,t'')\,g^{R}_{p^{}_1}(t'',t'),
\label{z6.27a}
\end{equation}
where $\varepsilon^{}_{p}=|{\bf p}|^2/2m$ is the free-particle
energy. The self-energy $\Sigma^{R}_{p^{}_1}$ is then
found in terms of $f^{}_{p}$,
$T^{R/A}_{12}$, and $g^{R/A}_{p}$ 
by using Eq.~(\ref{z5.31b}), the optical theorem~(\ref{z5.23}),
and the ansatz~(\ref{c5.zz}). 
We will not give this somewhat lengthy expression 
for $\Sigma^{R}_{p^{}_1}$.\\

\noindent
{\em B.\ Short-time kinetic equation in Born approximation}\\ 

To demonstrate the principal features of 
the collision integrals in the kinetic equation~~(\ref{z6.5}),
we shall consider the operators~(\ref{z6.6}) and~(\ref{z6.7})
in Born approximation for the $T$-matrices.
For weak interaction, Eq.~(\ref{7.18}) can be solved by iteration.
The result in a matrix notation up to second order in $V^{}_{12}$
reads
\begin{equation}
T^{R}_{12}(t,t')=
V^{}_{12}\,\delta(t-t') -
i\,V^{}_{12}\,g^{R}_{12}(t,t')
\left(\varrho^{\less}_1 \varrho^{\less}_2 -
\varrho^{\more}_1  \varrho^{\more}_2\right)^{}_{t'}
V^{}_{12}.
\label{z6.20}
\end{equation}
The analogous result for the advanced $T$-matrix is
\begin{equation} 
T^{A}_{12}(t,t')=
V^{}_{12}\,\delta(t-t') +
i\,V^{}_{12}
\left(\varrho^{\less}_1 \varrho^{\less}_2 
- \varrho^{\more}_1  \varrho^{\more}_2\right)^{}_{t}
g^{A}_{12}(t,t')\,
V^{}_{12}.
\label{z6.21}
\end{equation}
Now, to find the collision integrals in Born approximation,
these expressions are to be inserted into Eqs.~(\ref{z6.6})
and~(\ref{z6.7}). Note that in all terms, except for the first term
on the right-hand side of Eq.~(\ref{z6.7}), one can 
take the $T$-matrices to first order in the interaction
$V^{}_{12}$. As already discussed, the correlation 
functions $g^{\moreless}_{12}$ are eliminated 
in favor of the one-particle density matrices 
by means of the ansatz~(\ref{z6.10}).
Then, recalling Eqs.~(\ref{z6.2}) for the elements of
the one-particle density matrices,  
 a simple algebra gives for the collision integrals
\begin{eqnarray}
& &
\hspace*{-40pt}
I^{(B)}_{p^{}_1}(t)=
-\sum_{p^{}_2p'_1p'_2}\int_{t_0}^{t} dt'\,
W^{(B)}_{p^{}_1 p^{}_2,p'_1 p'_2}(t,t')
\left(
f^{}_{p^{}_1} f^{}_{p^{}_2}
{\bar f}^{}_{p^{\prime}_1}{\bar f}^{}_{p^{\prime}_2} 
-
{\bar f}^{}_{p^{}_1}{\bar f}^{}_{p^{}_2}
f^{}_{p^{\prime}_1} f^{}_{p^{\prime}_2}     
\right)^{}_{t'},
\label{z6.22}\\[8pt]
& &
\hspace*{-40pt}
I^{(C)}_{p^{}_1}(t)=
2\sum_{p^{}_2} {\rm Im}\big\{
\langle p^{}_1 p^{}_2|V^{}_{12}C^{}_{12}(t,t)|p^{}_1 p^{}_2\rangle
\big\}
\nonumber\\[6pt]
& &
\hspace*{20pt}
{}+\sum_{p^{}_2 p'_1p'_2}\int_{t_0}^{t} dt'\,
\Big\{
W^{(C)}_{p^{}_1 p^{}_2,p'_1 p'_2}(t,t')
\left(
f^{}_{p^{\prime}_1} f^{}_{p^{\prime}_2}
- {\bar f}^{}_{p^{\prime}_1}{\bar f}^{}_{p^{\prime}_2}
\right)^{}_{t'}
\nonumber\\[6pt]   
& &
\hspace*{100pt}
{}-    
W^{(C)}_{p^{\prime}_1 p^{\prime}_2,p^{}_1 p^{}_2}(t,t')
\left(
f^{}_{p^{}_1} f^{}_{p^{}_2}
- {\bar f}^{}_{p^{}_1}{\bar f}^{}_{p^{}_2}
\right)^{}_{t'} 
\Big\}.
\label{z6.23}   
\end{eqnarray} 
The quantities $W^{(B)}$ and $W^{(C)}$ play the role of the transition 
probabilities or the memory functions and are given by
\begin{eqnarray}
& &
W^{(B)}_{p^{}_1 p^{}_2,p'_1 p'_2}(t,t')=
\left| \langle p^{}_1 p^{}_2|\widetilde{V}^{}_{12}|p'_1 p'_2\rangle
\right|^2\,
{\rm Re}\left\{X^{(B)}_{p^{}_1 p^{}_2,p'_1 p'_2}(t,t')\right\},
\label{z6.24}\\[8pt]
& &
W^{(C)}_{p^{}_1 p^{}_2,p'_1 p'_2}(t,t')=
-2\,{\rm Re}
\left\{X^{(C)}_{p^{}_1 p^{}_2,p'_1 p'_2}(t,t')\right\},
\label{z6.25}
\end{eqnarray}
where
\begin{eqnarray}
& &
X^{(B)}_{p^{}_1 p^{}_2,p'_1 p'_2}(t,t')=
g^{R}_{p^{}_1 p^{}_2}(t,t')\,g^{A}_{p^{\prime}_1 p^{\prime}_2}(t',t),
\label{z6.24a}\\[8pt]
& &
X^{(C)}_{p^{}_1 p^{}_2,p'_1 p'_2}(t,t')=
\langle p^{}_1 p^{}_2| C^{}_{12}(t,t')\,V^{}_{12}
|p'_1 p'_2\rangle\,g^{A}_{p'_1 p'_2}(t',t)\,
\langle p'_1 p'_2|V^{}_{12}|p^{}_1 p^{}_2\rangle.
\label{z6.25a}
\end{eqnarray} 
Note that the memory function~(\ref{z6.24}) has the symmetry
property
\begin{equation}
W^{(B)}_{p^{}_1 p^{}_2,p'_1 p'_2}(t,t')=
W^{(B)}_{p'_1 p'_2,p^{}_1 p^{}_2}(t,t'),
\label{z6.26}
\end{equation}
while this is not the case for the memory function~(\ref{z6.25}).

A few comments should be made here about the results~(\ref{z6.22}) 
and~(\ref{z6.23}). The term $I^{(B)}$ is a  
Boltzmann-like collision integral, where the memory
effects are involved through the retarded and advanced Green's
functions. Collision integrals of this type are used extensively
in the delayed quantum kinetics (see, e.g.,~\cite{HaugJauho96} 
and references therein).
The extra collision integral, $I^{(C)}$, collects explicit 
contributions from
initial correlations and contains two different terms.
The first term in Eq.~(\ref{z6.23}) is linear in the 
interaction and does not depend
on the one-particle distribution function. 
Lee {\em et al.}~\cite{Lee70} were the first to derive 
the analogous term for a weakly coupled low-density gas. 
Recently the first-order correlation term in $I^{(C)}$ was 
re-derived within the density operator
theory~\cite{Bonitz98} and the real-time Green's function
formalism in which the initial correlation where specified
in the form of a cluster expansion~\cite{SemkatKremp99}.
The remaining integral term in Eq.~(\ref{z6.23})
has a ``gain-loss'' form and gives an account of a relaxation process 
caused by initial correlations.
To the authors knowledge, this term was not
considered before in short-time kinetics.
Physically, it can be relevant for describing the 
{\em quasiparticle formation\/} at the first
stage of evolution. 
The most interesting thing about the  
``gain-loss'' correlation term in Eq.~(\ref{z6.23})
is that it depends on the nonequilibrium one-particle distribution
functions, but its structure is quite different from
the structure of the Boltzmann-like collision integral~(\ref{z6.22}).      
Note that, for Fermi systems, the 
combination of one-particle distribution functions in this term,
$f^{}_{p^{}_1} f^{}_{p^{}_2}-{\bar f}^{}_{p^{}_1}
{\bar f}^{}_{p^{}_2}=
f^{}_{p^{}_1} f^{}_{p^{}_2} -(1- f^{}_{p^{}_1})(1- f^{}_{p^{}_2})$,
is nothing but the difference between particle-particle and hole-hole 
excitations, i.e., the Pauli blocking factor.\\ 
 
\noindent
{\em C. Retarded Green's function}\\   

To complete the kinetic equation~(\ref{z6.5}), we need to evaluate
the retarded Green's function from Eq.~(\ref{z6.27a}).
First of all we find 
the self-energy $\Sigma^{R}_{p^{}_1}(t,t')$,
given by Eq.~(\ref{z5.31b}), in Born approximation for the
$T$-matrices. To eliminate the correlation function $g^{\less}$,
we make use of the GKB ansatz~(\ref{c5.zz}). Then  we obtain
\begin{equation}
\Sigma^{R}_{p^{}_1}(t,t')=
\Sigma^{\rm HF}_{p^{}_1}(t)\,\delta(t-t')
+\sum_{p^{}_2} \Lambda^{}_{p^{}_1 p^{}_2}(t,t')\,
g^{A}_{p^{}_2}(t',t),
\label{z6.32}
\end{equation}
where
\begin{equation}
\Sigma^{\rm HF}_{p^{}_1}(t)=\sum^{}_{p^{}_2} 
\langle p^{}_1 p^{}_2|\widetilde{V}^{}_{12}|p^{}_1 p^{}_2\rangle\,
f^{}_{p^{}_2}(t)
\label{z6.33}
\end{equation}
is the  Hartree-Fock term and we have defined
\begin{eqnarray}
& &
\hspace*{-30pt}
\Lambda^{}_{p^{}_1 p^{}_2}(t,t')=
\frac{1}{2}
\sum_{p'_1p'_2} \left|
\langle p^{}_1 p^{}_2|\widetilde{V}^{}_{12}|p'_1 p'_2\rangle
\right|^2\,
g^{R}_{p'_1 p'_2}(t,t')
\left(f^{}_{p^{}_2} {\bar f}^{}_{p'_1}{\bar f}^{}_{p'_2}    
- {\bar f}^{}_{p^{}_2}f^{}_{p'_1}f^{}_{p'_2} 
\right)^{}_{t'}
\nonumber\\[6pt]
& &
\hspace*{80pt}
{}+\eta\, \langle p^{}_1 p^{}_2|V^{}_{12}\,C^{}_{12}(t,t')\,V^{}_{12}
|p^{}_1 p^{}_2\rangle.
\label{z6.34}
\end{eqnarray}
Insertion of Eq.~(\ref{z6.32}) into Eq.~(\ref{z6.27a}) leads to the 
integro-differential equation 
\begin{equation}
\left( i\,\frac{\partial}{\partial t} -E^{}_{p^{}_1}(t)\right)
g^{R}_{p^{}_1}(t,t')=
\delta(t-t')
+ \sum_{p^{}_2} 
\int_{t'}^{t} dt''\, \Lambda^{}_{p^{}_1 p^{}_2}(t,t'')
\big[g^R_{p^{}_2}(t,t'')\big]^* g^R_{p^{}_1}(t'',t'),
\label{z6.35y}
\end{equation}
where
\begin{equation}
E^{}_{p^{}_1}(t)=\varepsilon^{}_{p^{}_1} +\Sigma^{\rm HF}_{p_1}(t)
\label{z6.35a}
\end{equation} 
is the re-normalized particle energy. 
Equation~(\ref{z6.35y}) should to be solved together with the kinetic 
equation for the one-particle distribution function, but it seems to
be a very difficult task, even using numerical methods. We therefore
will construct an approximate solution to Eq.~(\ref{z6.35y}),
valid for a weak interaction.

First we note that the retarded Green's function $g^{R}_{p^{}_1}(t,t')$
can always be written in the form
\begin{equation}
g^{R}_{p^{}_1}(t,t')=
-i\,\theta(t-t')\,
\exp\left\{
-i\omega^{}_{p^{}_1}(t,t') -\Phi^{}_{p^{}_1}(t,t')
\right\}, 
\label{z6.27x}
\end{equation}
where
\begin{equation}
\omega^{}_{p^{}_1}(t,t')=\int_{t'}^{t} dt''\, E^{}_{p^{}_1}(t'')=
\varepsilon^{}_{p^{}_1}(t-t')
+\int_{t'}^{t} dt''\,\Sigma^{\rm HF}_{p^{}_1}(t''),
\label{z6.27xx}
\end{equation}   
and $\Phi^{}_{p^{}_1}(t,t')$ is some complex-valued function. It satisfies 
the obvious initial  condition $\Phi^{}_{p}(t,t)=0$ and has a simple physical 
interpretation. Its real part determines the quasiparticle damping, while
the imaginary part can be associated with higher-order corrections to the
quasiparticle energy. Inserting the expression~(\ref{z6.27x}) into
Eq.~(\ref{z6.35y}), one obtains for $\Phi^{}_{p^{}_1}(t,t')$ an 
integro-differential equation,
which can then  be recast into the following integral
equation:
\begin{equation}
\Phi^{}_{p^{}_1}(t,t')= -
\sum_{p^{}_2}
\int_{t'}^{t} d\tau^{}_1\int_{t'}^{\tau^{}_1} d\tau^{}_2\,
{\rm e}^{i\omega^{}_{p^{}_1 p^{}_2}(\tau^{}_1,\tau^{}_2)}\,
{\rm e}^{[\Phi^{}_{p^{}_1}(\tau^{}_1,t')
-\Phi^{}_{p^{}_1}(\tau^{}_2,t')
-\Phi^{}_{p^{}_2}(\tau^{}_1,\tau^{}_2)]}\,
\Lambda^{}_{p^{}_1 p^{}_2}(\tau^{}_1,\tau^{}_2),
\label{z6.38} 
\end{equation} 
where
\begin{equation}
\omega^{}_{p^{}_1 p^{}_2}(t,t')=
\omega^{}_{p^{}_1}(t,t')+\omega^{}_{p^{}_2}(t,t').
\label{z6.38x}
\end{equation}
Since the right-hand side of Eq.~(\ref{z6.38}) is 
already of second order in the
interaction, the leading approximation for $\Phi^{}_{p}(t,t')$
can be obtained by setting $\Phi=0$ in the integrand. In the weak
coupling case, the imaginary part of $\Phi^{}_{p}(t,t')$ 
is small compared with $\omega^{}_{p}(t,t')$, so that our interest is
in the real part of $\Phi^{}_{p}(t,t')$ describing the
quasiparticle damping. With Eq.~(\ref{z6.34}), we find 
\begin{eqnarray}
& &
\hspace*{-60pt}
\Gamma^{}_{p^{}_1}(t,t')= {\rm Re}\left\{\Phi^{}_{p^{}_1}(t,t')\right\}
\nonumber\\[6pt]
& &
\hspace*{-20pt}
{}={1\over2}\sum_{p^{}_2 p'_1 p'_2}
\int_{t'}^{t} d\tau^{}_1\int_{t'}^{\tau^{}_1}d\tau^{}_2
\left|
\langle p^{}_1 p^{}_2|\widetilde{V}^{}_{12}|p'_1 p'_2\rangle
\right|^2
\nonumber\\[6pt]
& &
\hspace*{30pt}
{}\times 
\,
\cos\!\left[\Delta\omega^{}_{p^{}_1 p^{}_2,p'_1p'_2}(\tau_1,\tau_2)\right]
\left(f^{}_{p^{}_2} {\bar f}^{}_{p'_1}{\bar f}^{}_{p'_2}    
- {\bar f}^{}_{p^{}_2}f^{}_{p'_1}f^{}_{p'_2} 
\right)^{}_{\tau^{}_2}
\nonumber\\[6pt]
& &
{}-\eta \sum^{}_{p^{}_2}
\int_{t'}^{t} d\tau^{}_1\int_{t'}^{\tau^{}_1}d\tau^{}_2\,
{\rm Re}
\left\{
{\rm e}^{i\omega^{}_{p_1 p_2}(\tau_1,\tau_2)}
\langle p^{}_1 p^{}_2|V^{}_{12} C^{}_{12}(\tau^{}_1,\tau^{}_2)
V^{}_{12}|p^{}_1 p^{}_2\rangle
\right\},
\label{z6.39}
\end{eqnarray}
where
\begin{equation}
\Delta\omega^{}_{p^{}_1 p^{}_2,p'_1 p'_2}=
\omega^{}_{p^{}_1 p^{}_2}(t,t')- \omega^{}_{p'_1 p'_2}(t,t').
\label{z6.39x}
\end{equation}  
The first term in the expression~(\ref{z6.39}) 
is long-lived because a substantial
contribution to the inner integral is made by the region
$(\tau^{}_1-\tau^{}_2)$ of the order of the collision duration time
$\tau^{}_c$ and, consequently, the first term grows with
$(t-t')$ if $(t-t')\gg\tau^{}_c$. 
The behavior of the second term 
depends critically on the time $\tau^{}_{in}$ required for damping
of the initial correlations. 
Since the matrix elements of $C^{}_{12}(t,t')$ according 
to Eq.~(\ref{z6.13}) are oscillating functions of $t-t^{}_0$ and 
$t'-t^{}_0$,  the second term in Eq.~(\ref{z6.39}) is 
expected to have a plateau at $(t-t^{}_0)\gg\tau^{}_{in}$ 
and $(t'-t^{}_0)\gg\tau^{}_{in}$. The interplay between the
``quasiparticle'' and ``correlation'' contributions to
$\Gamma^{}_{p}(t,t')$ at the initial stage of the evolution
can be of interest in short-time kinetics. However, a systematic analysis of 
this point requires more detailed information on the interaction 
operator $V^{}_{12}$ and the initial
two-particle correlation matrix $\chi^{}_{12}(t^{}_0)$.

To summarize, we may conclude that a reasonable approximation for the
Green's functions in Eqs.~(\ref{z6.13}), (\ref{z6.24a}), and~(\ref{z6.25a})
is given by
\begin{equation}
g^{R}_{p^{}_1}(t,t')=\left[g^{A}_{p^{}_1}(t',t)\right]^* =
-i\,\theta(t-t')\,
\exp\left\{-i\omega^{}_{p^{}_1}(t,t') -\Gamma^{}_{p^{}_1}(t,t')\right\}.
\label{z6.40}
\end{equation}
This retarded Green's  function satisfies the equation
\begin{equation}
\left( i\frac{\partial}{\partial t} -E^{}_{p^{}_1}(t) 
+i\gamma^{}_{p^{}_1}(t,t')\right) g^R_p(t,t')
=\delta(t-t')
\label{z6.41}
\end{equation} 
with the time-dependent damping
\begin{eqnarray}
& &
\hspace*{-15pt}
\gamma^{}_{p^{}_1}(t,t')= 
\frac{\partial}{\partial t}\,\Gamma^{}_{p^{}_1}(t,t')
\nonumber\\[6pt]
& &
\hspace*{-5pt}
{}=
{1\over2}\sum_{p^{}_2 p'_1 p'_2}
\int_{t'}^{t} d\tau
\left|
\langle p^{}_1 p^{}_2|\widetilde{V}^{}_{12}|p'_1 p'_2\rangle
\right|^2
\cos\!\left[\Delta \omega^{}_{p^{}_1 p^{}_2,p'_1 p'_2}(t,\tau)\right]
\left(f^{}_{p^{}_2} {\bar f}^{}_{p'_1}{\bar f}^{}_{p'_2}    
- {\bar f}^{}_{p^{}_2}f^{}_{p'_1}f^{}_{p'_2} 
\right)^{}_{\tau}
\nonumber\\[6pt]
& &
\hspace*{30pt}
{}-\eta \sum^{}_{p^{}_2}
\int_{t'}^{t} d\tau\,
{\rm Re}
\left\{
{\rm e}^{i\omega^{}_{p_1 p_2}(t,\tau)}
\langle p^{}_1 p^{}_2|V^{}_{12} C^{}_{12}(t,\tau)
V^{}_{12}|p^{}_1 p^{}_2\rangle\right\}.
\label{z6.42}
\end{eqnarray}
Equation~(\ref{z6.41}) can be considered as an approximate 
version of Eq.~(\ref{z6.35y}).\\ 
 
\noindent
{\em D.\ Memory effects and energy conservation}\\ 

The expressions~(\ref{z6.40}) for the retarded and advanced Green's functions
with $\omega^{}_{p^{}_1}(t,t')$ and $\Gamma^{}_{p^{}_1}(t,t')$
given respectively by Eq.~(\ref{z6.27xx}) and Eq.~(\ref{z6.39}) complete the 
kinetic equation since the memory functions in the collision 
integral, as well as the matrix
$C^{}_{12}(t,t')$, can now
be written in terms of the one-particle distribution function
and the initial correlation matrix $\chi^{}_{12}(t^{}_0)$.
The most important feature of the short-time kinetic equation
is the  non-Markovian structure of the memory functions.
Let us write, for instance, the explicit expression for
the memory function~(\ref{z6.24}) using Eq.~(\ref{z6.40}).
We have
\begin{equation}
W^{(B)}_{p^{}_1 p^{}_2,p'_1 p'_2}(t,t')=
\left| \langle p^{}_1 p^{}_2|\widetilde{V}^{}_{12}|p'_1 p'_2\rangle
\right|^2\,
{\rm e}^{-\Gamma_{p^{}_1 p^{}_2 p'_1p'_2}(t,t')}\,
\cos\!\left[\Delta\omega^{}_{p^{}_1 p^{}_2,p'_1 p'_2}(t,t')
\right],
\label{z6.43}
\end{equation}
where the following designation is used:
\begin{equation}
\Gamma_{p^{}_1 p^{}_2 p'_1p'_2}(t,t') =
\Gamma^{}_{p^{}_1}(t,t')
+\Gamma^{}_{p^{}_2}(t,t')
+\Gamma^{}_{p'_1}(t,t')
+\Gamma^{}_{p'_2}(t,t').
\label{z6.44}
\end{equation}
The structure of the memory function~(\ref{z6.43}) is typical for
the memory functions used in  the delayed quantum 
kinetics~(see, e.g.,~\cite{HaugJauho96}),
but the damping function $\Gamma^{}_{p}(t,t')$ is commonly taken in a 
simplified form $\Gamma^{}_{p}(t,t')=\gamma^{}_{p}(t-t')$ with a constant
$\gamma^{}_p$. 
In some sense this corresponds to the asymptotic form
of our result~(\ref{z6.39}). It should be noted, however, 
that at the first stage of evolution such an approximation 
seems to be very rough; we see from Eq.~(\ref{z6.39}) that 
the behavior of $\Gamma^{}_p(t,t')$
is rather complicated and involves the contribution 
from initial correlations.   
Another source of memory effects is the oscillating cosine term 
in Eq.~(\ref{z6.43}), which
replaces the energy-conserving delta function in Boltzmann-like kinetic
equations. We see that
to take into account the full structure of the memory functions,
even in performing the numerical evaluation of short-time quantum kinetic 
equations, is very complicated. Nevertheless, there are some necessary
conditions which must be checked for any kinetic equation. The most important
issue is whether or not the kinetic equation obeys the conservation laws.
The problem of conservation laws in the delayed quantum kinetics is
not trivial because of strong memory effects and the influence of
initial correlations.     
We shall discuss the conservation laws for the kinetic equation, where
the collision integrals are given by Eqs.~(\ref{z6.22}) and~(\ref{z6.23}).
Since there are no problems with
the conservation of the total number of particles and the total
momentum, even for the general form of $g^R$ and $g^A$,
we restrict our discussion to the conservation of energy,
which is a serious problem in kinetic theory. 

Since we are working in the Heisenberg picture, the average energy
of the system is written as ${\cal E}(t)=\langle \hat H(t)\rangle$.
In our case the Hamiltonian has the form~(\ref{2.1}).
Going over to  momentum representation and taking into account
that the system is assumed to be spatially homogeneous, we find that
\begin{equation}
{\cal E}(t)=\sum_{p^{}_1} \varepsilon^{}_{p^{}_1} f^{}_{p^{}_1}(t)
+{1\over 2}\sum_{p^{}_1 p^{}_2}
\langle p^{}_1 p^{}_2|V^{}_{12} \varrho^{}_{12}(t)|p^{}_1 p^{}_2\rangle,
\label{z6.45}
\end{equation}
where $\varrho^{}_{12}(t)$ is the nonequilibrium two-particle density
matrix with the elements given by Eq.~(\ref{zz6.14}). 
It is convenient to separate the correlated and non-correlated parts
of $\varrho^{}_{12}(t)$ by using Eq.~(\ref{z6.14}). Then the average
energy~(\ref{z6.45}) takes the form
\begin{equation}
{\cal E}(t)= 
{\cal E}^{}_{\rm kin}(t) + {\cal E}^{}_{\rm HF}(t) + 
{\cal E}^{}_{\rm corr}(t).
\label{z6.46} 
\end{equation}
The quantity ${\cal E}^{}_{\rm kin}(t)$ is the average kinetic energy;
this is just the first term in Eq.~(\ref{z6.45}). The Hartree-Fock
contribution to the average energy, ${\cal E}^{}_{\rm HF}(t)$, 
is given by 
\begin{equation}
{\cal E}^{}_{\rm HF}(t)={1\over 2}\sum_{p^{}_1 p^{}_2}
\langle p^{}_1 p^{}_2|\widetilde V^{}_{12}|p^{}_1 p^{}_2\rangle\,
f^{}_{p^{}_1}(t) f^{}_{p^{}_2}(t).
\label{z6.47}
\end{equation}   
Finally, the term
\begin{equation}
{\cal E}^{}_{\rm corr}(t)=
{1\over 2}\sum_{p^{}_1 p^{}_2}
\langle p^{}_1 p^{}_2|V^{}_{12} \chi^{}_{12}(t)|p^{}_1 p^{}_2\rangle
\label{z6.48}
\end{equation}  
may be called the {\em correlation energy}.
Energy conservation implies that the time derivative of
${\cal E}(t)$ is zero. Having a kinetic equation
for $f^{}_{p^{}_1}(t)$, one can calculate
\begin{equation}
\frac{d}{dt}
\Big( {\cal E}^{}_{\rm kin}(t) + {\cal E}^{}_{\rm HF}(t)\Big)=
\sum_{p^{}_1} E^{}_{p^{}_1} (t)\,
\frac{\partial f^{}_{p^{}_1}(t)}{\partial t}=
\sum_{p^{}_1} E^{}_{p^{}_1}(t) I^{}_{p^{}_1}(t),
\label{z6.49}
\end{equation}
where $E^{}_{p^{}_1}(t)$ is the re-normalized particle
energy, Eq.~(\ref{z6.35a}), and $I^{}_{p^{}_1}(t)$ is the collision integral.
Note that the above relation is exact.  
Thus, the necessary condition that a kinetic equation is consistent
with the energy conservation is that the sum on the right-hand side
of Eq.~(\ref{z6.49}) can be represented as the time derivative.
If this condition is fulfilled, then we have
\begin{equation}
\sum_{p^{}_1} E^{}_{p^{}_1}(t) I^{}_{p^{}_1}(t)=
- \frac{d {\cal E}^{}_{\rm corr}(t)}{dt}.
\label{z6.50}
\end{equation}
This formula allows one to evaluate the approximate correlation energy
from the kinetic equation.

It is interesting to check the condition~(\ref{z6.50}) for the
short-time kinetic equation, in which the collision integral 
$I^{}_{p^{}_1}(t)$
is the sum of the collision integrals $I^{(B)}_{p^{}_1}(t)$
and $I^{(C)}_{p^{}_1}(t)$ given by Eqs.~(\ref{z6.22}) and~(\ref{z6.23}).
The crucial point is the choice of the retarded and advanced Green's
functions entering into these collision integrals. It turns out that
the necessary condition for the energy conservation is satisfied if
$g^{R}_{p^{}_1}(t,t')$ and $g^{A}_{p^{}_1}(t',t)$ are taken in the 
form~(\ref{z6.40}), where the damping function 
$\Gamma^{}_{p^{}_1}(t,t')$ is set to be equal to zero. 
The corresponding manipulations are
straightforward but somewhat lengthy, so that we only explain the main
points. Multiplying Eqs.~(\ref{z6.22}) and~(\ref{z6.23}) by 
$E^{}_{p^{}_1}(t)$ and then summing over $p^{}_1$,
the products of the type $E^{}_{p}(t)g^{R}_{p}(t,t')$
or $g^{A}_{p}(t',t) E^{}_{p}(t)$ appear in each term.
These products are eliminated 
with the aid of Eq.~(\ref{z6.41}) 
and the corresponding equation for $g^{A}_{p}(t',t)$
(recall that $\Gamma^{}_{p}(t,t')=0$). It can then be shown that
the $\delta$-functions do not contribute to the sum 
$\sum_{p} E^{}_{p}(t) I^{}_{p}(t)$. 
The remaining terms form the time derivatives of the
contributions to the correlation energy.   
In this way, we arrive at Eq.~(\ref{z6.50}) with
the correlation energy 
\begin{eqnarray}
{\cal E}^{}_{\rm corr}(t)&=&
{1\over 4}\sum_{p^{}_1 p^{}_2 p'_1 p'_2}
\int_{t^{}_0}^{t} dt'
\left|
\langle p^{}_1 p^{}_2|\widetilde{V}^{}_{12}|p'_1 p'_2\rangle
\right|^2
 \sin\!\left[\Delta \omega^{}_{p^{}_1 p^{}_2,p'_1p'_2}(t,t')\right]
\nonumber\\[6pt]
& &
\hspace*{120pt}
{}\times
\left(f^{}_{p^{}_1} f^{}_{p^{}_2} {\bar f}^{}_{p'_1}{\bar f}^{}_{p'_2}    
- {\bar f}^{}_{p^{}_1} {\bar f}^{}_{p^{}_2}f^{}_{p'_1}f^{}_{p'_2} 
\right)^{}_{t'}
\nonumber\\[6pt]
{}&+&
{1\over 2}\sum_{p^{}_1p^{}_2}
{\rm Re}\Big\{
\langle p^{}_1 p^{}_2|V^{}_{12} C^{}_{12}(t,t)|p^{}_1 p^{}_2\rangle
\Big\}
\nonumber\\[6pt]
{}&-&\sum_{p^{}_1 p^{}_2 p'_1 p'_2}
\int_{t^{}_0}^{t} dt'\,
{\rm Im}\left\{X^{(C)}_{p^{}_1 p^{}_2,p'_1 p'_2}(t,t')\right\}
\left(f^{}_{p'_1} f^{}_{p'_2} -  
{\bar f}^{}_{p'_1}  {\bar f}^{}_{p'_2} \right)^{}_{t'}.
\label{z6.51}
\end{eqnarray}
Here the retarded and advanced Green's functions
entering into $C^{}_{12}(t,t)$ and 
$X^{(C)}(t,t')$ 
are understood to be taken in the approximation
described above (with $\Gamma^{}_p(t,t')=0$). 
An attractive feature of the expression~(\ref{z6.51}) is that it gives
the {\em exact\/} result for the initial correlation energy.
To show this, we recall Eq.~(\ref{z6.13}), from which it follows that
$C^{}_{12}(t^{}_0,t^{}_0)=\chi^{}_{12}(t^{}_0)$. Thus we see 
that, indeed, the initial values ${\cal E}^{}_{\rm corr}(t^{}_0)$ given by
Eqs.~(\ref{z6.48}) and~(\ref{z6.51}) are identical.

Some previously obtained results follow
from Eq.~(\ref{z6.51}) as special cases. 
If all the contributions from initial correlations are omitted and
one sets $\omega^{}_{p}(t,t')=\varepsilon^{}_p(t-t')$, 
the expression~(\ref{z6.51}) reduces to the correlation
energy calculated from the non-Markovian Boltzmann equation~\cite{Morawetz95}.
The second term in the correlation energy (linear in the interaction)
agrees with the density operator 
result~\cite{Bonitz98,BonitzKrempScott96} obtained by truncating
the quantum hierarchy for the reduced density matrices.
The last contribution to the correlation energy~(\ref{z6.51}) 
stems from the ``gain-loss'' term in the collision 
integral~(\ref{z6.23}).

It may appear tempting to improve the expressions for the collision
integrals and the correlation energy by taking the retarded and advanced
Green's functions in the form~(\ref{z6.40}) with 
$\Gamma^{}_{p^{}_1}(t,t')\not=0$. This means that one goes over to the 
memory functions involving effects of the quasiparticle damping
[see, e.g., Eq.~(\ref{z6.43})]. Such a procedure is often used
(with an empirical exponential damping factor)
in numerical calculations based on non-Markovian Boltzmann
kinetic equations~\cite{HaugJauho96} and in many cases is crucial
for the stability of the results. Unfortunately,
the inclusion of the quasiparticle damping into the memory
functions leads to collision integrals which {\em do not conserve\/} the
total energy of the system. The origin of this difficulty
is easy to see. We recall that the above proof of the energy
conservation rests heavily on Eq.~(\ref{z6.41}) for the retarded
Green's function. If $\gamma^{}_{p^{}_1}(t,t')=0$, then the product
$E^{}_{p^{}_1}(t) g^{R}_{p^{}_1}(t,t')$ can exactly
be expressed in terms of the time derivative of the retarded
Green's function. For a finite damping, however, this is not the
case, and the right-hand side of Eq.~(\ref{z6.50}) involves
extra terms violating the energy conservation. Clearly, 
the contribution of these extra terms to the energy balance equation 
is determined by the time behavior of the
retarded Green's function.
Strictly speaking, the total energy is conserved only if the 
product $\gamma^{}_{p^{}_1}(t,t') g^{R}_{p^{}_1}(t,t')$ 
is equal to zero for all $t-t'$. At the initial short-time stage
of evolution $\gamma^{}_{p^{}_1}(t,t')$ is completely negligible, 
so that the total energy is conserved. The real difficulties
appear in the long-time asymptotic behavior of the collision 
integrals, where $\gamma^{}_{p^{}_1}(t,t')$ 
is close to its stationary value. 
Then the decay of $g^{R}_{p^{}_1}(t,t')$ must be 
sufficiently fast for the energy non-conserving terms
in the collision integral to be negligible.     
For the electron scattering with
optical phonons, Haug and B\'anyai~\cite{HaugBanyai96}  
proposed a model retarded Green's function
which vanishes faster than the slowly decreasing 
function with the constant quasiparticle damping.
Their numerical results show that the collision integral 
with the improved $g^{R}_{p}(t,t')$ conserves the total energy
much better in comparison to the non-Markovian collision
integral with the exponential decay of the memory function. 
It is more natural, however, to describe the transition from the short-time
dynamics to the asymptotic Boltzmann regime 
by using approximate self-consistent solutions of the equation 
for the retarded Green's function. 
For instance, the result~(\ref{z6.42})
shows that the effective quasiparticle damping has a non-Markovian
structure. Furthermore, it depends on the nonequilibrium 
one-particle distribution
functions and involves contributions from initial correlations.
Nevertheless, even if one uses the self-consistent expression for
$g^{R}_p(t,t')$,  
the problem of energy conservation for all time
intervals still remains open since any 
quasiparticle damping function $\gamma^{}_{p}(t,t')$ 
(except for $\gamma^{}_{p}(t,t')=0$) leads
to the energy non-conserving collision integral. 
This is not surprising, because the extra terms in the energy
balance equation are formally beyond the Born approximation
used in the derivation of the kinetic equation. 
Thus, to obtain the collision integrals which 
involve the quasiparticle damping and are consistent
with the energy conservation, one has to 
calculate the $T$-matrices in the full collision
operators~(\ref{z6.6}) and~(\ref{z6.7})  
keeping terms of higher-order in the interaction.     

Concluding this section, we would like to touch upon
another important aspect of the short-time kinetics.
Suppose that the evolution of the system starts from 
the {\em equilibrium state}. Then a kinetic equation
must lead to the obvious result $\partial f^{}_{p}(t)/\partial t$
for all $t>t^{}_0$. In other words, the collision integral must be
equal to zero in thermal equilibrium. It is well known that
the stationary solution of the Markovian Boltzmann equation
corresponds to the equilibrium Fermi or Bose distribution
function for quasiparticles. If, however, the collision integral
takes account of initial correlation and memory effects, 
the question about the stationary solution of the kinetic equation
becomes nontrivial. We have shown that the kinetic equation
with the collision integrals~(\ref{z6.22}) and~(\ref{z6.23}) conserves the
total energy (if $\Gamma^{}_{p}(t,t')=0$ in the memory functions)
and that the initial value of the correlation energy given by
Eq.~(\ref{z6.51}) coincides with the {\em exact result}. Due to this fact,
it is reasonable to expect the kinetic equation to have the
correct stationary solution. Nevertheless, we should show
strictly that the collision integral vanishes in thermal equilibrium,
which is a stronger condition than the conservation of energy.
In Appendix~F we calculate the collision integrals~(\ref{z6.22})
and~(\ref{z6.23}) up to terms of second order in the interaction 
using the equilibrium two-particle correlation
matrix $\chi^{({\rm eq})}_{12}$. The resulting kinetic
equation has the form
\begin{eqnarray}
& &
\hspace*{-35pt}
\frac{\partial}{\partial t}\,f^{}_{p^{}_1}(t)=
-\sum_{p_2 p'_1 p'_2}
\left|\langle p^{}_1 p^{}_2|
\widetilde{V}^{}_{12}|p'_1 p'_2\rangle\right|^2\,
\int_{t^{}_0}^{t} dt'\,
 \cos\!\left[\Delta\omega^{}_{p^{}_1 p^{}_2,p'_1 p'_2}(t,t')\right]
{\cal F}^{}_{p^{}_1p^{}_2,p'_1 p'_2 }\!\left(\{f(t')\}\right)
\nonumber\\[6pt]
& &
{}+
\sum_{p^{}_2 p'_1 p'_2}
\left|\langle p^{}_1 p^{}_2|
\widetilde{V}^{}_{12}|p'_1 p'_2\rangle\right|^2\,
\frac{\sin\!\left[\Delta\omega^{}_{p^{}_1 p^{}_2,p'_1 p'_2}(t,t^{}_0)\right]}
{E^{}_{p^{}_1}+E^{}_{p^{}_2}- E^{}_{p'_1} -E^{}_{p'_2}}\,
{\cal F}^{}_{p^{}_1p^{}_2,p'_1 p'_2 }\!\big(\{f^{(\rm eq)}\}\big),
\label{z6.52}
\end{eqnarray} 
where $E^{}_p$ is the particle energy in thermal equilibrium,
including the Hartree-Fock contribution, and the function ${\cal F}$
is defined as
\begin{equation}
{\cal F}^{}_{p^{}_1p^{}_2,p'_1 p'_2 }\!\left(\{f\}\right)=
f^{}_{p^{}_1} f^{}_{p^{}_2} {\bar f}^{}_{p'_1}{\bar f}^{}_{p'_2}    
- {\bar f}^{}_{p^{}_1} {\bar f}^{}_{p^{}_2}f^{}_{p'_1}f^{}_{p'_2}.
\label{z6.53}
\end{equation}
The first term on the right-hand side of Eq.~(\ref{z6.52})
is the non-Markovian Boltzmann collision integral~(\ref{z6.22}),
where the quasiparticle damping is omitted. The second term
is due to the equilibrium correlations.
In this term $f^{({\rm eq})}_{p}$ is the equilibrium 
one-particle distribution function. 
Since, in equilibrium, 
\begin{equation}
\Delta\omega^{}_{p^{}_1 p^{}_2,p'_1 p'_2}(t,t')=
\left(
E^{}_{p^{}_1}+E^{}_{p^{}_2}- E^{}_{p'_1} -E^{}_{p'_2}
\right)(t-t'),
\label{z6.54}
\end{equation} 
it can easily  be seen that $f^{}_{p^{}_1}=f^{({\rm eq})}_{p^{}_1}$
is a stationary solution of Eq.~(\ref{z6.52}). Physically, this means
that in equilibrium the changes in the one-particle
distribution function caused by collisions and correlations 
exactly cancel each other. Lee {\em et al.\/}~\cite{Lee70} 
were the first to demonstrate this fact for a weakly
interacting low-density classical gas.  
The kinetic equation~(\ref{z6.52}) shows that the same situation
holds in quantum systems. It should also be noted that in the latter
case the Hartree-Fock corrections to the particle energies
have to be included, in order to assure the conservation of energy.        
  
The fact that in equilibrium the  collision and correlation
effects cancel each other is analogous to the 
fluctuation-dissipation theorem
and suggests that correlations and fluctuations are
closely related.    
The interplay between collisions and
correlations becomes particularly clear if one neglects the
Hartree-Fock terms in the memory functions. Then 
both terms on the right-hand side of Eq.~(\ref{z6.52})
can be combined to give
\begin{eqnarray}
& &
\hspace*{-60pt}   
\frac{\partial}{\partial t}\,f^{}_{p^{}_1}(t)=
-\sum_{p^{}_2 p'_1 p'_2}
\left|\langle p^{}_1 p^{}_2|
\widetilde{V}^{}_{12}|p'_1 p'_2\rangle\right|^2\,
\int_{t^{}_0}^{t} dt'\,
 \cos\!\left[\big(\varepsilon^{}_{p^{}_1}+\varepsilon^{}_{p^{}_2} 
-\varepsilon^{}_{p'_1} -\varepsilon^{}_{p'_2}
\big)(t-t')\right]
\nonumber\\[6pt]
& &
\hspace*{90pt}
{}\times\left[
{\cal F}^{}_{p^{}_1p^{}_2,p'_1 p'_2 }\!\left(\{f(t')\}\right)
-{\cal F}^{}_{p^{}_1p^{}_2,p'_1 p'_2 }\!\big(\{f^{(\rm eq)}\}\big)
\right].
\label{z6.55}
\end{eqnarray}
Neglecting here the correlation term we recover the well-known
Levinson equation~\cite{HaugJauho96,Levinson69}.  
The advantage of Eq.~(\ref{z6.55}) in comparison with
the Levinson equation is that it has the right equilibrium
solution. Although the correlation term was introduced
in the equilibrium form, Eq.~(\ref{z6.55}) can be considered
as a rather simple and a reasonable kinetic equation
describing short-time quantum kinetics in the presence
of initial correlations.
      
\setcounter{equation}{0}
\renewcommand{\theequation}{8.\arabic{equation}}

\vspace*{15pt}

\begin{center}
VIII. {\large C}ONCLUSION 
\end{center}

\vspace*{5pt}

We  now summarize the main features of the outlined
approach to short-time quantum kinetics:

1)
The formalism is applicable to a wide range of
correlated initial states described by nonequilibrium 
statistical thermodynamics. 

2)
Initial correlations and the time evolution
are incorporated on the same footing through the  
many-component (``mixed'') Green's functions. 
The essential point is that 
the introduction of the mixed Green's functions  
makes only minor changes in the standard real-time
Green's function method and allows the use of the  Dyson
equation for correlated initial states.
The existence of the  Dyson equation for the mixed Green's
function leads to some exact relations which play a crucial role in
the derivation of  kinetic equations. 
In particular, the generalized Kadanoff-Baym ansatz can be
formulated for short-time processes in the presence of
initial correlations.
 
3) 
The well-known approximations, originally developed
within the standard real-time Green's method, say the
$T$-matrix approximation, are conveniently extended
to short-time kinetics with correlated initial states.    
  
4) 
The method given in this paper is non-perturbative in
external fields due to the self-consistency of the
generalized ansatz for the real-time and cross
correlation functions. This is important when considering
a direct effect of a strong external field on collisions.

\vspace*{10pt}

\begin{center}
  {\large ACKNOWLEDGMENTS} 
\end{center}

\vspace*{5pt}

The authors wish to express  their
appreciation to D.~Kremp, Yu.~A.~Kukharenko, and K.~Morawetz 
for valuable discussions.

\setcounter{equation}{0}
\renewcommand{\theequation}{A.\arabic{equation}}

\vspace*{10pt}

\begin{center}
{\large APPENDIX A:}\\[4pt] 
{\large T}HERMODYNAMIC {\large G}REEN'S {\large F}UNCTIONS
\end{center}

\vspace*{5pt}

Here we briefly discuss some properties of the thermodynamic 
Green's functions in terms of which initial correlations are
described. Owing to the structure of the 
evolution operator~(\ref{2.6}), these properties are in many ways
similar to those of the Matsubara-Green's 
functions~\cite{FetterWalecka71}. 
In particular, the one-particle
thermodynamic Green's function
${\cal G}(1,1')\equiv {\cal G}(r^{}_1x^{}_1,r'_1x'_1)$ 
depends on $x^{}_1$ and $x'_1$ through the difference
$x^{}_1-x'_1$, as follows at once from the cyclic invariance
of the trace in Eq.~(\ref{2.8}). Note also that 
${\cal G}(1,1')$ is defined for $-1\leq (x^{}_1-x'_1)\leq 1$, 
which is a necessary condition for convergence of the trace. 
Clearly this
condition is satisfied if the ``imaginary'' Heisenberg 
picture~(\ref{2.7}) is introduced for $x^{}_0\leq x \leq x^{}_0+1$,
where $x^{}_0$ is an arbitrary parameter. 
Then it is easy to verify that ${\cal G}(1,1')$ has the property
\begin{equation}
{\cal G}(1,1')\Big|^{}_{x^{}_1=x^{}_0+1}=
\eta\,
{\cal G}(1,1')\Big|^{}_{x^{}_1=x^{}_0},
\label{A.1}
\end{equation} 
which is  a generalization of the well-known
Kubo-Martin-Schwinger boundary condition for the equilibrium
Green's functions~\cite{FetterWalecka71}.

Perturbation expansions of the thermodynamic Green's functions can be
constructed by introducing the interaction picture
\begin{equation}
\hat A^{}_I(x)={\rm e}^{x\hat S^{0}}\hat A\, 
{\rm e}^{-x\hat S^{0}},
\label{A.2}
\end{equation} 
where $x^{}_0\leq x\leq x^{}_0+1$, and $\hat S^0$
is the entropy operator for a non-correlated state [cf. Eq.(\ref{2.4})].
The interaction picture evolution
operator is defined as
\begin{equation}
{\cal U}^{}_I(x,x')=
{\rm e}^{x\hat S^0}
{\rm e}^{-(x-x')\hat S}
{\rm e}^{-x'\hat S^0}=
T^c_x\,\exp\left\{
-\int_{x'}^{x} dx''\,\hat S'_I(x'')
\right\}.
\label{A.3}
\end{equation}
Then, using the identity
\begin{equation}
{\rm e}^{-\hat S}=
{\cal U}^{}_I(0,x^{}_0)\,{\rm e}^{-\hat S^0}\,
{\cal U}^{}_I(x^{}_0+1,0),
\label{A.4}
\end{equation}
which follows directly from the definition of ${\cal U}^{}_I(x,x')$,
the one-particle thermodynamic Green's function~(\ref{2.8})
can be written in the form
\begin{equation}
{\cal G}(1,1')=
-\,\frac{
\left\langle
T^c_x
\left(
{\cal U}^{}_I(x^{}_0+1,x^{}_0)\,
\psi^{}_I(1)\,\psi^{\dagger}_I(1')
\right)
\right\rangle^{}_0
}
{
\left\langle
{\cal U}^{}_I(x^{}_0+1,x^{}_0)
\right\rangle^{}_0
},
\label{A.5}
\end{equation}  
where the symbol $\langle \ldots \rangle^{}_0$ stands for
averages calculated with the statistical operator
\begin{equation}
\varrho^0(t^{}_0)=
{\rm e}^{-\hat S^0}\left/
{\rm Tr}\left\{{\rm e}^{-\hat S^0}\right\}.
\right.
\label{A.6}
\end{equation}
Expression~(\ref{A.5}) has the structure typical for a diagram
technique. Since $\hat S^0$ is  bilinear in
the field operators, the statistical operator $\varrho^0$
admits Wick's decomposition and, consequently, 
expectation values of products of the interaction-picture field 
operators factorize into one-particle thermodynamic Green's
functions.  The corresponding Feynman rules 
depend on the specific form of the ``correlation'' term 
$\hat S'$ in the entropy operator.     

In the diagram language, the thermodynamic 
self-energy ${\cal K}$ is introduced as an irreducible part   
of ${\cal G}$.
It can also be defined without reference to diagrams.
To do this, let us consider equations of motion for the
field operators in the Heisenberg picture~(\ref{2.7}). 
They read
\begin{eqnarray}
& &
\frac{\partial}{\partial x^{}_1}\,\psi^{}_{H}(1)
= -\int d1'\,
\lambda^{}_{1}(1,1')\,\psi^{}_{H}(1')
+ \left[
\hat S'_{H}(x_1),\psi^{}_{H}(1)\right],
\nonumber\\[5pt]
& &
\frac{\partial}{\partial x^{}_1}\,\psi^{\dagger}_{H}(1)
= \int d1'\,
\psi^{\dagger}_{H}(1')
\lambda^{}_{1}(1',1)
+ \left[
\hat S'_{H}(x_1),\psi^{\dagger}_{H}(1)\right],
\label{A.8}
\end{eqnarray}
where $d1'=dx'_1\,dr'_1$, and the function $\lambda^{}_{1}(1,1')$
is defined as
\begin{equation}
\lambda^{}_{1}(1,1')=\lambda^{}_{1}(r^{}_1,r'_1)\,
\delta(x^{}_1- x'_1).
\label{A.9}
\end{equation} 
Then, differentiating the Green's function~(\ref{2.8}) with respect to
$x^{}_1$ and~$x'_1$, and using the above equations of motion for the
field operators, we arrive at the Dyson equations~(\ref{2.12}) with
the self-energy 
\begin{eqnarray}
{\cal K}(1,1')&=&
\int d1''\,
\left\langle
T^c_x\left\{
\left[\hat S'_H(x_1),\psi^{}_H(1)\right]
\psi^{\dagger}_{H}(1'')
\right\}
\right\rangle
{\cal G}^{-1}(1'',1')
\nonumber\\[5pt]
{}&=&
\int d1''\,
{\cal G}^{-1}(1,1'')
\left\langle
T^c_x\left\{
\psi^{}_{H}(1'')
\left[\psi^{\dagger}_{H}(1'),\hat S'_H(x'_1)\right]
\right\}
\right\rangle.
\label{A.10}
\end{eqnarray}
These relations can be used to express the self-energy in terms of
higher-order thermodynamic Green's functions. 
If the entropy operator is given by 
Eq.~(\ref{2.4}), we have (integration over repeated arguments is
implied)
\begin{eqnarray}
{\cal K}(1,1')&=&
\eta\,
{\cal V}(12,1''2'')\,
{\cal G}^{(2)}(1''2'',1'''2^+)\,{\cal G}^{-1}(1''',1'),
\nonumber\\[5pt]
{}&=&
\eta\,
{\cal G}^{-1}(1,1''')\,
{\cal G}^{(2)}(1''' 2^-,1''2'')\,
{\cal V}(1''2'',1'2),
\label{A.11}
\end{eqnarray} 
where the amplitude ${\cal V}$ is given by Eq.~(\ref{m4.11}).

\setcounter{equation}{0}
\renewcommand{\theequation}{B.\arabic{equation}}

\vspace*{10pt}

\begin{center}
{\large APPENDIX B:}\\[4pt] 
{\large I}NTERACTION {\large P}ICTURE ON THE {\large E}XTENDED 
{\large C}ONTOUR
\end{center}

\vspace*{5pt}

Let the Heisenberg picture on the contour $\underline{C}$ of Fig.~2
be defined as
\begin{equation}
\underline{{\hat A}}^{}_{\,H}(\xi)=\underline{U}(\xi^{-}_0,\xi)\,
\hat A\,\underline{U}(\xi,\xi^{-}_0),
\label{B.1}
\end{equation}
where the variable $\xi=(t,x)$ specifies a point on $\underline{C}$ and
$\underline{U}(\xi^{}_1,\xi^{}_2)$ is
the evolution operator~(\ref{3.11}). For the contour $\underline{C}$
shown in Fig.~2, the ``effective Hamiltonian''
$\underline{{\hat{\cal H}}}(\xi)$ coincides with $\hat H$ on
the Keldysh part $C$ and with $-i\hat S$ on the parts 
$C'_x$ and $C''_x$. 

The definition~(\ref{B.1}) of the Heisenberg picture on $\underline{C}$
allows us to introduce the path-ordered (``mixed'') Green's functions,
as given by Eqs.~(\ref{3.14}) and~(\ref{3.15}). 
Due to the fact that the evolution over the Keldysh contour $C$ is 
described
by the identity operator $\underline{U}(\xi^{+}_0,\xi^{-}_0)=1$, the mixed
Green's functions coincide, up to a factor, with the thermodynamic
Green's functions discussed in 
Sec.~II and Appendix~A, if all the arguments correspond to
the parts $C^{\prime}_x$ and $C^{\prime\prime}_x$ of $\underline{C}$.
On the other hand, if all the arguments of the mixed Green's 
function correspond to the Keldysh contour, then this function
coincides with the real-time (path-ordered) Green's function.
The latter property follows directly from the cyclic invariance of the
trace in Eqs.~(\ref{3.14}) and~(\ref{3.15}).

To formulate perturbation theory for the mixed Green's
functions defined on the contour $\underline{C}$ of Fig.~2, 
let us introduce the ``interaction'' picture 
through the relation
\begin{equation}
\underline{{\hat A}}^{}_{I}(\xi)=\underline{U}^{0}(\xi^{-}_0,\xi)\,
\hat A\,\underline{U}^0(\xi,\xi^{-}_0).
\label{B.3}
\end{equation}
The evolution operator is 
\begin{equation}
\underline{U}^0(\xi^{}_1,\xi^{}_2)=
T^{}_{\underline{C}}\,\exp\left\{
-i\int_{\xi^{}_2}^{\xi^{}_1}
\underline{{\hat{\cal H}}}^0(\xi)\,d\xi
\right\},
\label{B.4}
\end{equation}
where the unperturbed  ``effective Hamiltonian''
$\underline{{\hat{\cal H}}}^0(\xi)$ coincides with $\hat H^0$
on the Keldysh contour $C$ and with $-i\hat S^0$
on the parts $C'_x$ and $C''_x$ of $\underline{C}$.    
Then an operator in the Heisenberg picture~(\ref{B.1}) can
be written as
\begin{equation}
\underline{{\hat A}}^{}_{\,H}(\xi)=
\underline{U}^{}_I(\xi^{-}_0,\xi)\,
\underline{{\hat A}}^{}_{I}(\xi)\,
\underline{U}^{}_I(\xi,\xi^{-}_0),
\label{B.6}
\end{equation} 
where
\begin{equation}
\underline{U}^{}_I(\xi^{}_1,\xi^{}_2)=
\underline{U}^0(\xi^{-}_0,\xi^{}_1)\,
\underline{U}(\xi^{}_1,\xi^{}_2)\,
\underline{U}^{0}(\xi^{}_2,\xi^{-}_0)
\label{B.7}
\end{equation}
is the interaction picture evolution operator.
This operator is represented by a path-ordered exponent
\begin{equation}
\underline{U}^{}_I(\xi^{}_1,\xi^{}_2)=
T^{}_{\underline{C}}\,\exp\left\{
-i\int_{\xi^{}_2}^{\xi^{}_1}
\underline{{\hat{\cal H}}}^{\prime}_{I}(\xi)\,d\xi
\right\}
\label{B.8}
\end{equation}
with the effective ``interaction Hamiltonian''
\begin{equation}
\underline{{\hat{\cal H}}}^{\prime}_{I}(\xi)=
\left\{
\begin{array}{ll}
\hat H'_{I}(t) &\quad \xi\in C,\\[3pt]
-i\hat S^{\prime}_{I}(x) &\quad \xi\in 
C^{\prime}_x, C^{\prime\prime}_x.
\end{array}
\right.
\label{B.9}
\end{equation} 
Here $\hat H'_{I}(t)$ is an operator in the real-time
interaction picture and $\hat S^{\prime}_{I}(x)$
is the correlation part of the entropy operator in
the thermodynamic interaction picture introduced in
Appendix~A. 
The representation~(\ref{B.8}) can be derived by 
solving the equation 
\begin{equation}
i\,\frac{\partial}{\partial\xi^{}_1}\,
\underline{U}^{}_I(\xi^{}_1,\xi^{}_2)=
\underline{{\hat{\cal H}}}^{\prime}_I(\xi^{}_1)\,
\underline{U}^{}_I(\xi^{}_1,\xi^{}_2),
\label{B.10}
\end{equation}
which follows from Eq.~(\ref{B.7}).

Let us now consider the product of Heisenberg operators
$\underline{{\hat A}}^{}_{\,1H}(\xi^{}_1)\cdots
\underline{{\hat A}}^{}_{\,kH}(\xi^{}_k)$, where the points
$\xi^{}_i$ are arranged in a certain order on the contour 
$\underline{C}$ of Fig.~2.
Making use of Eq.~(\ref{B.6}), 
we can write this product in terms of the
interaction picture operators:
\begin{eqnarray}
& &
\hspace*{-40pt}
\underline{{\hat A}}^{}_{\,1H}(\xi^{}_1)\cdots
\underline{{\hat A}}^{}_{\,kH}(\xi^{}_k)
\nonumber\\[5pt]
& &
{}=
\underline{U}^{}_I(\xi^{-}_0,\xi^{}_1)\,
\underline{{\hat A}}^{}_{\,1I}(\xi^{}_1)\,
\underline{U}^{}_I(\xi^{}_1,\xi^{}_2)\cdots
\underline{U}^{}_I(\xi^{}_{k-1},\xi^{}_k)\,
\underline{{\hat A}}^{}_{\,kI}(\xi^{}_k)\,
\underline{U}^{}_I(\xi^{}_{k},\xi^{-}_0).
\label{B.11}
\end{eqnarray}  
As a next step to perturbation expansions of the mixed Green's
functions, we use the relation
\begin{equation}
{\rm e}^{-\hat S}=
\underline{U}^{}_{I}(\xi^-_0,\xi^{}_{\rm in})\,
{\rm e}^{-\hat S^0}\,
\underline{U}^{}_{I}(\xi^{}_{\rm end},\xi^-_0),
\label{B.12}
\end{equation}
which is nothing but the identity~(\ref{A.4}) since
\begin{equation}
\underline{U}^{}_{I}(\xi^-_0,\xi^{}_{\rm in})=
{\cal U}^{}_I(0,x^{}_0),
\qquad
\underline{U}^{}_{I}(\xi^{}_{\rm end},\xi^-_0)=
{\cal U}^{}_I(x^{}_0+1,0),
\label{B.13}
\end{equation}
as is evident from Fig.~2. Now, combining Eqs.~(\ref{B.11}) 
and~(\ref{B.12}), the one-particle mixed Green's
function~(\ref{3.14}) takes the form
\begin{equation}
\underline{G}(1,1')=
-i\,
\frac{\left\langle
T^{}_{\underline{C}}
\left\{
\exp\left[
-i
\displaystyle{\int_{\underline{C}} d\xi\,
\underline{{\hat{\cal H}}}^{\prime}_I(\xi)
}
\right]
\underline{\psi}^{}_{\,H}(1)\,
\underline{\psi}^{\dagger}_{\,H}(1')
\right\}
\right\rangle^{}_0
}
{
\left\langle
T^{}_{\underline{C}}
\exp\left[
-i
\displaystyle{\int_{\underline{C}} d\xi\,
\underline{{\hat{\cal H}}}^{\prime}_I(\xi)
}
\right]
\right\rangle^{}_0
},
\label{B.14}
\end{equation} 
where averages are calculated over the non-correlated initial
ensemble described by the statistical operator~(\ref{A.6}).

Note that the representation~(\ref{B.14}) is valid for any value of
the parameter $x^{}_0$ in the interval $-1\leq x^{}_0\leq 0$. It
is convenient, however, to take $x^{}_0=-1$ or $x^{}_0=0$.
To illustrate this point, let us take $x^{}_0=0$
(see Fig.~3), as in the main
body of the paper. Then, expanding Eq.~(\ref{B.14}) in terms of 
$\underline{{\hat{\cal H}}}^{\prime}_I(\xi)$ 
and applying Wick's decomposition to each
term in this expansion, one will only obtain the cross Green's
functions~(\ref{3.17}) and~(\ref{3.18}), in which the $x$-arguments
are always  later on $\underline{C}$ than the $t$-arguments.
In particular, the expansion of the real-time component of
$\underline{G}$ can be derived from formula
\begin{equation}
G(1,1')=
-i\,
\frac{
\left\langle
{\cal U}^{}_I(1,0)\,
T^{}_C
\left\{
\exp\left[
-i\displaystyle{
\int_{C} dt\,\hat{H}'_I(t)
}
\right]
\psi^{}_{H}(1)\, \psi^{\dagger}_{H}(1')
\right\}
\right\rangle^{}_0
}{
\left\langle\,
{\cal U}^{}_I(1,0)
\right\rangle^{}_0
},
\label{B.15}
\end{equation}    
where all the operators with the imaginary evolution 
are arranged to the left of operators with the real-time evolution.
For a more general contour $\underline{C}$ shown in Fig.~2, one has to introduce
not only the cross Green's functions~(\ref{3.17}) 
and~(\ref{3.18}), but also two functions
\begin{eqnarray}
& &
{\cal F}^{\more}(1,1')=
{\cal F}^{\more}(r^{}_1t^{}_1,r'_1 x'_1)=
-i\left\langle
\psi^{}_{H}(r^{}_1t^{}_1)\,\psi^{\dagger}_{H}(r'_1x'_1)
\right\rangle,
\nonumber\\[6pt]
& &
{\cal F}^{\less}(1,1')=
{\cal F}^{\less}(r^{}_1 x^{}_1,r'_1t'_1)=
-i\eta
\left\langle
\psi^{}_{H}(r^{\prime}_1 t^{\prime}_1)\,\psi^{\dagger}_{H}(r^{}_1 x^{}_1)
\right\rangle,
\label{B.16}
\end{eqnarray}
which appear when the $x$-argument is taken on the part 
$C^{\prime}_x$ of $\underline{C}$. It should be emphasized, however, that of
the four cross functions, 
${\cal G}^{\moreless}$ and ${\cal F}^{\moreless}$, only two are
independent of each other due to relations
\begin{equation}
\left[ {\cal G}^{\more}(1,1')\right]^*=
-{\cal F}^{\more}(1',\bar{1}),
\qquad
\left[ {\cal G}^{\less}(1,1')\right]^*=
-{\cal F}^{\more}(\bar{1}',1),
\label{B.17}
\end{equation}
where $(\bar{k})=(r^{}_k,-x^{}_k)$.

\setcounter{equation}{0}
\renewcommand{\theequation}{C.\arabic{equation}}

\vspace*{15pt}

\begin{center}
{\large APPENDIX C:}\\[4pt] 
{\large A}N {\large A}LTERNATIVE {\large F}ORM OF {\large G}ENERALIZED 
{\large K}ADANOFF-{\large B}AYM {\large E}QUATIONS
\end{center}

\vspace*{5pt}

We shall discuss  the connection between
our equations~(\ref{m4.24}), (\ref{m4.26}), 
and those derived by
the diagram technique~\cite{Danielewicz84a}.
First of all we note that Eqs.~(\ref{m5.1}) yield the following
expressions for the correlation terms appearing in the
generalized Kadanoff-Baym equations:
\begin{equation}
{\cal K}^{\less}\,{\cal G}^{\more}=
i\,{\cal K}^{\less}\,{\cal G}\,{\cal K}^{\more}\,g^A
+ {\cal K}^{\less}\,{\cal G}^{\more}_{\rm hom},
\qquad
{\cal G}^{\less}\,{\cal K}^{\more}=
i\,g^R\,{\cal K}^{\less}\,{\cal G}\,{\cal K}^{\more}
+{\cal G}^{\less}_{\rm hom}\,{\cal K}^{\more}. 
\label{C.1}
\end{equation}
We now make use of Eqs.~(\ref{m5.4}) to write
\begin{equation}
{\cal G}^{\less}_{\rm hom}{\cal K}^{\more}=
g^R \Sigma^C,
\qquad
{\cal K}^{\less}{\cal G}^{\more}_{\rm hom}=
\Sigma^{}_C g^A,
\label{C.2}
\end{equation}  
where we have introduced the singular self-energies
\begin{eqnarray}
& &
\Sigma^C(t_1,t_2)=
i\delta(t_1-t_0-0)\,
{\cal G}^{\less}(t_0)\,{\cal K}^{\more}(t_2).
\label{C.3}\\[5pt]
& &
\Sigma^{}_C(t_1,t_2)=
-i{\cal K}^{\less}(t_1)\,{\cal G}^{\more}(t_0)\,
\delta(t_2 -t_0-0).
\label{C.4}
\end{eqnarray}
Combining Eqs.~(\ref{C.1}) with~(\ref{C.2}), we see that
Eqs.~(\ref{m4.24}) and~(\ref{m4.26})
can be rewritten in the form
\begin{eqnarray}
& &
\left(g^{-1}_0 -\Sigma^R\right) g^{\moreless}=
\left(\overline{\Sigma}^{\,\moreless} +\Sigma^{}_C\right) g^A,
\label{C.5}\\[5pt]
& &
g^{\moreless}
\left(g^{-1}_0 -\Sigma^A\right)=
g^R
\left(\overline{\Sigma}^{\,\moreless} +\Sigma^{C}\right),
\label{C.6}
\end{eqnarray}
where
\begin{equation}
\overline{\Sigma}^{\,\moreless}=\Sigma^{\moreless} +
i {\cal K}^{\less}\, {\cal G}\, {\cal K}^{\more} 
\label{C.7}
\end{equation}
are the re-normalized real-time components of the self-energy.

Equations~(\ref{C.5}) and~(\ref{C.6}) are similar
to the equations derived by
Danielewicz from diagram expansions of Green's functions~\cite{Remark}.  
In the diagram language, 
$\overline{\Sigma}^{\,\moreless}$ are  
irreducible parts of the real-time Green's function, which begin
and end with an interaction amplitude in the Hamiltonian 
$\hat H$~\cite{Danielewicz84a}.
On the other hand, the singular self-energy $\Sigma^C$ 
($\Sigma^{}_C$) is represented by diagrams which begin
(end) with a correlation matrix and end (begin)
with the interaction amplitude $V$. 
In our approach, such a structure of the 
above self-energies follows directly from equations of 
motion for ${\cal G}^{\moreless}$, Eqs.~(\ref{m4.31})
and~(\ref{m4.34}) , since  
the function ${\cal K}^{\less}$ (${\cal K}^{\more}$)
contains the interaction
amplitude $V$ as a left-side (right-side) multiplier.

\setcounter{equation}{0}
\renewcommand{\theequation}{D.\arabic{equation}}

\vspace*{15pt}

\begin{center}
{\large APPENDIX D:}\\[4pt]
{\large S}ELF-{\large E}NERGY ON THE {\large E}XTENDED 
{\large C}ONTOUR
\end{center}

\vspace*{5pt}

If the entropy
operator~(\ref{2.3}) contains a finite number of the correlation
terms, one  can derive a hierarchy of equations 
for the mixed Green's functions, which is analogous to the
Martin-Schwinger
hierarchy in the standard real-time Green's function
formalism~\cite{Botermans90}. Truncation of this new
hierarchy at some order allows one to formulate reasonable
approximations for the self-energy on the extended contour $\underline{C}$,
as much as truncation of the Martin-Schwinger hierarchy is used for
evaluation of the self-energy on the Keldysh contour $C$. 
Here we will discuss briefly the connection
between the self-energy $\underline{\Sigma}$ and higher-order mixed Green's
functions in the model where the entropy operator includes
only the two-particle correlation term, Eq.~(\ref{2.4}).   
      
The hierarchy of equations  for the mixed Green's functions
follows directly from equations of motion for the field operators
$\underline{\psi}^{}_{\,H}(1)$ and $\underline{\psi}^{\dagger}_{\,H}(1)$
in the Heisenberg picture on the extended contour $\underline{C}$:
\begin{equation}
i\frac{\partial}{\partial \xi^{}_1}\,
\underline{\psi}^{}_{\,H}(1)=
\left[
\underline{\psi}^{}_{\,H}(1),\underline{{\hat{\cal H}}}(\xi^{}_1)
\right],
\qquad
i\frac{\partial}{\partial \xi^{}_1}\,
\underline{\psi}^{\dagger}_{\,H}(1)=
\left[
\underline{\psi}^{\dagger}_{\,H}(1),\underline{{\hat{\cal H}}}(\xi^{}_1)
\right].
\label{D.1} 
\end{equation}
Recalling the definition~(\ref{3.12}) of the ``effective Hamiltonian''
on the contour $\underline{C}$,
the above equations can be transformed to (with integration over repeated
arguments) 
\begin{eqnarray}
& &
\hspace*{-15pt}
\underline{G}^{-1}_{\,0}(1,1')\,\underline{\psi}^{}_{\,H}(1')=
\underline{V}(12,1'2')\,
\underline{\psi}^{\dagger}_{\,H}(2)\,
\underline{\psi}^{}_{\,H}(2')\,
\underline{\psi}^{}_{\,H}(1'),
\nonumber\\[6pt]
& &
\hspace*{-15pt}
\underline{\psi}^{\dagger}_{\,H}(1')\,
\underline{G}^{-1}_{\,0}(1',1)=
\underline{V}(1'2',12)\,
\underline{\psi}^{\dagger}_{\,H}(1')\,
\underline{\psi}^{\dagger}_{\,H}(2')\,
\underline{\psi}^{}_{\,H}(2),
\label{D.2} 
\end{eqnarray}
where the operator $\underline{G}^{-1}_{\,0}$ and
the ``interaction amplitude'' $\underline{V}$
on the contour $\underline{C}$ are  defined by 
Eq.~(\ref{m4.8}) and Eq.~(\ref{m4.9}), respectively.
Now $\underline{G}^{-1}_{\,0}$ operating on the Green's 
function~(\ref{3.14})
from the left and right and then using Eqs.~(\ref{D.2}) results in the 
following equations:
\begin{eqnarray}
& &
\hspace*{-40pt}
\underline{G}^{-1}_{\,0}(1,1'')\,\underline{G}(1'',1')=
\underline{\delta}(1,1')
+i\eta\,
\underline{V}(12,1''2'')\,\underline{G}^{(2)}(1''2'',1'2^+),
\nonumber\\[6pt]
& &
\hspace*{-40pt}
\underline{G}(1,1'')\,\underline{G}^{-1}_{\,0}(1'',1')=
\underline{\delta}(1,1')
{}+i\eta\,
\underline{G}^{\,(2)}(1 2^-,1''2'')\,
\underline{V}(1''2'',1'2).
\label{D.3}  
\end{eqnarray}
These are the first equations of the hierarchy
for the mixed Green's functions on the contour $\underline{C}$,
which is the analogue of the Martin-Schwinger hierarchy 
in the standard 
real-time Green's function formalism~\cite{Botermans90}.
Assuming that $\underline{G}$ has an inverse on the contour
$\underline{C}$, Eqs.~(\ref{D.3}) can be written as Dyson
equations~(\ref{m4.12}) and~(\ref{m4.13}) with the matrix
self-energy $\underline{\Sigma}$ given by Eq.~(\ref{m4.14}).

We emphasize that it has been possible to 
express the self-energy $\underline{\Sigma}$ in terms of the 
two-particle mixed Green's
function $\underline{G}^{(2)}$  due to the structure of the entropy 
operator~(\ref{2.4}). For a more general form of the
entropy operator, the self-energy $\underline{\Sigma}$ 
will depend on higher-order mixed Green's functions.

\vspace*{15pt}

\setcounter{equation}{0}
\renewcommand{\theequation}{E.\arabic{equation}}

\begin{center}
{\large APPENDIX E:}\\[4pt]
{\large T}HE {\large T}WO-{\large P}ARTICLE {\large T}IME 
{\large C}ORRELATION {\large M}ATRIX
\end{center}

\vspace*{5pt}

To calculate the time correlation matrix~(\ref{z6.8}), we use the
ansatz~(\ref{z6.11}) for the cross Green's functions and find that
\begin{equation}
C^{}_{12}(t,t')=g^{R}_{12}(t,t^{}_0)\,
{\cal G}^{\less}_{12}(t^{}_0)\,\widetilde{\cal T}^{}_{12}\,
{\cal G}^{\more}_{12}(t^{}_0)\,g^{A}_{12}(t^{}_0,t').
\label{E.1}
\end{equation}
Let us now turn to Eqs.~(\ref{z5.14}) for the thermodynamic $T$-matrix.
Recalling Eq.~(\ref{z5.16x}) and the boundary conditions~(\ref{3.19}), 
we write
\begin{equation}
{\cal G}^{\less}_{12}(t^{}_0)\,\widetilde{\cal T}^{}_{12}\,
{\cal G}^{\more}_{12}(t^{}_0)=
\int_{0}^1 dx\,dx'\,
{\cal G}^{}_{12}(0,x)\,
 \widetilde{\cal T}^{}_{12}(x,x')\,
{\cal G}^{}_{12}(x',0),
\label{E.3}
\end{equation} 
where we have introduced the notation
\begin{equation}
{\cal G}^{}_{12}(x,x')=
{\cal G}^{}_{1}(x,x')\,
{\cal G}^{}_{2}(x,x').
\label{E.4}
\end{equation}
We now wish to show that the right-hand side of Eq.~(\ref{E.3})
can be expressed in terms of the two-particle thermodynamic
Green's function ${\cal G}^{(2)}$ [see Eq.~(\ref{2.9})].
To that end we use Eq.~(\ref{z5.4}) which, when written
for the thermodynamic component of $\underline{G}^{(2)}$,
reads
\begin{eqnarray}
& &
\hspace*{-40pt}
{\cal G}^{(2)}(12,1'2')=
\Big({\cal G}(1,1')\,{\cal G}(2,2')\Big)^{}_{\rm ex}
\nonumber\\[6pt]
& &
\hspace*{40pt}
{}+{\cal G}(1,1'')\, {\cal G}(2,2'')\,
\widetilde{\cal T}(1''2'',1'''2''')\,
{\cal G}(1''',1')\,{\cal G}(2''',2').
\label{E.5}
\end{eqnarray}
It is convenient here to go to the matrix notation with respect to
the single-particle quantum numbers. Defining 
the matrix ${\cal G}^{(2)}_{12}(x^{}_1 x^{}_2,x'_1 x'_2)$ by
\begin{equation}
{\cal G}^{(2)}(12,1'2')=
\langle r^{}_1 r^{}_2|\,{\cal G}^{(2)}_{12}(x^{}_1 x^{}_2,x'_1 x'_2)
|r'_1 r'_2\rangle,
\label{E.6}
\end{equation}
we have
\begin{eqnarray}
& &
\hspace*{-30pt}
{\cal G}^{(2)}_{12}(x^{}_1 x^{}_2,x'_1 x'_2)=
\Big({\cal G}^{}_1(x^{}_1,x'_1)\,{\cal G}^{}_2(x^{}_2,x'_2)
\Big)^{}_{\rm ex}
\nonumber\\[6pt]
& &
{}+
\int_{0}^1 dx''\,dx'''\,
{\cal G}^{}_1(x^{}_1,x'')\,{\cal G}^{}_2(x^{}_2,x'')\,
\widetilde{\cal T}^{}_{12}(x'',x''')\,
{\cal G}^{}_1(x''',x'_1)\,{\cal G}^{}_2(x''',x'_2).
\label{E.7}
\end{eqnarray} 
A comparison with Eq.~(\ref{E.3}) shows that
\begin{equation}
{\cal G}^{\less}_{12}(t^{}_0)\,\widetilde{\cal T}^{}_{12}\,
{\cal G}^{\more}_{12}(t^{}_0)=
\lim^{}_{x\to 0\atop{x'\to 0}}
\left\{
{\cal G}^{(2)}_{12}(xx,x'x') -
\Big({\cal G}^{}_1(x,x')\, {\cal G}^{}_2(x,x')\Big)^{}_{\rm ex}
\right\}.
\label{E.8}
\end{equation}
Using the obvious relation
\begin{eqnarray}
& &
\hspace*{-40pt}
\langle r^{}_1 r^{}_2|\,{\cal G}^{(2)}_{12}(xx,x'x')
|r'_1 r'_2\rangle=
\theta(x-x')\,
\langle \psi^{}_H(r^{}_1 x)\psi^{}_H(r^{}_2 x)
\psi^{\dagger}_H(r'_2 x')\psi^{\dagger}_H(r'_1 x')\rangle
\nonumber\\[6pt]
& &
\hspace*{40pt}
{}+\theta(x'-x)\, 
\langle
\psi^{\dagger}_H(r'_2 x')\psi^{\dagger}_H(r'_1 x')
\psi^{}_H(r^{}_1 x)\psi^{}_H(r^{}_2 x)
\rangle,
\label{E.9}
\end{eqnarray}
it is easy to verify that the order of the limits in 
Eq.~(\ref{E.8}) is of no significance, as it should be.
In both cases we obtain the same result 
\begin{equation}
{\cal G}^{\less}_{12}(t^{}_0)\,\widetilde{\cal T}^{}_{12}\,
{\cal G}^{\more}_{12}(t^{}_0)=
\chi^{}_{12}(t^{}_0),
\label{E.10}
\end{equation} 
where $\chi^{}_{12}(t^{}_0)$
is the initial two-particle correlation matrix~(\ref{z6.14}).
Relations~(\ref{E.1}) and~(\ref{E.10}) complete the derivation
of Eq.~(\ref{z6.13}).

\vspace*{15pt}

\setcounter{equation}{0}
\renewcommand{\theequation}{F.\arabic{equation}}

\begin{center}
{\large APPENDIX F:}\\[4pt]
{\large D}ERIVATION OF THE 
{\Large K}INETIC {\Large E}QUATION WITH
{\large E}QUILIBRIUM  {\large C}ORRELATIONS
\end{center}

\vspace*{5pt}

The first (Boltzmann) term on the right-hand side of Eq.~(\ref{z6.52})
follows immediately from Eq.~(\ref{z6.22}) if we take the
memory function $W^{(B)}$ in the form~(\ref{z6.43}) 
with $\Gamma^{}_p=0$. Thus it remains to evaluate the correlation
term given by Eq.~(\ref{z6.23}). Since in the case of
equilibrium correlations the matrix  $C^{}_{12}(t,t')$ is proportional
to $V^{}_{12}$ and we wish to obtain $I^{(C)}$ 
to order $V^{2}_{12}$, we can restrict
our discussion to the first term in Eq.~(\ref{z6.23}).  
We start by evaluating the equilibrium correlation matrix
\begin{equation}
\langle p^{}_1 p^{}_2|\chi^{(\rm eq)}_{12}|p'_1 p'_2\rangle=
\langle a^{\dagger}_{p'_2} a^{\dagger}_{p'_1} 
a^{}_{p^{}_1} a^{}_{p^{}_2}\rangle^{}_{\rm eq}
- f^{({\rm eq})}_{p^{}_1} f^{({\rm eq})}_{p^{}_2}
\left(\delta^{}_{p^{}_1p'_1}  \delta^{}_{p^{}_2 p'_2}
+\eta \delta^{}_{p^{}_1p'_2} \delta^{}_{p^{}_2 p'_1} 
\right),
\label{F.0}
\end{equation} 
where the symbol $\langle \cdots\rangle^{}_{\rm eq}$
stands for averages calculated with the statistical
operator~(\ref{1.1}), and 
$f^{({\rm eq})}_p=\langle a^{\dagger}_{p} a^{}_{p}\rangle^{}_{\rm eq}$
is the equilibrium one-particle distribution function.  
The second-quantized Hamiltonian in momentum representation is given by
\begin{equation}
\hat H= \hat H^0 +\hat H'=
\sum_{p^{}_1}
\varepsilon^{}_{p^{}_1} a^{\dagger}_{p^{}_1} a^{}_{p^{}_1}
+ {1\over2}\sum_{p^{}_1 p^{}_2 p'_1 p'_2}
 \langle p^{}_1 p^{}_2|{V}^{}_{12}|p'_1 p'_2\rangle\,
a^{\dagger}_{p'_2} a^{\dagger}_{p'_1} a^{}_{p^{}_1} a^{}_{p^{}_2}.
\label{F.1}
\end{equation}
Recall that we need to evaluate the correlation matrix to first order
in the interaction. This can be done by elementary methods by
expanding the statistical operator~(\ref{1.1}) in $\hat H'$.
It is convenient, however, to redefine the unperturbed 
Hamiltonian by taking the Hartree-Fock term into the
particle energies. Then, instead of Eq.~(\ref{1.1}) we now have
\begin{equation}
\varrho^{}_{\rm eq}=
{\rm e}^{-\beta(\hat{\cal H}^{0}+\hat{\cal H}')}
\left/
{\rm Tr}\,{\rm e}^{-\beta(\hat{\cal H}^{0}+\hat{\cal H}')},
\right.
\label{F.2}
\end{equation}
where we have defined
\begin{eqnarray}
& &
\hat{\cal H}^{0}=
\sum_{p^{}_1}
(E^{}_{p^{}_1}-\mu)\, a^{\dagger}_{p^{}_1} a^{}_{p^{}_1},
\label{F.3}\\[6pt]
& &
\hat{\cal H}'=
{1\over4}\sum_{p^{}_1 p^{}_2 p'_1 p'_2}
 \langle p^{}_1 p^{}_2|\widetilde{V}^{}_{12}|p'_1 p'_2\rangle\,
a^{\dagger}_{p'_2} a^{\dagger}_{p'_1} a^{}_{p^{}_1} a^{}_{p^{}_2}
- \sum^{}_{p^{}_1} \Sigma^{{\rm HF}}_{p^{}_1}\, 
a^{\dagger}_{p^{}_1} a^{}_{p^{}_1}.
\label{F.4}
\end{eqnarray}
The re-normalized particle energies  are given by
\begin{equation}
E^{}_{p^{}_1}=\varepsilon^{}_{p^{}_1}+  \Sigma^{{\rm HF}}_{p^{}_1}
=\varepsilon^{}_{p^{}_1}+  
\sum^{}_{p^{}_2}
 \langle p^{}_1 p^{}_2|\widetilde{V}^{}_{12}|p^{}_1 p^{}_2\rangle\,  
f^{({\rm eq})}_{p^{}_2}.
\label{F.5}
\end{equation}
Up to terms linear in ${\cal H}'$ Eq.~(\ref{F.2}) reads
\begin{equation}
\varrho^{}_{\rm eq}=
\left[1-
\int_{0}^{\beta} d\lambda\,
{\rm e}^{-\lambda \hat{\cal H}^0}
\left(
\hat{\cal H}' -\langle\hat{\cal H}'\rangle^{}_0 
\right){\rm e}^{\lambda \hat{\cal H}^0}
\right]\varrho^0_{\rm eq},
\label{F.6}
\end{equation} 
where the statistical operator
\begin{equation}
\varrho^{0}_{\rm eq}=
{\rm e}^{-\beta \hat{\cal H}^{0}}
\left/
{\rm Tr}\,{\rm e}^{-\beta\hat{\cal H}^{0}}
\right.
\label{F.7}
\end{equation}
describes the ideal quantum gas with the quasiparticle 
energies~(\ref{F.5}). The symbol $\langle\cdots\rangle^{}_0$
means the average with $\varrho^0_{\rm eq}$. 
Making use of Eq.~(\ref{F.6}), it is easy to see  
that to first order in the interaction the one-particle
distribution can be replaced by the Fermi or Bose distribution
\begin{equation}
f^{({\rm eq})}_{p}=\left[{\rm e}^{\beta(E^{}_{p}-\mu)}-\eta\right]^{-1}.
\label{F.8}
\end{equation} 
We now can evaluate the average value in Eq.~(\ref{F.0})
with the aid of Eq.~(\ref{F.6}). Using the relations
\begin{equation} 
{\rm e}^{\lambda \hat{\cal H}^0} a^{}_{p}{\rm e}^{-\lambda \hat{\cal H}^0} 
={\rm e}^{-\lambda(E^{}_p-\mu)} a^{}_{p} ,
\qquad
{\rm e}^{\lambda \hat{\cal H}^0} a^{\dagger}_{p} 
{\rm e}^{-\lambda \hat{\cal H}^0}
={\rm e}^{\lambda(E^{}_p-\mu)} a^{\dagger}_{p},
\label{F.8a}
\end{equation}
we find that
\begin{equation}
\langle p^{}_1 p^{}_2|\chi^{(\rm eq)}_{12}|p'_1 p'_2\rangle=
\frac{1}{E^{}_{p^{}_1 p^{}_2} - E^{}_{p'_1 p'_2}}
\left({\rm e}^{-\beta\big(E^{}_{p^{}_1 p^{}_2} - E^{}_{p'_1 p'_2}\big)}
-1\right) 
\left\langle a^{\dagger}_{p'_2} a^{\dagger}_{p'_1} 
a^{}_{p^{}_1} a^{}_{p^{}_2} \hat{\cal H}'\right\rangle^{(c)}_{0},
\label{F.9}
\end{equation}
where $E^{}_{p^{}_1 p^{}_2}=E^{}_{p^{}_1} -E^{}_{p^{}_2}$.
In the above expression the superscript $(c)$ shows that only
the connected part of the average must be taken. This means that,
in applying Wick's theorem, all the creation
and annihilation operators in ${\cal H}'$ must be contracted with 
the ``free'' operators. Further manipulations are very simple.
After some algebra we find
\begin{equation}
\langle p^{}_1 p^{}_2|\chi^{(\rm eq)}_{12}|p'_1 p'_2\rangle=
\frac{{\cal F}^{}_{p^{}_1p^{}_2,p'_1 p'_2 }\!\left(\{f^{({\rm eq})}\}\right)}
{E^{}_{p^{}_1 p^{}_2} - E^{}_{p'_1 p'_2}}\,
\langle p^{}_1 p^{}_2|
\widetilde{V}^{}_{12}|p'_1 p'_2\rangle,
\label{F.10}
\end{equation}
where the function ${\cal F}(\{f\})$ is defined by Eq.~(\ref{z6.53}).
In writing Eq.~(\ref{F.10}), the temperature dependent factors
have been eliminated by means of Eq.~(\ref{F.8}).

Now we substitute the expression~(\ref{F.10}) into Eq.~(\ref{z6.13})
and take the retarded and advanced Green's functions in the 
form~(\ref{z6.40}) with $\Gamma^{}_p=0$. This gives us the
time-correlation function $C^{}_{12}(t,t)$ which is to be used
for evaluating the first term in the collision integral~(\ref{z6.23}),
i.e., the correlation contribution into the kinetic
equation~(\ref{z6.52}).

\newpage
\thispagestyle{empty}

\begin{center}
{\Large\bf FIGURES}
\end{center}

\begin{center}
\unitlength=3pt  
\special{em:linewidth 0.4pt}
\linethickness{0.4pt}
\hspace*{-150pt}
\begin{picture}(200.00,70.00)
\put(80.00,25.00){\vector(1,0){80.00}}
\put(90.00,28.00){\line(1,0){45.00}}
\put(135.00,28.00){\line(1,0){5.00}}
\put(110.00,22.00){\vector(1,0){00.00}}
\put(140.00,22.00){\line(-1,0){40.00}}
\put(100.00,22.00){\line(-1,0){10.00}}
\bezier{32}(140.00,22.00)(142.00,22.00)(142.00,25.00)
\bezier{32}(142.00,25.00)(142.00,28.00)(140.00,28.00)
\put(90.00,22.00){\circle*{1.40}}
\put(90.00,28.00){\circle*{1.40}}
\put(90.00,32.00){\makebox(0,0)[cb]{\large$t^{}_0$}}
\put(123.00,28.00){\vector(-1,0){00.00}}
\put(123.00,31.00){\makebox(0,0)[cb]{\large$C^-$}}
\put(113.00,20.00){\makebox(0,0)[lt]{\large$C^+$}}
\put(142.00,21.00){\makebox(0,0)[lt]
{$t^{}_{\rm max}$}}
\put(160.00,27.00){\makebox(0,0)[rb]{\large$t$}}
\end{picture}

\hspace*{-100pt}
{\large\sf Figure~1}
\end{center}

\vspace*{50pt}

\begin{center}
\unitlength=3pt  
\special{em:linewidth 0.4pt}
\linethickness{0.4pt}
\hspace*{-100pt}
\begin{picture}(200.00,80.00)
\put(60.00,35.00){\vector(1,0){105.00}} 
\put(70.00,4.00){\vector(0,1){75.00}} 
\put(67.00,77.00){\makebox(0,0)[cb]{\large$x$}}
\put(165.00,37.00){\makebox(0,0)[rb]{\large$t$}}
\put(90.00,37.00){\line(1,0){45.00}}
\put(135.00,37.00){\line(1,0){5.00}}
\put(140.00,33.00){\line(-1,0){40.00}}
\put(100.00,33.00){\line(-1,0){10.00}}
\bezier{32}(140.00,33.00)(142.00,33.00)(142.00,35.00)
\bezier{32}(142.00,35.00)(142.00,37.00)(140.00,37.00)
\put(90.00,33.00){\line(0,-1){25.00}}
\put(90.00,37.00){\line(0,1){38.00}}
\put(128.00,33.00){\vector(1,0){00.00}}
\put(112.00,37.00){\vector(-1,0){00.00}}
\put(90.05,21.00){\vector(0,1){0.00}}
\put(90.05,57.00){\vector(0,1){0.00}}
\put(93.00,5.00){\makebox(0,0)[lb]{\large$\xi^{}_{\rm in}
=(t_0,x_0)$}}
\put(93.00,73.00){\makebox(0,0)[lb]{\large$\xi^{}_{\rm end}
=(t_0,x_0+1)$}}
\put(93.00,40.00){\makebox(0,0)[lb]{\large$\xi^-_0$}}
\put(93.00,27.00){\makebox(0,0)[lb]{\large$\xi^+_0$}}
\put(90.05,37.05){\circle*{1.40}}
\put(90.05,75.05){\circle*{1.40}}
\put(90.05,33.05){\circle*{1.40}}
\put(90.05,8.05){\circle*{1.40}}
\put(115.00,44.00){\makebox(0,0)[lt]{\large$C^-$}}
\put(130.00,31.00){\makebox(0,0)[lt]{\large$C^+$}}
\put(84.00,63.00){\makebox(0,0)[cb]{\large$C^{\prime\prime}_x$}}
\put(84.00,17.05){\makebox(0,0)[cb]{\large$C^{\prime}_x$}}
\end{picture}

\vspace*{40pt}

\hspace*{-100pt}
{\large\sf Figure~2}
\end{center}

\newpage
\thispagestyle{empty}

\vspace*{2cm}

\begin{center}
\unitlength=3pt  
\special{em:linewidth 0.4pt}
\linethickness{0.4pt}
\hspace*{-100pt}
\begin{picture}(200.00,70.00)
\put(60.00,35.00){\vector(1,0){100.00}} 
\put(160.00,37.00){\makebox(0,0)[rb]{\large$t$}}
\put(70.00,20.00){\vector(0,1){50.00}} 
\put(67.00,67.00){\makebox(0,0)[cb]{\large$x$}}
\put(90.00,37.00){\line(1,0){45.00}}
\put(135.00,37.00){\line(1,0){5.00}}
\put(140.00,33.00){\line(-1,0){40.00}}
\put(100.00,33.00){\line(-1,0){10.00}}
\bezier{32}(140.00,33.00)(142.00,33.00)(142.00,35.00)
\bezier{32}(142.00,35.00)(142.00,37.00)(140.00,37.00)
\put(90.00,37.00){\line(0,1){30.00}}
\put(128.00,33.00){\vector(1,0){00.00}}
\put(112.00,37.00){\vector(-1,0){00.00}}
\put(90.05,57.00){\vector(0,1){0.00}}
\put(92.00,66.00){\makebox(0,0)[lb]{\large$\xi^{}_{\rm end}
=(t_0,1)$}}
\put(92.00,40.00){\makebox(0,0)[lb]{\large$\xi^{}_0$}}
\put(92.00,27.00){\makebox(0,0)[lb]{\large$\xi^{}_{\rm in}$}}
\put(90.05,37.05){\circle*{1.40}}
\put(90.05,67.05){\circle*{1.40}}
\put(90.05,33.05){\circle*{1.40}}
\put(115.00,43.00){\makebox(0,0)[lt]{\large$C^-$}}
\put(130.00,31.00){\makebox(0,0)[lt]{\large$C^+$}}
\put(85.00,54.00){\makebox(0,0)[cb]{\large$C^{}_x$}}
\end{picture}

\hspace*{-150pt}
{\large\sf Figure~3}
\end{center}

\newpage
\thispagestyle{empty}

\begin{center}
{\Large\bf FIGURE CAPTIONS}
\end{center}

\vspace*{10pt}

{\large\bf Figure~1}:

The Keldysh contour $C$ with
the lower (chronological) branch $C^+$ and the upper
(anti-chronological) branch $C^{-}$

\vspace*{20pt}

{\large\bf Figure~2}:

The extended  contour $\underline C$ 
with the real-time evolution
on the Keldysh contour $C$ and the ``imaginary''
evolution on the parts $C^{\prime}_x$ and $C^{\prime\prime}_x$.
The parameter $x_0$ 
satisfies  $-1\leq x_0\leq 0$.

\vspace*{20pt}

{\large\bf Figure~3}:

A special case ($x^{}_0=0$) of the contour shown in Fig.~2

\end{document}